\documentclass[11pt,a4paper]{article}
\usepackage{jcappub}

\usepackage{bm}
\usepackage{epsfig}
\usepackage{graphicx}
\usepackage{amssymb}
\usepackage{color}
\usepackage{subfigure}

\renewcommand\({\left(}
\renewcommand\){\right)}

\renewcommand\({\left(}
\renewcommand\){\right)}

\newcommand{\be}{\begin{equation}}
\newcommand{\ee}{\end{equation}}
\newcommand{\bea}{\begin{eqnarray}}
\newcommand{\eea}{\end{eqnarray}}

\newcommand{\mgg}{{m^2_\gamma}}

\newcommand{\muu}{m}

\begin{document}



\subheader{\hfill MPP-2014-33}

\title{Atlas of solar hidden photon emission}

\author{Javier~Redondo}

\affiliation{Departamento de F\'isica Te\'orica, Universidad de Zaragoza,\\ Pedro Cerbuna 12, E-50009, Zaragoza, Espa\~na, and }

\affiliation{Max-Planck-Institut f\"ur Physik, 
(Werner-Heisenberg-Institut)\\
F\"ohringer Ring 6,  80805 M\"unchen, Germany}

\emailAdd{redondo@mpp.mpg.de}

\abstract{
Hidden photons, gauge bosons of a U(1) symmetry of a hidden sector, can constitute the dark matter of the universe and a smoking gun for large volume compactifications of string theory. 
In the sub-eV mass range, a possible discovery experiment consists on searching the copious flux of these particles emitted from the Sun in a helioscope setup \`a la Sikivie.  
In this paper, we compute the flux of transversely polarised HPs from the Sun, a necessary ingredient for interpreting such experiments. 
We provide a detailed exposition of photon-HP oscillations in inhomogenous media,  with special focus on resonance oscillations, which play a leading role in many cases. 
The region of the Sun emitting HPs resonantly is a thin spherical shell for which we justify an averaged-emission formula and which implies a distinctive morphology of the angular distribution of HPs on Earth in many cases. 
Low mass HPs with energies in the visible and IR have resonances very close to the photosphere where the solar plasma is not fully ionised and requires building a detailed model of solar refraction and absorption. 
We present results for a broad range of HP masses (from 0-1 keV) and energies (from the IR to the X-ray range), the most complete atlas of solar HP emission to date. 
}

\maketitle

\section{Introduction}                        \label{sec:introduction}

The existence of hidden photons (HP)~\cite{Okun:1982xi}-- hypothetical vector particles that mix kinetically very weakly with ordinary photons~\cite{Holdom:1985ag,Foot:1991kb,Foot:1991bp}-- has attracted recently much attention~\cite{Jaeckel:2013ija} in the low energy frontier of particle physics~\cite{Jaeckel:2010ni}. Phenomenologically, they give rise to a number of puzzling (and yet unobserved) phenomena: distortions of the cosmic microwave background spectrum~\cite{Georgi:1983sy,Axenides:1983pd,Nordberg:1998wn,Mirizzi:2009iz,Jaeckel:2008fi} and radio sources~\cite{Lobanov:2012pt}, deviations of the Coulomb law~\cite{Bartlett:1970js,Williams:1971ms} and distortions of planetary magnetic fields, atomic level shifts~\cite{Jaeckel:2010xx} and light-shining through walls~\cite{Ahlers:2007rd,Ahlers:2007qf,Ehret:2010mh,Redondo:2010dp}. HPs can be also produced in the early universe contributing to the dark radiation~\cite{Jaeckel:2008fi,Vogel:2013raa} or dark matter (DM) of the universe~\cite{Pospelov:2008jk,Redondo:2008ec,Nelson:2011sf,Arias:2012az,Horns:2012jf,Jaeckel:2013sqa,Jaeckel:2013eha}, in which case they can be searched in dark matter direct detection experiments~\cite{An:2014twa} or through a small relic background of photons, which they produce because they are (slowly) decaying dark matter~\cite{Pospelov:2008jk,Redondo:2008ec}. When the HP mass is above twice the electron mass, their fast decay into two electrons precludes them from being directly accessible DM candidates but they have become the prototypical force mediator in particle models where DM is self-interacting~\cite{ArkaniHamed:2008qn}, which has triggered enormous attention in the intensity frontier of particle physics~\cite{Essig:2013lka}. 
These particles appear as gauge bosons of new U(1) symmetries, which are the simplest extensions of the standard model of particle physics and appear copiously in its most ambitious extension, string theory~\cite{Dienes:1996zr,Lukas:1999nh,Abel:2003ue,Blumenhagen:2005ga,Abel:2006qt,Abel:2008ai,Goodsell:2009pi,Goodsell:2009xc,Goodsell:2010ie,Heckman:2010fh,Bullimore:2010aj,Cicoli:2011yh,Goodsell:2011wn,Cicoli:2012sz}. 

\begin{figure}[b]
\begin{center}
\includegraphics{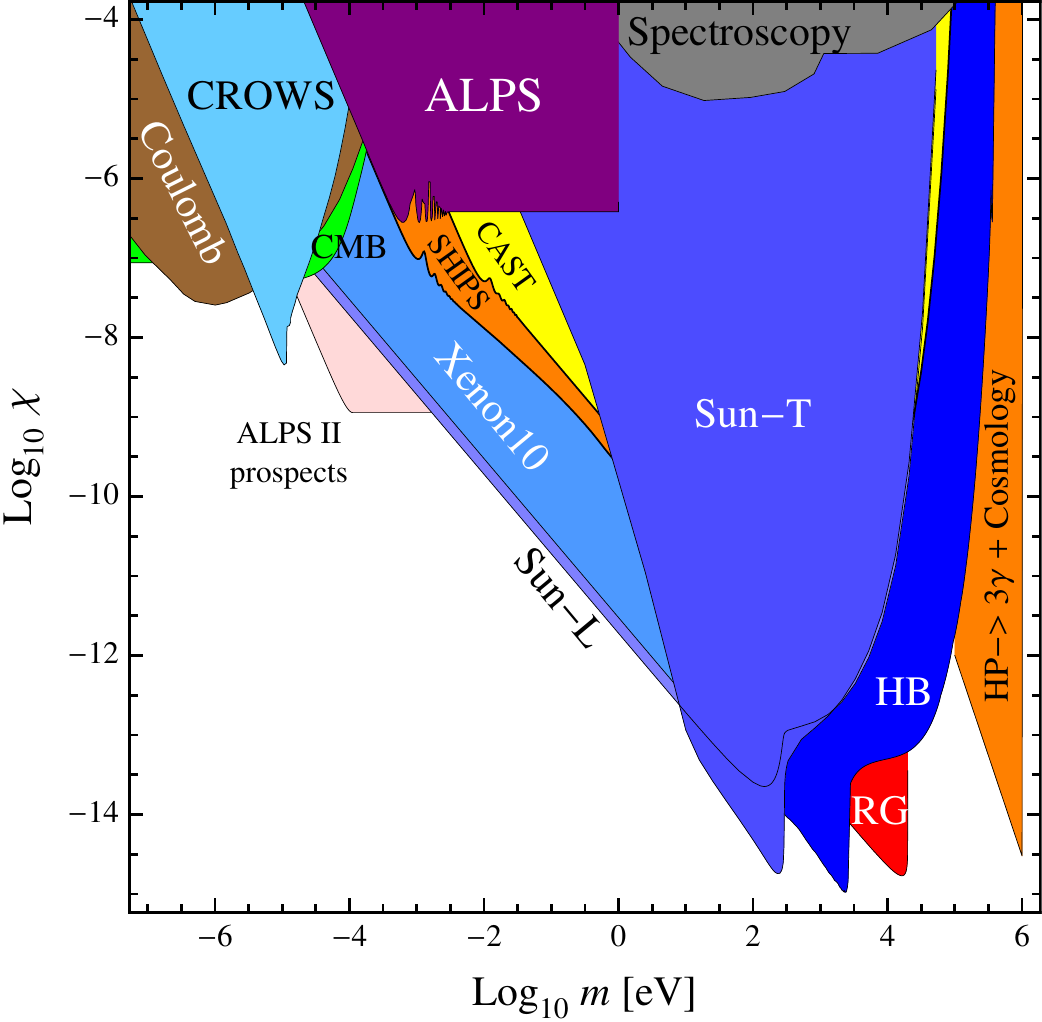}
\caption{Constraints on the kinetic mixing, $\chi$, of hypothetical hidden photons with ordinary photons, as a function of the hidden photon mass. References are CMB~\cite{Jaeckel:2008fi,Mirizzi:2009iz}, Coulomb~\cite{Williams:1971ms}, ALPS~\cite{Ehret:2010mh}, Spectroscopy~\cite{Jaeckel:2010xx}, Sun-T, Sun-L, HB, RG~\cite{An:2013yfc,Redondo:2013lna,Vinyoles:2015aba}, CAST~\cite{Redondo:2008aa}, Xenon10~\cite{An:2013yua}, SHIPS~\cite{newSHIPS} and HP DM decay and cosmology~\cite{Redondo:2008ec}. Also shown are the prospects for the ALPS-IIb experiment~\cite{Bahre:2013ywa}. }
\label{fig:constraints}
\end{center}
\end{figure}

In a huge region of parameter space, the most powerful laboratory to constrain the HP properties is the Sun, see exclusion regions labelled Sun-L and Sun-T in Fig. \ref{fig:constraints}.   
Photons inside of the Sun can oscillate into HPs  and leave it unimpeded, draining energy very efficiently. 
Since this energy must come from nuclear reactions, the temperature of the solar core tends to be hotter in models with HP emission, with the consequently higher neutrino flux from nuclear reactions and distorted sound speed profile.  The agreement between solar models and the observations of the flux of Boron neutrinos can be used to constrain the emission of low mass bosons~\cite{Gondolo:2008dd} and in particular the HP and its parameters~\cite{Redondo:2013lna}. Even stronger constraints arise from a recent global fit of helioseismology data and neutrino fluxes with realistic solar models perturbed by HP emission~\cite{Vinyoles:2015aba}. It is intriguing that constraints from other stelar systems, which are stronger for axions, are not as strong for small mass HPs~\cite{Redondo:2013lna,Pospelov:2008jk,An:2013yfc}. 

Very powerful constraints come from trying to detect the flux of solar HPs in Earth-bound experiments, which we generically call helioscopes. 
In the sub-keV mass range, the emission of HPs proceeds mostly through resonant oscillations~\cite{Redondo:2008aa} and is dominated by longitudinally polarized HPs~\cite{An:2013yfc,Redondo:2013lna}. 
The detection on Earth of longitudinal HPs can be done by measuring the low-energy ionization events in Dark Matter detectors such as XENON10, technique with which one obtains the most stringent laboratory constraints on HPs in this mass range~\cite{An:2013yua}. 

The detection of transversely polarized HPs can be devised in the same way, first proposed and performed with the HPGe detector~\cite{Horvat:2012yv}. But a more efficient technique is 
to exploit the phenomenon of HP$\leftrightarrow$photon  oscillations in vacuum~\cite{Goncharenko:1993jw}, similarly to the way one looks for solar axions in the original helioscope experiment proposed by Sikivie~\cite{Sikivie:1983ip}. 
Indeed, the results of the first axion helioscope by the BFRT collaboration~\cite{Lazarus:1992ry} were used to set constraints on the flux of solar HPs in the keV range~\cite{Popov:1999}. Also an experiment in the UV was performed~\cite{Popov:1999,Baillon}, although their results were not published.
Later, a more detailed calculation of the keV range flux~\cite{Redondo:2008aa} was used  
to benefit from the constraints of the more powerful axion helioscope to date, the CERN Axion Solar Telescope CAST~\cite{Zioutas:2004hi,Andriamonje:2007ew}. 

However, it was noted that low mass transverse HPs are more efficiently produced near the surface and thus the flux peaks at lower energies~\cite{Redondo:2008aa} where CAST was not optimal\footnote{Indeed, this fact was already implicit in the experimental sensitivities presented in~\cite{Popov:1999} where the low-energy searches appear more advantageous at the lowest energies.  }. 
New experiments were then proposed to maximize the chances of a discovery~\cite{Gninenko:2008pz,Troitsky:2011rx}. 
The CAST~\cite{oai:arXiv.org:0809.4581} and the SUMICO~\cite{Mizumoto:2013jy,Inoue:2013jpa} collaborations performed dedicated runs at low energies and the dedicated SHIPS experiment~\cite{SHIPS} releases these days their first results~\cite{SHIPS,newSHIPS}. 
The experiments consist on long light-tight vacuum pipes that track the Sun, where HPs from the Sun can oscillate into photons that can be detected at the far side. 
As detectors, they use photomultipliers sensitive to photon frequencies in the visible range.  
In this paper we calculate the solar flux of transversely polarised HPs at low energies, focusing on visible frequencies $\omega\sim 1.5-3.5$ eV. This ingredient is necessary to interpret the null searches of CAST,  SUMICO and SHIPS and to estimate the size of plausible scaled-up versions that could compete with the results of DM detectors~\cite{An:2013yua}. 

This calculation has already some history. 
The integrated flux was first estimated by Okun to constrain the HP flux with the solar luminosity~\cite{Okun:1982xi}. Here, the photon mean-free-path in the Sun was roughly estimated by Thomson scattering. Later, Popop and Vasil'ev introduced the photon refraction term due to the plasma in the oscillation formula, used free-free transitions for the solar opacity and integrated over a solar model~\cite{Popov:1991}. Popov used also this formalism in his 1999 paper~\cite{Popov:1999} where he first discussed helioscopes for detecting the flux of HPs produced inside the Sun -- reference~\cite{Goncharenko:1993jw} proposed searching for the solar HP flux due to oscillations of surface photons -- but the flux itself was not shown. 
Much later, a more careful calculation introduced absorption (decoherence) in the oscillation probability~\cite{Redondo:2008aa} and a more detailed treatment of Thomson and free-free processes in the photon opacity.  
This work discussed the keV spectrum in some detail and also discussed the resonant production of longitudinally polarised HPs, which was underestimated due to a mistake in the residue of the plasmon propagators.  
This mistake was pointed out in a recent paper~\cite{An:2013yfc}, which affected much of the low-mass parameter space, and that was corroborated in~\cite{Redondo:2013lna}, where also the Sun-L and Sun-T constraints shown in Fig. \ref{fig:constraints} were carefully reevaluated. 
Now that the formalism is set, it is timely to consider the low and intermediate energy flux of transversely polarised HPs.  

The main difficulty in computing the low energy flux is that for very low HP masses, the resonant photon-HP conversions in the Sun happen in a tiny shell very close to the photosphere. The solar plasma is not fully ionised in this region and one has to consider the role of neutral atoms and metallic donors in the index of refraction for it influences strongly the conversion probability. Indeed, the presence of neutral hydrogen is responsible for displacing the locus of the resonance slightly inwards to the Sun with respect to the case in which all the plasma would be considered to be ionised. First studies along this direction were taken already some time ago and first results were presented in~\cite{oai:arXiv.org:1202.4932,oai:arXiv.org:1010.4689}, which turn out to be quite accurate as order of magnitude estimates. 
In this paper, we ratify these estimates with a more detailed model of refraction in the Sun and exclude a possible strong influence of metals and excited states of hydrogen (with some caveats). 
Another complication is that, close to the photosphere, photons can travel a long distance before being absorbed, so long that the characteristics of the plasma change. In previous works, the photon-HP conversions where computed assuming that the solar plasma is homogenous in a mean-free-path but this is not the case close to the photosphere. 
Therefore, we also have to deal with the problem of photon-HP resonant oscillations in a non-homogenous plasma, which so far has not been discussed in this context. 
Perhaps, the most remarkable finding of this paper is that the total flux of HPs averaged over the region where resonant conversions happen is independent of this fact (as long as the resonance region is thin) and the previously used formulas for the Sun averaged HP emission are still correct (!). 
 One last complication with the resonance emission is that near the surface, the Sun is not spherically symmetric. 
Inhomogeneities in the temperature and density of the plasma arise because the energy transport is convective and outgoing hot cells and incoming cool flows have differences in their index of refraction. 
The resonance region is no longer a perfect spherical cell, it becomes corrugated and time-dependent. 
Nevertheless, the techniques developed for a spherically symmetric Sun with an effective 1D atmosphere can be applied also in this case. We find that the volume of the resonance region is slightly increased due to corrugation and the spectrum somewhat distorted because the 3D temperature profiles are steeper close to the surface and shallower inside. The result is a moderate O(1) increase of the HP flux coming from inner resonances and a decrease of the outer ones. 

But the resonance emission does not explain the HP spectrum in the whole energy range. Low mass HPs have resonance regions close to the solar surface where the temperature is O(eV) and emission of UV or X-ray energy HPs is exponentially suppressed. In these cases, the HP flux emitted from the bulk of the Sun dominates. At these energies, bound-bound and bound-free processes of helium and metals do contribute notably to the photon opacity and have to be included in the calculation. In this paper we use monochromatic opacities from the Opacity Project to compute this flux and find a significant contribution.   

The plan of the paper is as follows. In section \ref{sec:oscillations} we recall the physics of photon-HP oscillations in vacuum, homogenous and in homogenous plasmas and we derive the formulas useful for the resonant and non-resonant emission of HPs in the Sun. Section \ref{sec:plasma} is devoted to compute the ingredients needed to build a model for the refraction and absorption of photons in the solar plasma. 
In section \ref{sec:flux} we present our results for the flux of transversely polarised HPs emitted from the Sun from the near infrared to the keV range, with an emphasis on the visible part of the spectrum. 
We first train the intuition of the reader treating the Sun as an spherically symmetric plasma and only later we discuss the implications of the convection-driven inhomogeneities close to the surface.  
Finally, in section \ref{sec:conclusions} we use our results to set new constraints on the HP properties and speculate on future experimental searches for the transversely polarised HP flux. 
Note that henceforth we will refer exclusively to {\em transversely polarised} HPs merely as HPs for the sake of simplicity. When we speak about {\em longitudinally polarised} HPs we will do so explicitly. 

\section{Photon$\leftrightarrow$HP oscillations and the solar HP flux}
\label{sec:oscillations}

\subsection{Hidden photons}
Consider hidden photons as gauge bosons of a hidden U(1) gauge symmetry, i.e. a symmetry under which standard model particles do not transform. New particles or string excitations charged under the new U(1) and under the hypercharge of the standard model will generate kinetic mixing  between the hidden photon and the hypercharge boson~\cite{Holdom:1985ag,Foot:1991kb,Foot:1991bp,Dienes:1996zr,Lukas:1999nh,Abel:2003ue,Blumenhagen:2005ga,Abel:2006qt,Abel:2008ai,Goodsell:2009pi,Goodsell:2009xc,Goodsell:2010ie,Heckman:2010fh,Bullimore:2010aj,Cicoli:2011yh,Goodsell:2011wn,Cicoli:2012sz} or, at energies below electroweak symmetry breaking,  with the photon and $Z$ boson.  
Typically, integrating out these mediator new physics, also introduces an extra infinite tower of new operators, which in contrast to the kinetic mixing, are generically suppressed by the energy scales of the new physics (usually particle masses). 
Neglecting these feeble interactions and $Z$ mixing, at low energies the HP interacts with the standard model particles only though kinetic mixing with the ordinary photon. The lagrangian density describing this low energy theory is 
\be
{\cal L} = -\frac{1}{4}F_{\mu\nu}F^{\mu\nu}-\frac{1}{4}X_{\mu\nu}X^{\mu\nu} -\frac{\chi}{2}F_{\mu\nu}X^{\mu \nu} +\frac{1}{2}m^2 X_\mu X^\mu+ j^\mu A_\nu , 
\ee
where $A_\mu, X_\mu$ are the photon and the HP fields, $F_{\mu\nu},X_{\mu\nu}$ their field strengths and $j^\mu$ the ordinary electromagnetic current\footnote{In this paper we always work in Lorentz-Heaviside natural units $\hbar=c=k_B=1$ and $\alpha=e^2/4\pi$.}, which for the purposes of this paper we can take as $-e \bar \psi_e \gamma^\mu \psi_e$ ($\psi_e$ the electron field). 
The HP mass can arise either from the St\"uckelberg mechanism (see~\cite{Goodsell:2009xc,Cicoli:2011yh}) or from a hidden Higgs field developing a vacuum expectation value~\cite{Ahlers:2008qc}. 
The latter case reduces to the former when the mass of the hidden Higgs field is much larger than the energies under consideration, but it is very different if the hidden Higgs is as light as the HP~\cite{Ahlers:2008qc,An:2013yua}. In this paper we focus on the HP phenomenology, which is equivalent to the 
St\"uckelberg case or the heavy hidden Higgs case. 
 
The $A_\mu$ and $X_\mu$ fields are non-orthogonal because of the kinetic mixing. 
The field redefinition 
\be
X_\mu\to S_\mu-\chi A , 
\ee
diagonalises the kinetic part of the lagrangian, getting rid of the kinetic mixing,  
\be
\label{eq:lala}
{\cal L} = -\frac{1}{4}F_{\mu\nu}F^{\mu\nu}-\frac{1}{4}S_{\mu\nu}S^{\mu\nu}  +\frac{1}{2}m^2 (S_\mu-\chi A_\mu)^2+ j^\mu A_\nu +{\cal O}(\chi^2) . 
\ee
and reveals the interaction states $A_\mu$ (photon-like, the eigenstate interacting with the electric charge) and $S_\mu$ (sterile state that does not interact with the electric charge). 

\subsection{Photon ``flavor'' oscillations}

The lagrangian \eqref{eq:lala} makes explicit that $A,S$ are not propagation eigenstates in vacuum because of the non-diagonal mass term. 
The state radiated by electrons is pure photon-like ($A$) but, since it is not a propagation eigenstate, it will develop a $S$-component after some time and its magnitude is modulated by the time lapse. 
The picture is very much equivalent to neutrino oscillations: neutrinos produced in beta decay are produced in a flavour state (electron-flavor) and they oscillate into muon-flavor neutrinos, for example. 
In complete analogy with 2-flavor neutrino oscillations, the probability of finding a HP after a distance $L$ from the photon production region is given by 
\be
\label{eq:vacuumP}
P(A\to S)= (2\chi)^2\sin^2\(\frac{m^2 L}{4\omega}\)
\ee   
where $\omega$ is the photon/HP energy and we have assumed that HP are relativistic $\omega\gg m$. 
The role of the mixing angle in neutrino oscillations is played in this case by the kinetic mixing, $\chi$. Since this has been already constrained to be very small, see Fig. \ref{fig:constraints}, the probabilities of $A\to S$ transitions are minute and we always work at first order in $\chi$. Since the difference between the states $X$ and $S$ is of order $\chi$, we don't need to distinguish between them. When we compute oscillation probabilities, we refer to both simply as HPs. 

The situation changes in a medium because photons (the $A$ state) refract and get absorbed. The refraction and absorption properties in a linear medium can be casted in the form of an effective photon mass, which changes the definition of propagation states and thus alters the oscillation probability. This is again analogous to the case of neutrino oscillations, although in this context one speaks of ``matter potential'' instead of effective mass.  

The probability of a photon of frequency $\omega$ to oscillate into a HP in a \emph{homogeneous} absorbing medium has been computed in a number of papers in different frameworks: using the equations of motion in~\cite{Redondo:2008aa}, Feyman diagrams and thermal propagators in~\cite{An:2013yfc} and thermal field theory in~\cite{Redondo:2013lna}. 
All these calculations lead to the same result in the small mixing regime, 
\be
\label{eq:Prob1}
P(\gamma\to {\rm HP}) \simeq \frac{\chi^2 m^4}{(\Pi_r-m^2)^2+\Pi_i^2} , 
\ee
where the complex polarisation tensor $\Pi=\Pi_r+i \Pi_i$ (photon self-energy in the medium) depends on $\omega$ and can be understood as an effective photon mass $m_\gamma^2$ and an absorption coefficient $\Gamma$ 
\be
\Pi\equiv m_\gamma^2+i \omega \Gamma , 
\ee
which account for refraction and absorption (together with stimulated emission) of photons in the medium under consideration.  Alternatively it can be expressed as a function of the complex index of refraction, $N=n-i\kappa$, $
\Pi=\omega^2 (1-N^2)\simeq 2\omega^2(1-n)+i 2\omega^2 \kappa$. 

\subsection{Inhomogeneous medium}
In an inhomogeneous plasma, the probability amplitude of $\gamma\to$HP conversion after a length $L$ can be written as an integral over the putative photon trajectory ${\bf  r}= {\bf r}(l)$ as 
\be
\label{eq:Ageneral}
i {\cal A}(\gamma\to {\rm HP})(L)=i \chi \frac{\muu^2}{2\omega}\int_0^L  e^{i \varphi(l)-\tau(l)/2}dl
\ee
where 
\be
\label{eq:Phitau}
\varphi(l)=\int_0^l \frac{\mgg({\bf r}(l'))-\muu^2}{2 \omega}dl' \quad ; \quad  \tau(l)=\int_0^l \Gamma({\bf r}(l'))dl' .
\ee
Here, $\varphi$ is the the phase difference between the HP and photon waves integrated along the line of sight, in short, the number of $\gamma\to $HP oscillations, and $\tau$ is essentially the optical depth. 
The physical interpretation of this formula is described in some detail in~\cite{Redondo:2010dp}. It is based on the perturbative expansion for photon-axion oscillations derived by Raffelt and Stodolsky~\cite{Raffelt:1987im} but applied onto the photon-HP system that was presented in~\cite{Redondo:2013lna} (kinetic approach of Sec.~3). 
At first order in $\chi$, photons and HPs are propagation-eigenstates which can convert into each other with a a probability (amplitude) per unit length given by $\chi \muu^2/2\omega$. 
The transition can happen after any length $l$ between the photon source and the (hypothetical) HP detector. The integral over the path-length $l$ reflects this fact. 
The $e^{i\varphi(l)/2}$ factor is the phase difference between the photon and HP waves accumulated up to the length $l$. Conversions after different lengths can interfere constructively or destructively. The factor $e^{-\tau(l)/2}$ reflects the absorption (and stimulated emission) of the photonic wave before reaching the conversion point at a distance $l$. See~\cite{Redondo:2010dp} for more details.  

From \eqref{eq:Ageneral} and \eqref{eq:Phitau}, it is straightforward to derive the two formulas for the probability in homogeneous media shown before. In vacuum, we have $\mgg,\Gamma\to 0$ and thus  
\be
\nonumber
P(\gamma\to {\rm HP})(L)=|{\cal A}(\gamma\to {\rm HP})(L)|^2=\left|\left[\chi e^{-i\frac{m^2l}{2\omega}}\right]^{L}_{0}\right|^2=4\chi^2\sin^2\(\frac{\muu^2}{4\omega}L\) ,  
\ee 
while for a homogenous medium
\bea
\nonumber
P(\gamma\to {\rm HP})(L)&=&\left|\left[\frac{\chi m^2 e^{\frac{i(m_\gamma^2-m^2)-\omega\Gamma}{2\omega} l}}{i (m_\gamma^2-m^2)-\omega\Gamma} \right]^{L}_{0}\right|^2 \\
&=& \frac{\chi^2 m^2  \(1+e^{-\Gamma L}-2\cos\(\frac{(m_\gamma^2-m^2)L}{2\omega}\)e^{-\frac{\Gamma L}{2}}\)}{(m_\gamma^2-m^2)^2+(\omega\Gamma)^2} \\
&\xrightarrow{\Gamma L\to\infty}& \frac{\chi^2 m^4}{(m_\gamma^2-m^2)^2+(\omega\Gamma)^2} . 
\eea

This formula has to be a good approximation to media in which $m_\gamma,\Gamma$ change very little in an absorption length. 
The general formula \eqref{eq:Ageneral} allows us to evaluate its limit of validity. 
Let us then consider a linearly changing medium,  
\be
\Pi(l)=\Pi_0+\Pi'_{0}l + ... 
\ee
Expanding \eqref{eq:Ageneral} up to linear order in the derivative we find 
\be
{\cal A}(L\to \infty)\sim \frac{i\chi \muu^2}{\Pi_0-\muu^2}
\(1-\frac{2 \omega \Pi'}{\(\Pi_0-\muu^2\)^2}+...\)
\ee
so \eqref{eq:Prob1} is valid as long as 
\be
\label{criterion}
 \frac{2 \omega |\Pi'_{0}|}{\(\mgg_0-\muu^2\)^2+(\omega \Gamma)^2} \ll 1 . 
\ee

Defining the mean-free-path as $\lambda=\Gamma_0^{-1}$ and an oscillation length as 
$\lambda_{\rm osc}= \omega/ (m^2_\gamma-m^2)$ the criterium reads approximately
\be
\label{eq:crit}
\frac{2 |\Pi'_{0}|}{\omega} {\rm min}\{ \lambda^2,\lambda_{\rm osc}^2\} \ll 1
\ee

In this case $\varphi(l)=\frac{m_\gamma^2l}{2\omega}+\frac{m_{\gamma 0}^{2'}l^2}{4 \omega}$, so the condition requires that $\varphi$ is a linear function of $l$ during a mean-free-path or an oscillation length, whatever it is smaller, up to corrections smaller than 1. The same of course, has to be satisfied by the optical depth function $\tau$. 

As it turns out, the condition \eqref{eq:crit} is satisfied almost in all the conditions of our interest because 
the typical length scale of an oscillation in the parameter range of our interest is extremely small, due to the large values of $m_\gamma^2$ in the solar interior. Even the vacuum oscillation length, 
\be
\lambda_{\rm osc,vac}\sim 0.01\, {\rm km} \frac{\omega}{2\, \rm eV} \(\frac{10^{-3}\, \rm eV}{m}\)^2
\ee
is much smaller than the characteristic density and temperature scale heights in the Sun, hundreds of kilometers at the surface and much longer in the solar interior.   

\subsection{Resonances}

\subsubsection{Optically thick}

The only obvious exception to the validity of \eqref{eq:Prob1} happens in regions where $m_\gamma^2\simeq m^2$ to good accuracy and the oscillation probability \eqref{eq:Prob1} is resonant.
Close to a resonance, $\lambda_{\rm osc} $ grows very large and eventually larger than the mean-free-path (which is typically longer than $\lambda_{\rm osc}$). 
In the deep Sun, $\lambda$ is extremely small and thus \eqref{eq:crit} is nevertheless satisfied. 
Resonances for which \eqref{eq:crit} holds can be called {\em optically thick} because their typical extent 
is 
\be
\Delta r_{\rm thick}\sim \omega\Gamma \left|\frac{d m_\gamma^2}{dr} \right|^{-1} , 
\ee
and therefore \eqref{eq:crit} implies $\Gamma \Delta r \ll 1$.    

Thus, \eqref{eq:Prob1} is an excellent approximation for the solar conditions except for what we shall call optically-thin resonant regions ($\Pi'_{0} \lambda^2 /\omega\ll 1$). 
Fortunately, for these cases we can also also derive a simple analytical formula. 

\subsubsection{Optically thin }

The mean free path $\lambda$ grows increasingly large as we exit the Sun, so the criterion \eqref{criterion} is eventually violated. The effect is exacerbated by the fact that near the solar surface, the density decreases much faster than in the interior and therefore so does $\Pi'$. 

Consider \eqref{eq:Ageneral} and focus on a photon trajectory ${\bf r}(l)$ which intersects a resonance, i.e. a point where  
$\mgg({\bf r} (l_s))\equiv \mgg(l_s)=m^2$. In this point $d\varphi/dl=\varphi'=0$ and the integral has a saddle point, which typically gives the dominant contribution to the probability amplitude. A classical saddle point approximation gives 
\bea
{\cal A}_{\rm sad}&\sim& \chi \frac{\muu^2}{2\omega}e^{i \varphi(l_s)-\tau(l_s)/2} \int_{0}^\infty e^{i \varphi''(l_s)(l-l_s)^2/2} \\
\label{eq:2p/p}
&\sim& \chi \frac{\muu^2}{2\omega} e^{i \varphi(l_s)-\tau(l_s)/2}\sqrt{\frac{2\pi i}{\varphi''(l_s)}}  {\cal C}(\Delta)
=
\chi \frac{\muu^2}{2\omega} e^{i \varphi(l_s)-\tau(l_s)/2}\sqrt{\frac{4\omega \pi i}{\mgg'(l_s)}} {\cal C}(\Delta), 
\eea
where prime denotes differentiation with respect to $l$, $\Delta \equiv l_s\sqrt{|\varphi''|}$ and ${\cal C}(\Delta)$ is a smooth step-like function $\sim \Theta(\Delta)$,  which can be better approximated as
\be 
{\cal C}(\Delta)\approx \(\frac{1+{\rm sign} (\Delta)}{2}-\frac{e^{i\Delta^2/2}}{\sqrt{\pi}(1-i)\Delta+2{\rm sign}(\Delta)} \) .  \ee
The conversion probability can then be estimated as
\be
\label{eq:Prob2}
P(\gamma\to {\rm HP})_{\rm sad}\sim \frac{\pi \chi^2 m^4}{\omega |m_\gamma^{2'}(l_s)|}e^{-\tau(l_r)}|{\cal C}(\Delta)|^2 .
\ee

The interpretation of this formula is very interesting. Most of the oscillations cancel out in the integral but there is a region of size $(\delta l)^2 \sim (\varphi'')^{-1}$ around $l_s$ where the phase has almost a constant value $e^{i \varphi(l_s)}$ and the amplitude has decreased by a factor $e^{-\tau(l_s)/2}$. The saddle point approximation singles out this region around the resonance as the most important contribution to the amplitude. 
In the particle conversion language, the photon$\to$HP conversions along the path interfere destructively due to the fast varying (but smooth) function $\varphi(l)$, except those happening in the region $\delta l$, which interfere constructively and thus dominate the probability. 

\begin{figure}[t]
\begin{center}
\includegraphics[width=0.45\textwidth]{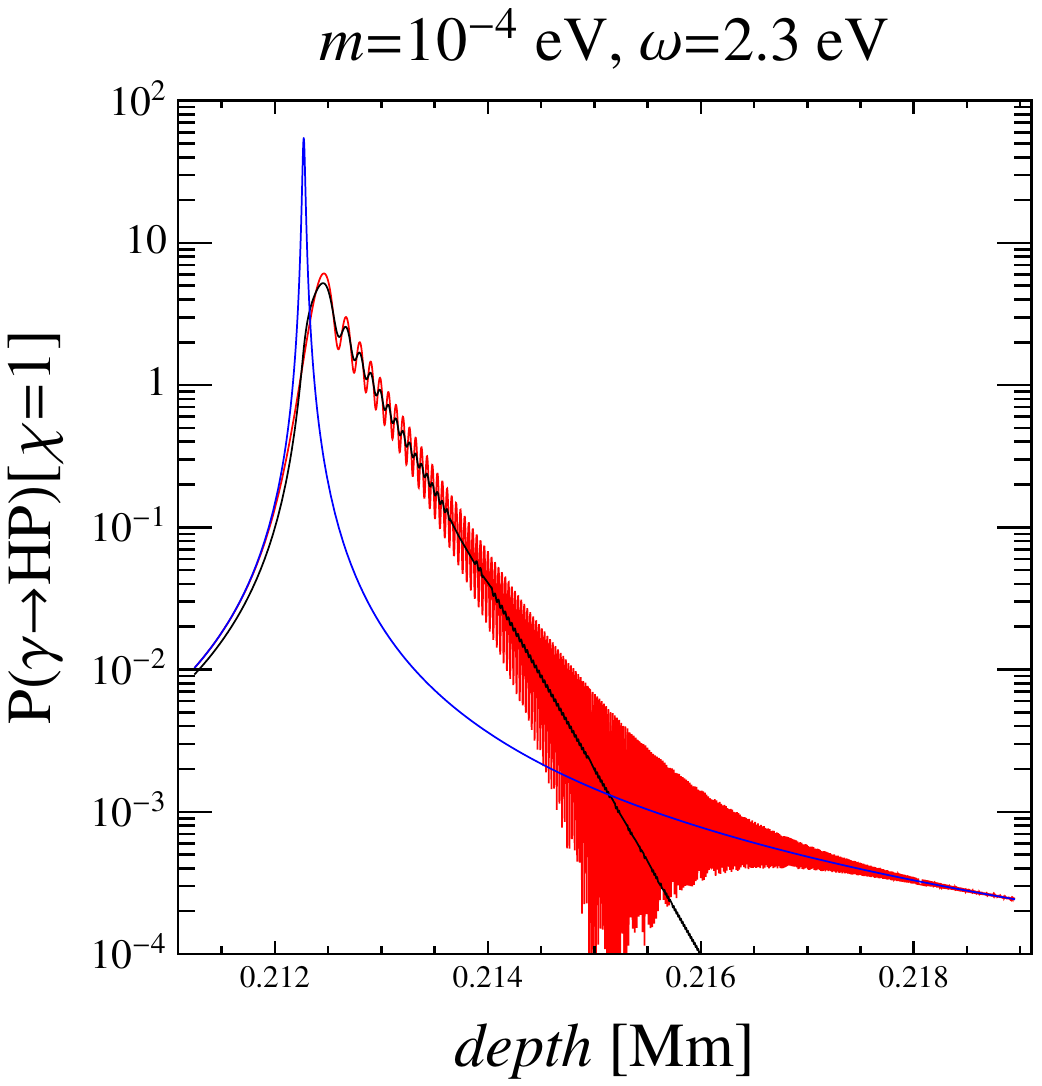}
\includegraphics[width=0.45\textwidth]{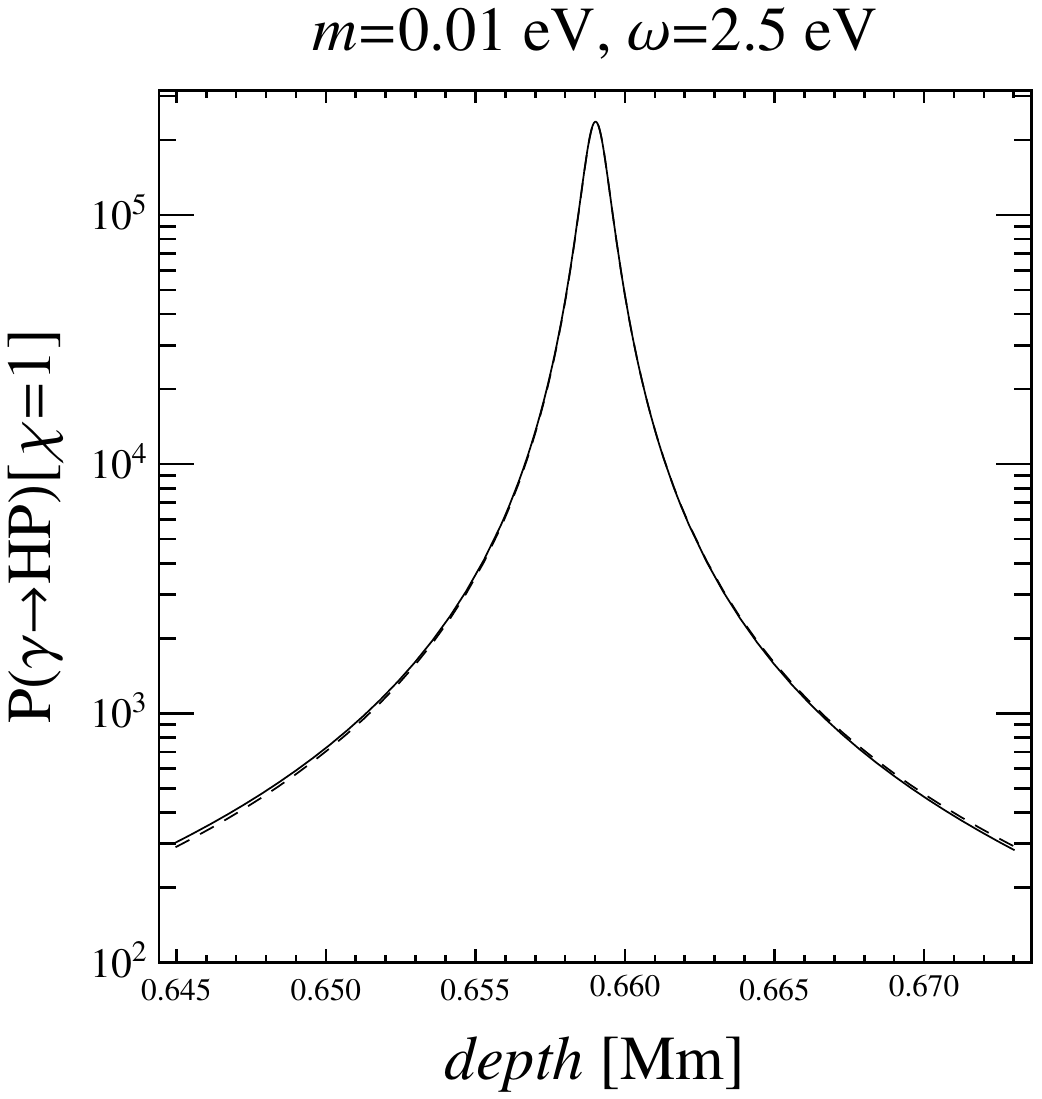}
\caption{ 
Photon$\to$HP oscillation probabilities as a function of the solar depth below the surface, zoomed into  the resonance regions. LEFT: Example of an optically thin resonance. The red line shows the conversion probability of a photon originating at the given depth and moving in the radial outwards direction (leftwards in the plot, towards smaller depths). 
The black line is our approximation formula \eqref{eq:Prob2} and the blue line the homogeneous plasma approximation, \eqref{eq:Prob1}. 
RIGHT: Example of an optically thick resonance region. The black line shows the homogeneous approximation \eqref{eq:Prob1} and the dashed the small corrections arising by taking into account inhomogeneities.}
\label{fig:proba}
\end{center}
\end{figure}

Formula \eqref{eq:Prob2} is extremely simple and useful for our purposes. For the sake of completeness let us briefly discuss its limitations.  
The formula implicitly assumes that the saddle point dominates the integral.  
However, it is clear that if we separate enough from the resonance, at some point the exponential suppression due to the optical depth $\tau(l_s)$ will suppress the contribution from the saddle point below the contribution from the first mean-free-paths. In other words, for sufficiently large $l_s$ we are so far away from the resonance that $\lambda_{\rm osc}$ is again small and the probability should again tend to formula \eqref{eq:Prob1} because \eqref{eq:Prob2} is exponentially suppressed at large distances. We expect a smooth transition between the two formulas. 
The situation is exemplified in Fig. \ref{fig:proba} where we show examples of thin and thick resonance regions in the Sun.  
The left figure shows an example of photon$\to$HP conversion probabilities around a thin resonance region. 
In this example, the photons originate at the given depth and move leftwards towards the solar surface. 
The red line shows a realistic approximation of the full result. In the deep Sun and far from the resonance it coincides with the homogenous approximation \eqref{eq:Prob1} (shown in blue) but as the production point gets closer, the contribution from the resonance region starts to dominate and grows exponentially (here the mean-free-path around the resonance is $\lambda\sim 300$ m). Our probability formula \eqref{eq:Prob2} captures this exponential growth. 
Once the resonance is past, the probability returns quickly to the homogeneous result. 
The oscillations seen at the beginning of the exponential growth originate from interference of the local and saddle point contributions and they are never relevant as they tend to average out when frequency averaged or when the probability is integrated over some region of the Sun as when we compute the HP flux of the entire Sun. 

In contrast, the right figure shows a thick resonance region where the criterion \eqref{eq:crit} is satisfied and photons oscillate into HPs before realising any inhomogeneity in their index of refraction or absorption (at the resonance centre $\lambda\sim 3$ m). 

Note finally that in order to use \eqref{eq:Prob2} for $\Delta<0$ ($l_s<0$), i.e. for photons moving away from the resonance, we have to taylor a small modification. The required change is to use $\tau(l_s)=0$ for $l_s<0$ because the region expected to contribute to the integral is the closest to the photon emission, thus $\tau(l_s)\to \tau(l_s)\Theta(l_s)$.    
As $\Delta$ becomes negative the photon path contains less and less of the maximal mixing region and the amplitude decreases as $\propto 1/\Delta$. In this regime we can recover explicitly \eqref{eq:Prob1} in yet a different fashion. Aside from irrelevant phases we find
\bea
P(\gamma\to {\rm HP})_{\rm sad}(\Delta< 0)&\approx& 
\frac{\pi \chi^2 m^4}{\omega |m_\gamma^{2'}(l_s)|} \left|\frac{1}{\pi(1-i)\Delta} \right|^2=
\frac{\pi \chi^2 m^4}{\omega |m_\gamma^{2'}|}\frac{1}{2\pi l^2_s |\varphi''|} =
\frac{\chi^2 m^4}{|m_\gamma^{2'}|^2 l_s^2}\\
&=&\frac{\chi^2 m^4}{(m^2_{\gamma0}-m^2)^2}, 
\eea
where for the last expression we have used the Taylor expansion of $m_\gamma^2$ around the resonance $m_\gamma^2(l)=m^2+m_\gamma^{2'}(l_s) (l-l_s)+...$.
We obtain this expression because in this limit, most of the conversion probability comes from the first oscillation, as in the homogeneous case.  

\subsection{Solar averaged emission of hidden photons}

\subsubsection{General aspects}

The emission rate of transversely polarised HPs of energy $\omega$ per unit volume at a given position, ${\bf r}_0$, of an inhomogenous plasma in local thermal equilibrium (LTE) can be written as the 
photon production rate $\Gamma_{\rm P}$ times the conversion probability $P(\gamma\to {\rm HP})$~\cite{Redondo:2008aa} integrated over the different directions in which a photon can be produced,  
\bea
\frac{d N}{dV dt}({\bf r}_0) &=& 2\int \frac{d^3 {\bf k}}{(2\pi)^3} \Gamma_{\rm P}(\omega,{\bf r}_0) P(\omega,{\bf r}(l))  .
\eea 
Note that the conversion probability $P=P(\gamma\to {\rm HP})$ depends in principle upon the whole trajectory ${\bf r}(l)\sim {\bf r}_0+l \hat{\bf k}$ and in particular on the direction of the momentum $\bf k$.  The photon production rate only depends on the creation point, as follows from our assumption of LTE. The factor of 2 accounts for the two transverse polarisations. 
In LTE, the photon production rate is related to the absorption rate $\Gamma_{\rm A}$ by detailed balance, $\Gamma_{\rm P}=e^{-\omega/T}\Gamma_{\rm A}$, and the imaginary part of the self energy, $\Gamma = \Gamma_{\rm A}-\Gamma_{\rm P}$, so that 
\be
\Gamma_{\rm P}(\omega) = \frac{\Gamma(\omega)}{e^{\omega/T}-1} . 
\ee
with $T=T({\bf r})$ being a smoothly-varying plasma temperature. 

The specific HP flux on Earth is the integral of the volume emission of the Sun divided by the surface of a sphere of radius the Sun-Earth distance, $R_{\rm Earth}\simeq 149.6\times 10^6$ km, 
\be
\frac{d \Phi}{d\omega}= \frac{1}{4\pi R_{\rm Earth}^2}
\int_{\rm sun} dV \frac{d N}{dV d\omega dt} , 
\ee
where we have used the HP dispersion relation $\omega^2=|{\bf k}|^2+m^2$. 

\begin{figure}[t]
\begin{center}
\vspace{-2cm}
\includegraphics[width=0.7\textwidth]{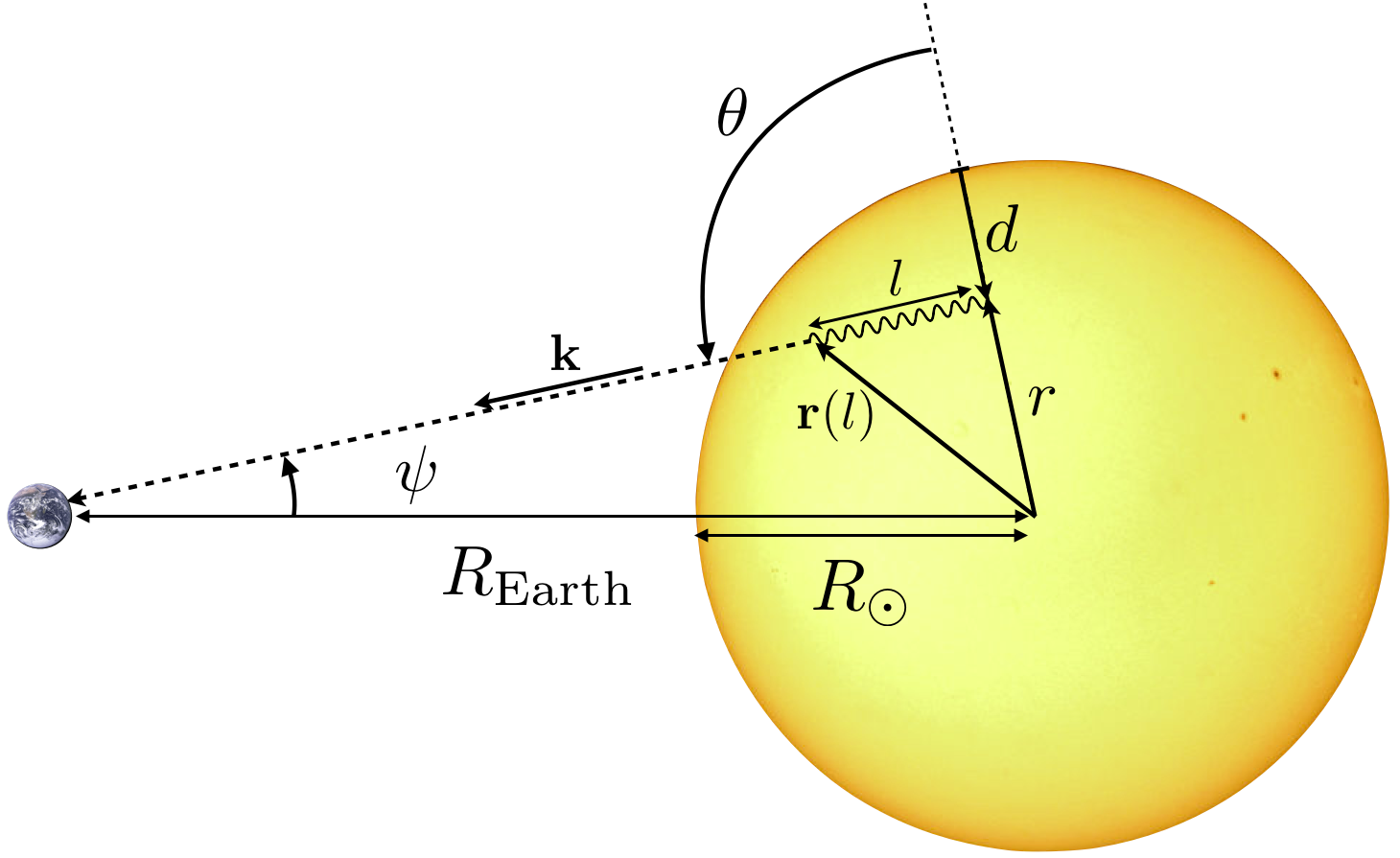}
\caption{Geometry of the solar HP flux calculation. Photons created at position ${\bf r}(l=0)$ with three-momentum ${\bf k}$ follow a trajectory ${\bf r}(l)$ (wavy line) towards the Earth and can be converted into collinear hidden photons (dashed line) after any length $l$. The solar plasma properties, such as temperature and density are function of the radial coordinate $r=|{\bf r}|$ or the depth $d=R_{\rm Sun}-r$. The photon/HP trajectories depend on the azimuth angle $\theta$, 
$r(l)\sim r_0+l\cos\theta$.  
 }
\label{fig:geo}
\end{center}
\end{figure}

Let us now assume the Sun to be spherically symmetric and postpone the discussion on inhomogeneities. Then the probability $P$ only depends on the radial position, $r$, and the azimuthal angle, $\theta$, which defines the photon/HP trajectory, see Fig. \ref{fig:geo}. Integrating over the remaining angular variables we find
\be
\label{eq:generalflux}
\frac{d \Phi}{d\omega}= \frac{1}{R_{\rm Earth}^2}\frac{\omega\sqrt{\omega^2-m^2}}{2\pi^2}  
\int_{\rm sun} r^2 dr \int d\cos\theta \frac{\Gamma(r)}{e^{\omega/T(r)}-1} P(\omega,r,\theta) . 
\ee
In order to compute this integral, we need to know the functions $\Gamma(\omega,r),T(r)$ and $P(\omega,r,\theta)$, for which we also need $m_\gamma^2(\omega,r)$. In Sec. \ref{sec:plasma} we describe how to build these expressions from a solar model and in Sec. \ref{sec:flux} we perform the relevant computations to obtain the solar HP flux and discuss the results.

\subsubsection{Resonance domination}
\label{sec:resonancedomination}
It turns out that for values of $\omega$ in the visible range, the conversion probability is very peaked around a spherical shell located at radius $r_*$ defined by 
\be
m_\gamma^2(r_*)=m^2
\ee 
where the photon-HP oscillations are resonantly enhanced. The emission from this region can easily dominate the emission from the rest of the Sun, making possible a humongous simplification of the calculations. Indeed we can perform analytically the remaining two integrals of \eqref{eq:generalflux} to obtain a surprisingly simple result. The derivation will take the rest of this section. 

We have simple probability formulas which apply when the resonance region is optically thick or optically thin. 
The criterion we shall use to identify them is 
\bea
\frac{1}{\omega}\left|\frac{d m_\gamma^{2}({\bf r}(l))}{dl}\right|_{r_*} \ll \Gamma^2(r_*)  && (\rm thick) \\
\frac{1}{\omega}\left|\frac{d m_\gamma^{2}({\bf r}(l))}{dl}\right|_{r_*} \gg \Gamma^2(r_*)  && (\rm thin)
\eea
Note that the gradient depends on the azimuth of the trajectory
\be
\label{eq:cosdep}
\left.\frac{d m_\gamma^{2}}{dl}\right|_{l=l_s}=\left.\frac{d m_\gamma^{2}}{dr}\right|_{r=r_*} \frac{dr}{dl} \simeq 
\left.\frac{d m_\gamma^{2}}{dr}\right|_{r=r_*}\cos\theta, 
\ee
and thus resonances are thin or thick depending on $\cos\theta$. 
It is interesting thus to focus first on the $r$-integral, which we do in the following.  

\subsubsection*{Optically-thick resonance}
If the resonance is optically thick, the $\gamma\to$HP conversion probability is well approximated by \eqref{eq:Prob1} and the $r$-integral is
\be
\label{eq:Ithick}
I(\theta)_{\rm thick} =\int_{\rm sun} r^2 dr \frac{\Gamma(r)}{e^{\omega/T(r)}-1} \frac{\chi^2 m^4}{(m_\gamma^2(r)-m^2)^2+(\omega\Gamma(r))^2} ,    
\ee
which actually does not depend on $\cos\theta$ itself.
In the resonance, the probability is enhanced by a factor $m^4/(\omega\Gamma)^2$ with respect to the vacuum case and much more if we compare it with very dense regions where $m_\gamma^2(r)\gg m^2$. 
Since usually we have $m_\gamma^2\gg \omega\Gamma$, the integral so strongly peaked that we can approximate the Lorenzian shape of the probability by a Dirac delta to obtain 
\be
I(\theta)_{\rm thick} \approx \frac{\pi r_*^2}{\omega}\frac{\chi^2 m^4}{e^{\omega/T(r_*)}-1} \left|\frac{d m_\gamma^{2}}{dr}\right|^{-1}_{r=r_*}  . 
\ee
The explicit dependency on $\Gamma$ drops out~\cite{Redondo:2008aa,An:2013yfc,Redondo:2013lna} and the resulting flux only depends upon our solar modelling through $m_\gamma^2(\omega,r)$ and $T (r_*)$.  

\subsubsection*{Optically-thin resonance}
If the resonance is optically thin we can use \eqref{eq:Prob2} for the $r$-integral, which now depends explicitly on $\theta$ because of the $\cos\theta$ factor in \eqref{eq:cosdep} and the optical depth to the resonance, 
\be
\tau(l_s)=\tau (r,r_*)=\frac{1}{\cos\theta} \int_r^{r_*}dr' \Gamma(\omega,r') . 
\ee
The $r$-integral is
\bea
I(\theta)_{\rm thin}= \int_{\rm res} r^2 
\frac{\Gamma(\omega,r)}{e^{\omega/T}-1} 
\frac{\pi \chi^2 m^4}{\omega }\frac{1}{|\cos\theta|}\left|\frac{d m_\gamma^{2}}{dr}\right|^{-1}_{r=r_*} e^{-\tau(r,r_*)} \Theta.   
\eea
where we have taken ${\cal C}(\Delta)=\Theta(\Delta)$, which we write as $\Theta\equiv \Theta\left((r_*-r)\cos\theta\right)$, ensuring that only photon trajectories that cross the resonance are included. If the photon is produced at $r<r_*$ only trajectories with $\cos\theta>0$ will cross the resonance, the opposite case being when $r>r_*$.  
The integral is dominated by the region where the optical depth to the resonance is O(1), typically a few mean-fee-paths.  
Assuming that $r$ and $T$ do not change much in that narrow region we can substitute for the values at the resonance, $r_*,T_*$ and perform the integral explicitly
\bea
\int_{\rm res}  \frac{dr}{|\cos\theta|} 
\Gamma(\omega,r) e^{-\tau(r,r_*)}\Theta &=&\\
\Theta(\cos\theta) \int_{\rm in}^{r_*}  \frac{dr}{\cos\theta} 
\Gamma(\omega,r) e^{-\tau(r,r_*)}
&+&
\Theta(-\cos\theta)  \int_{r_*}^{\rm out}  \frac{dr}{-\cos\theta} 
\Gamma(\omega,r) e^{-\tau(r,r_*)}\\
=\Theta(\cos\theta)+\Theta(-\cos\theta)(1-e^{-\tau_*})&=&1-\Theta(-\cos\theta)e^{-\tau_*}
\eea
where $\tau_*$ is the optical depth of the resonance with respect to the surface. 
If the resonance lies deep enough inside the Sun ($\tau_*\sim 3-4$ or so), HPs converted from photons travelling through it from the inside or from the outside contribute an equal amount to the total HP flux. If it lies close or in the photosphere, the amount of photons produced outside and traveling inwards decreases very much and we are left only with the HP converted from outgoing photons. 
Remarkably, again the resulting flux does not depend explicitly on $\Gamma$ or other properties of the solar model other than $T$ and $m_\gamma^2$.
We have then 
\be
I(\theta)_{\rm thin} \approx \frac{\pi r_*^2}{\omega}\frac{\chi^2 m^4}{e^{\omega/T(r_*)}-1} \left|\frac{d m_\gamma^{2}}{dr}\right|^{-1}_{r=r_*}  \left(1- \Theta(-\cos\theta)e^{-\tau_*}\right)
\ee
Remarkably, it is given by exactly the same expression than for the optically thick resonance (i.e. as long as $\tau_*\gtrsim 3-4$).  

\subsubsection*{Master formula for resonance emission}

If we assume that the resonance region is either thick or thin or, more specifically, that the values of $\cos\theta$ for which the resonance is neither are not quantitative relevant for the $\cos\theta$ integral, the $\cos\theta$ integral is trivial 
\be
\( \int_{\rm thick} d\cos\theta I_{\rm thick}+\int_{\rm thin} d\cos\theta I_{\rm thin}\) = I_{\rm thick}\(1-\delta\theta\frac{e^{-\tau_*}}{2}\) . 
\ee
where $\delta\theta=\int_{\rm thin}d\cos\theta\Theta(-\cos\theta)$ corrects for the HPs produced from photons originated at $r>r_*$ which travel inwards the Sun. The 1 in the formula would be symmetric result, which is to be corrected when $e^{-\tau_*}$ is not negligible. In practice, $\delta\theta\simeq1$ for all practical purposes.  

Our final formula for the HP emission from the entire resonance region is 
\bea
\label{eq:resonantflux}
\frac{d\Phi}{d\omega}&\approx& \frac{r_*^2}{\pi R_{\rm Earth}^2}\frac{\chi^2 m^4\sqrt{\omega^2-m^2}}{e^{\omega/T(r_*)}-1} \left|\frac{d m_\gamma^{2}}{dr}\right|^{-1}_{r=r_*} \(1-\frac{e^{-\tau_*}}{2}\) . 
\eea

The formula was already found in~\cite{Redondo:2008aa,An:2013yfc,Redondo:2013lna} under the assumption that the resonance region is optically thick (extremely good approximation in the solar interior, where these references were focused). 
In this paper we have shown explicitly that its validity extends to optically thin resonances and for the mixed case where the resonance region is optically thin for $\cos\theta \sim \pm 1$ and thick for $\cos\theta\sim 0$. 
For that, we needed to assume that the regions in $\theta$ for which the resonances are neither thick or thin do not change the behaviour. I have performed many explicit calculations and cross-checks  that show that this is the case, but at the moment I have no general proof. However, in the simple case of constant $r,T,\Gamma,dm_\gamma^2/dr$ in the resonance region (which is not a bad approximation in the Sun) one can perform analytically all the integrals and prove that there is no funny behaviour when the resonance is neither thin or thick. The calculation is shown in  appendix \ref{sec:prueba}. 
 
\subsection{Angular distribution of the signal}
\label{sec:angular}

The specific HP flux at earth (HPs per unit area, time, energy and stereoradian) along a line of sight intercepting the Sun is 
\be
\frac{d \Phi(\omega,\psi)}{d\omega d\Omega}=\frac{\omega \sqrt{\omega^2-\muu^2}}{4\pi^3}  \int_0^\infty {ds} \frac{\Gamma(\omega,r)}{e^{\omega/T(r)}-1}
P(\omega,r,\theta(r))  , 
\ee
where $r=r(s)$ with $s$ the line-of-sight distance from the Earth to the production point (dashed line in Fig.~\ref{fig:geo}).
Defining $r_{\rm min}=R_{\rm Earth}\sin\psi$ as the impact parameter of the trajectory we change the integration variable to the radial coordinate $r$ 
\be
\frac{d \Phi (\omega,\psi)}{d\omega d\Omega}\approx \frac{\omega \sqrt{\omega^2-\muu^2}}{4\pi^3 }\int^\infty_{r_{\rm min}} \frac{2rdr}{\sqrt{r^2-r_{\rm min}^2}} \frac{\Gamma}{e^{\omega/T}-1} P(\omega,r,\theta) . 
\ee

If the trajectory intercepts the resonance region, the bulk of the emission can be readily evaluated by similar calculations which took us to \eqref{eq:resonantflux}. 
The key simplification is to use the optically thick resonance formula \eqref{eq:Prob1} for the probability because either $\sqrt{r^2-r_{\rm min}^2}$ is relatively constant during the integral (and then optically thin and thick resonances give the same result) or $r_*\sim r_{\rm min}$ and then $\cos\theta\sim 0$ and  \eqref{eq:Prob1} is justified again.   
Using $r\sim r_*$ in the numerator and $r^2-r_{\rm min}\sim (r_*+r_{\rm min})(r-r_{\rm min})$ in the denominator, the integral can be readily evaluated as 
\be
\label{angdist}
\frac{d\Phi (\omega,\psi)}{d\omega d\Omega}\simeq 
\left.\frac{d\Phi}{d\Omega }\right|_{\rm total} \frac{d X}{d \Omega}
\ee
where the angular distribution is given by 
\be
\frac{d X}{d \Omega} \simeq \frac{1}{2\pi}\frac{R_{\rm Earth}}{r_*}\sqrt{\frac{\sqrt{(\psi_*-\psi)^2+\Delta\psi_*^2}+\psi_*-\psi}{2[(\psi_*-\psi)^2+\Delta\psi_*^2]}}\sqrt{\frac{\psi_*}{\psi_*+\psi}}
\ee
where the angle at which $\psi$ is tangential to the resonance shell is $\psi_*\simeq r_*/R_{\rm Earth}$ and the width, which we have assumed to be small, is given by 
\be
\Delta\psi_*\simeq \frac{\Delta r_{\rm thick}}{R_{\rm Earth}}=\frac{\omega\Gamma_*}{R_{\rm Earth}}
\left|
\frac{d m_\gamma^2}{dr}\right|^{-1}_{r_*} . 
\ee
Moving from the solar center outwards, the angular distribution grows as $\psi\to \psi_0$  and peaks at  $\psi_*-\psi\simeq \Delta\psi_*$, where the line-of-sight is tangential to the resonance shell and thus benefits from more resonant emission, and then decrease extremely fast. The peak has a $1/\sqrt{\psi-\psi_*}$ behaviour and of course does not dominate the integral. 

Since the resonance can be extremely sharp it might turn out that cannot be resolved by a telescope. In this case, the resonance will be broaden by the finite resolution of the apparatus. The angular integral of the resonance is independent of the resonance width as long as it is very small
\be
{\rm lim}_{\Delta\psi_*\to0}2\pi\int_{\psi_*-\delta}^{\psi_*+\delta} d\psi \psi \frac{dX}{d\Omega} \approx  2\sqrt{2\delta } 
\ee
so in practice one can use 
\be
\frac{d X}{d \Omega} \simeq \frac{1}{2\pi}\sqrt{\frac{R_{\rm Earth}}{r_*}}
\sqrt{\frac{1}{\psi_*-\psi}} . 
\ee

\section{Refraction and absorption in the Sun}
\label{sec:plasma}

\subsection{Basics}

In order to compute the solar HP emission, we need to model the refraction and absorption properties of light in the solar plasma. For the latter, very exhaustive studies exist since absorption determines the radiative energy transfer inside of the Sun, which in turn determines the solar structure. We highlight the Opacity Project ~\cite{OP,Seaton1994,Badnell2004,Seaton:2004uz} and Los Alamos opacity code LEDCOP opacities~\cite{LEDCOP,LEDCOPref,TOPS}.  
To the best of our knowledge, no study of the refractive index throughout  the solar plasma exists in the literature. In principle, the real part of the index of refraction can be obtained from the imaginary part with the Kramers-Kronig relation.  Unfortunately, the existing data is often smoothed over frequencies and it is only available for a small number of density and temperature points. A rough interpolation would introduce large errors in the determination of the resonance region so we have to follow a different procedure. 

We can calculate explicitly the most relevant contributions to refraction and absorption as a function of temperature, density and composition so that we have a very smooth map of them inside of the Sun (as smooth as the solar model we might use). These are the contributions from electrons either free or bound in H atoms. The effects of electrons bound in Helium and heavier atoms are subdominant for refraction, but can be relevant for absorption around frequencies corresponding to the strongest atomic transitions. This contribution is then taken from existing opacity calculations to avoid the extremely involved atomic physics. Indeed, we can use OP opacities tables, conveniently provided for each metal separately and thus allowing arbitrary mixtures. 

The solar plasma consists on electrons (free or bound in ions) and nuclei, which being much heavier interact weaker with light than the former and can be neglected. 
Unbound (free) electrons contribute to the polarisation tensor with a simple term
\be
\Pi_{\rm free} = \frac{4\pi\alpha}{m_e}n_e^{\rm free} - i\, \omega \frac{8\pi\alpha^2}{3m_e^2}n_e^{\rm free}
\ee 
whose real and imaginary parts we recognise as the plasma frequency squared and the absorption coefficient due to Thomson scattering. 

Electrons bound in H atoms can be in first approximation modelled by a set of oscillators whose contribution of the polarisation tensor is 
\be
\label{eq:Pibb}
\Pi_{\rm bb}=\frac{4 \pi \alpha}{m_e}
 n_{\rm H^0} \sum_{n}  {\cal Z}_{n}\, \sum_{n'} f_{nn'} \frac{\omega^2}{(\omega^2-\omega_r^2)^2+ (\omega\gamma_r)^2}\(\omega^2-\omega_r^2-i \, \omega\gamma_r\)
\ee
where $n_{\rm H^0}$ is the number density of neutral H atoms, ${\cal Z}_{n}$ the probability of finding the bound electron with principal quantum number $n$ and the last sum is over resonant transitions $n\to n'$, which happen at frequencies $\omega_r={\rm Ry}\(\frac{1}{n^2}-\frac{1}{n'^2}\)$. The oscillator strength $f_{nn'}$ is summed over final states and orbital quantum numbers $l$ and $l'=l\pm 1$ for electric dipole transitions. We neglect fine-structure corrections as we are not interested in fine details of the spectrum. 
The width of the resonance is given by the natural line width $2\alpha \omega \omega_r /3 m_e$. 
Impact broadening due to proton collisions through the Stark effect is taken into account in the quasi-static approximation by folding each resonance contribution with a Holtsmark distribution following~\cite{Griem1960}. We don't consider collisional broadening from electron collisions nor Doppler broadening, which are not relevant at the level of accuracy required. 
Broadening is relevant close to the resonant transitions and in the wings of absorption lines for $\Gamma$, but not for $m_\gamma^2$.  
Note that free electrons behave as an additional bounded species with $f=1$ and $\omega_r=0$. 

The above expressions account for the most important contributions to refraction but are not enough for absorption. 
The leading contribution to photon opacity in the deep Sun comes from photon absorption during electron-proton scattering, $\gamma +e^-+p^+\to e^-+p^+$. In outer shells, also the photoelectric effect, $\gamma+{\rm H}^*\to e^-+p^+$, is important. Usually these reactions are known as  free-free and bound-free processes, respectively. 
Their contribution to $\Pi_i$ can be casted in the following compact form
\bea
\Pi_{i,\rm ff}=\omega\Gamma_{\rm ff} &=& \omega\frac{64\pi^2\alpha^3}{3 m_e^2\omega^3}\sqrt{\frac{m_e}{2\pi  T}}
\(1-e^{-\omega/T}\)n^{\rm free}_e n _p\, {F}_{\rm ff}  \\
\Pi_{i,\rm bf}=\omega\Gamma_{\rm bf} &=& \omega\frac{8\pi m_e \alpha^5}{3\sqrt{3} \omega^3}\(1-e^{-\omega/T}\)   n_{\rm H^0} \sum_{n} {\cal Z}_n \frac{1}{n^5}{F}_{\rm bf} \Theta(\omega-E_n) \\
&+& \omega\(1-e^{-\omega/T}\)   n_{\rm H^-} \sigma(\gamma+{\rm H}^-\to H+e^-)
\eea
where $F_{\rm ff},F_{\rm bf}$ are thermally averaged Gaunt-like factors, introduced as corrections to the classical result which depend mildly on frequency and atomic details and can be found for instance in~\cite{karzas1961}. 
Close to the threshold $F_{\rm bf}\sim 1$ so we neglect it. As for $F_{\rm ff}$, for our purposes it is sufficient to consider the Born-Elwert approximation~\cite{Elwert1939} with a simple screening prescription 
\be
F_{\rm ff}(w=\omega/T)= \int_0^{\infty}dx \frac{x\, e^{-x^2}}{2}\frac{\sqrt{x^2+w}}{x}\frac{1-\exp\(-\frac{2\pi\alpha}{\sqrt{x^2+w}}\sqrt{\frac{m_e}{2T}}\)}{1-\exp\(-\frac{2\pi\alpha}{x}\sqrt{\frac{m_e}{2T}}\)}\int_{\sqrt{x^2+w}-x}^{\sqrt{x^2+w}+x}\frac{t^3 dt}{(t^2+y^2)^2}
\ee
where $y=k_{\rm D}\sqrt{2 m_e T}$ and we take $k_{\rm D}$ to be the Debye-screening scale given by 
$k_{\rm D}^2=4\pi\alpha\sum_\alpha Q_\alpha^2 n_\alpha/T$ where the sum extends to all charged particles (mostly electrons, protons and some He atoms).  
In the bound-free expression, $E_n={\rm Ry}/n^2$ is the energy of the $n$-th energy level. 

Despite its low density, the negative H ion, H$^-$, plays an important role in the opacity near the photosphere at near IR and visible frequencies~\cite{rau}.  
The photoionisation cross section of H$^-$ does not have a sharp edge at its ionisation threshold $\omega=E_-=0.75$ eV because the electrons are ejected in a p-wave~\cite{rau,Andersen2004} due to the neutrality of the H$^0$ final state. Instead of the $\propto\Theta(\omega-E_n)/\omega^3$ behaviour of other contributions to $\Pi_{\rm bf}$ one gets $\propto(\omega-E_-)^{3/2}/\omega^3$. 
We have taken the photonization cross section from~\cite{Ohmura1959}. 

These contributions have associated refractive parts, which can be computed through the Kramers-Kronig relations, neglecting the $\omega$ dependence of the Gaunt factors. 
The bound-free contribution of H$^0$ turns out to be of similar size to the bound-bound contributions and has to be included. Due to the low density and the smoothness of the threshold the 
refractive part associated with the photionization of H$^-$ turns out to be negligible. 
Neglecting the frequency dependence of $F_{\rm bf}$ we find
\be
\Pi_{r,\rm bf}=m^2_{\gamma\rm bf} \simeq 
\frac{8\pi m_e \alpha^5}{3\sqrt{3} \omega^2} n_{\rm H^0}
 \sum_{n} {\cal Z}_n \frac{1}{n^5}
\(\frac{\omega^2}{E_e^2}-\log\(\frac{E_e^2}{|E^2-\omega^2|}\)\)
\ee

Finally, we want to include the absorption coefficient due to metals. In principle, suitable generalisation of the above formulas is possible and regularly performed for opacity calculations. 
For the scope of this paper it is enough to use monochromatic opacities present in the literature. We chose the detailed opacities of the Opacity Project from~\cite{Seaton:2004uz}, which are available in tables for different temperatures and electron densities that can be interpolated at will. 
The absorption coefficient is tabulated for each element as an effective cross section that includes the contribution from scattering, bound-bound, bound-free and free-free reactions involving the element $Z$, 
$\sigma_{Z}(\omega)=\Gamma_Z/n_Z(1-e^{-\omega/T})$ where $n_Z$ is the density of the corresponding nucleus. 
We thus build
\be
\Pi_{i,Z} = \omega \sum_Z \Gamma_Z=\omega (1-e^{-\omega/T})\sum_Z \sigma_Z n_Z . 
\ee

The polarisation tensor $\Pi$ is the sum of all the above contributions, 
\be
\Pi= \Pi_{\rm free}+\Pi_{\rm bb}+\Pi_{\rm bf}+\Pi_{\rm ff} +\Pi_Z. 
\ee
\subsection{A model for refraction and absorption in the Sun}

The mass density $\rho$, chemical composition and temperature $T$ as a function of solar radius can be taken from a standard solar model calculation. The state of the art on solar modelling is able to fit all available solar data from helioseismology and neutrino flux detectors with typical percent accuracy. Small discrepancies with observations still exist~\cite{oai:arXiv.org:1104.1639} after the recent revision of solar abundances of CNO species~\cite{oai:arXiv.org:0909.0948}, which lowered slightly the opacity, but they do not compromise the degree of precision of our calculation. 
Interestingly, the authors of~\cite{Vincent:2012bw} proved that the existence of hidden photons cannot possibly influence the determination of the light elements in the Sun.  
 
\begin{figure}[t]
\begin{center}
\includegraphics[width=0.45\textwidth]{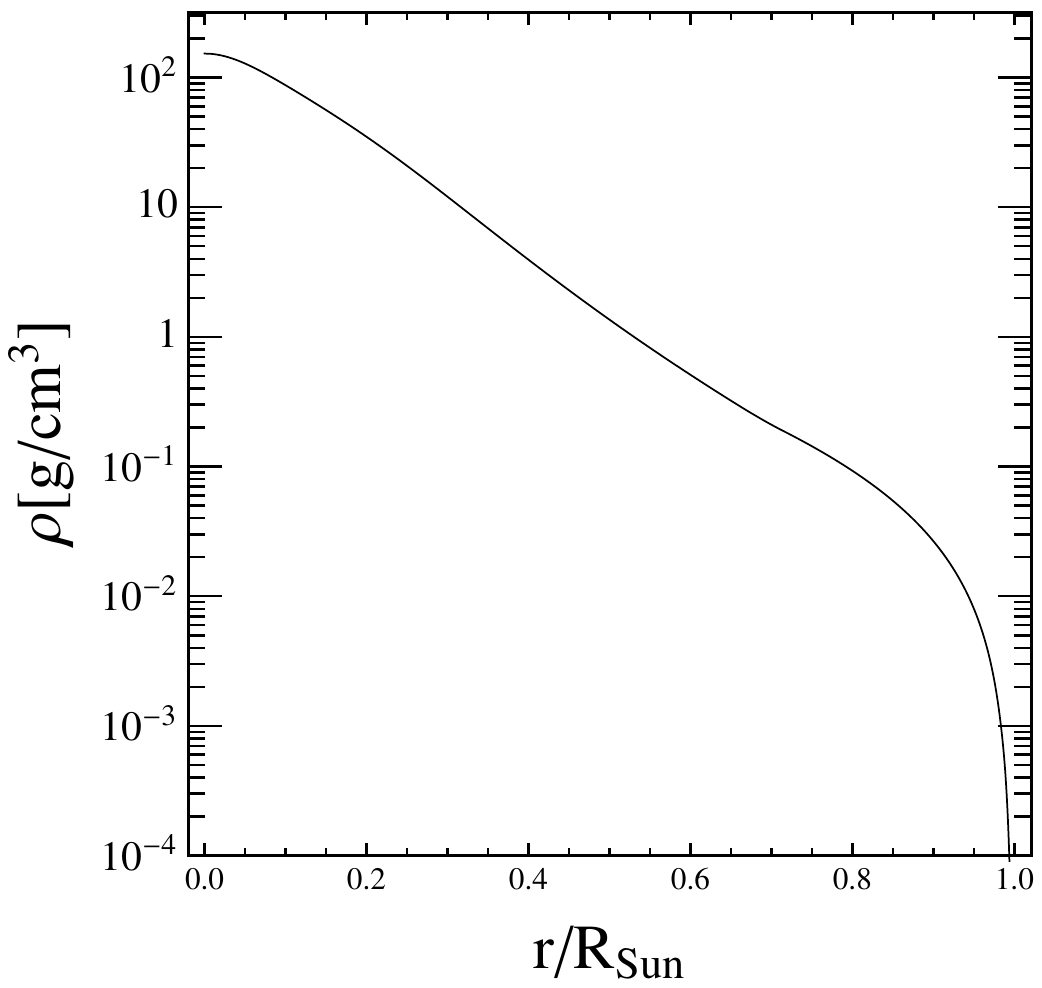}
\includegraphics[width=0.45\textwidth]{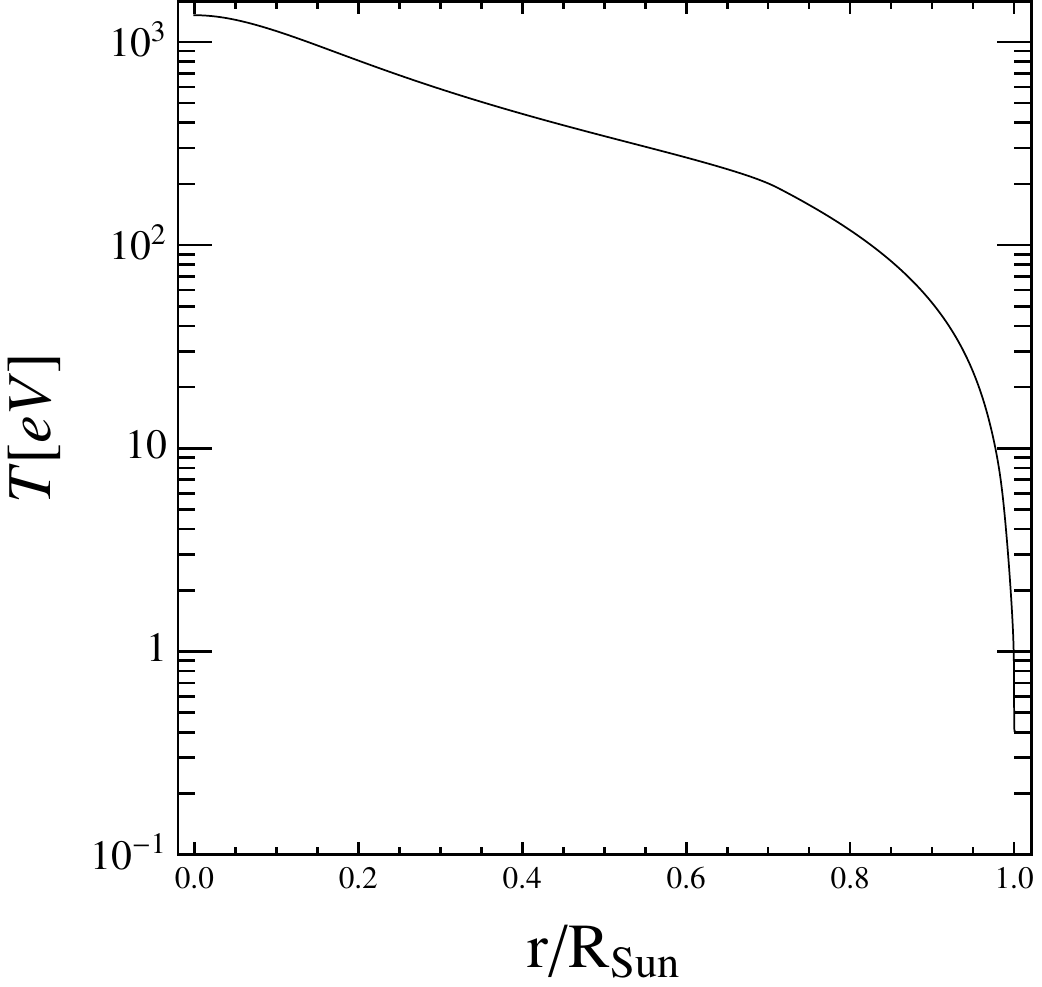}
\caption{Mass density, $\rho$, and temperature, $T$, of the solar model of ~\cite{saclay} as a function of the solar radial coordinate $r$ measured in units of the solar radius, $R_{\rm Sun}=0.6955\times 10^6$ km. }
\label{fig:solarmodel}
\end{center}
\end{figure}

\begin{figure}[t]
\begin{center}
\includegraphics[width=0.45\textwidth]{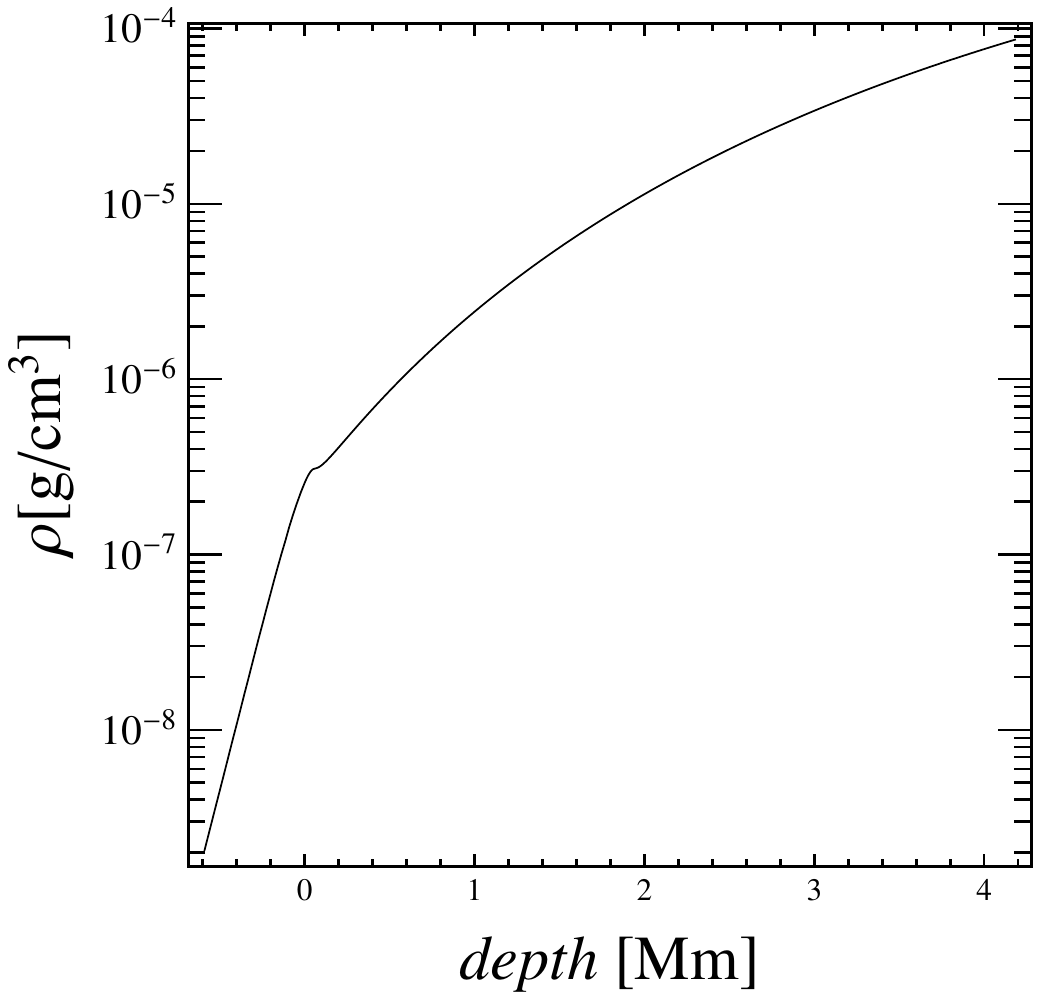}
\includegraphics[width=0.42\textwidth]{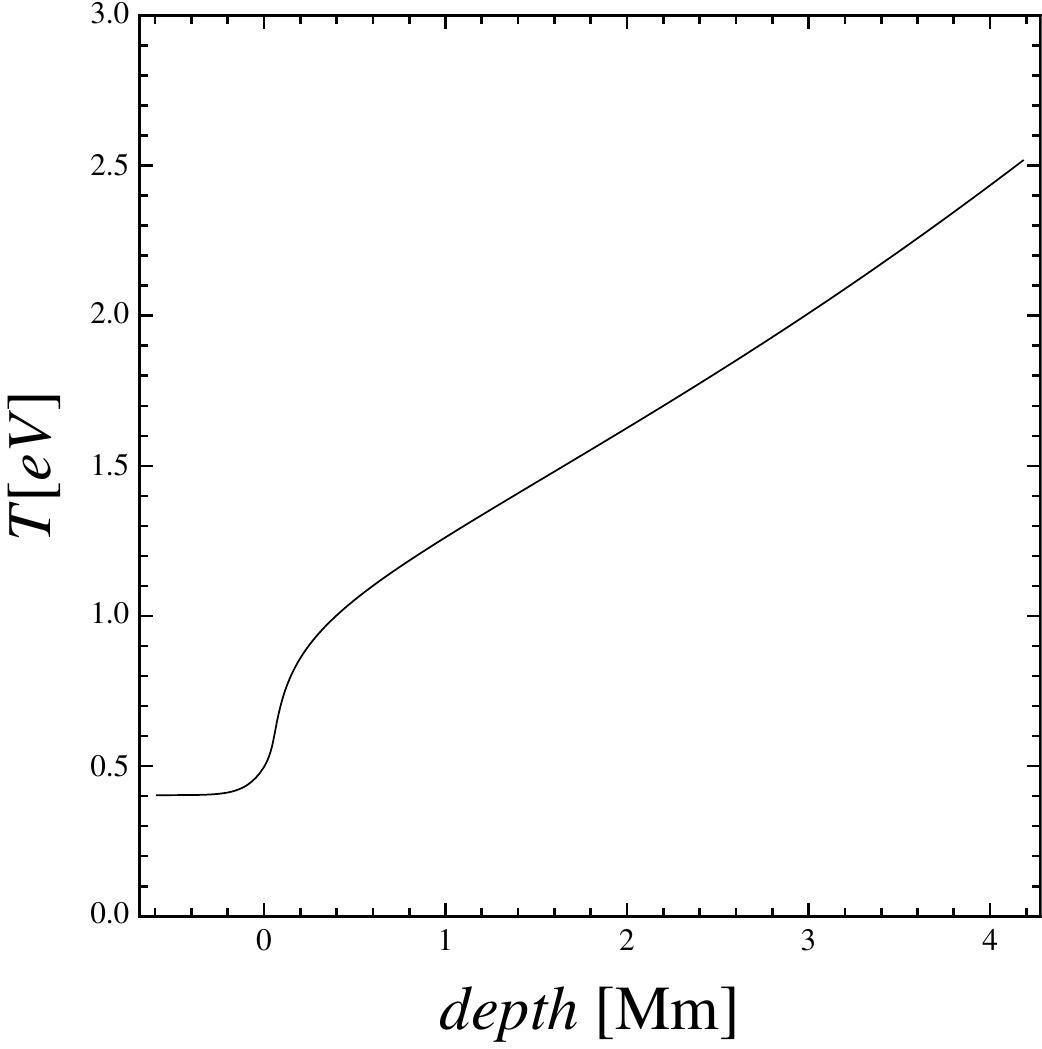}
\caption{Mass density, $\rho$, and temperature, $T$, of the solar model of ~\cite{saclay} as a function of the depth inside the solar surface measured in Megameters. }
\label{fig:solarsurface}
\end{center}
\end{figure}

In this paper we make use of the Saclay seismic solar model~\cite{saclay,Couvidat:2003ba} available at~\cite{saclayURL} which has excellent detail in the outer layers of the Sun, our primary concern. The mass density and temperature profiles are shown in Fig. \ref{fig:solarmodel}. 
For future reference, Fig. \ref{fig:solarsurface} shows the outer layers of the Sun in more detail as a function of the depth towards the solar core.

\begin{figure}[t]
\begin{center}
\includegraphics[width=0.43\textwidth]{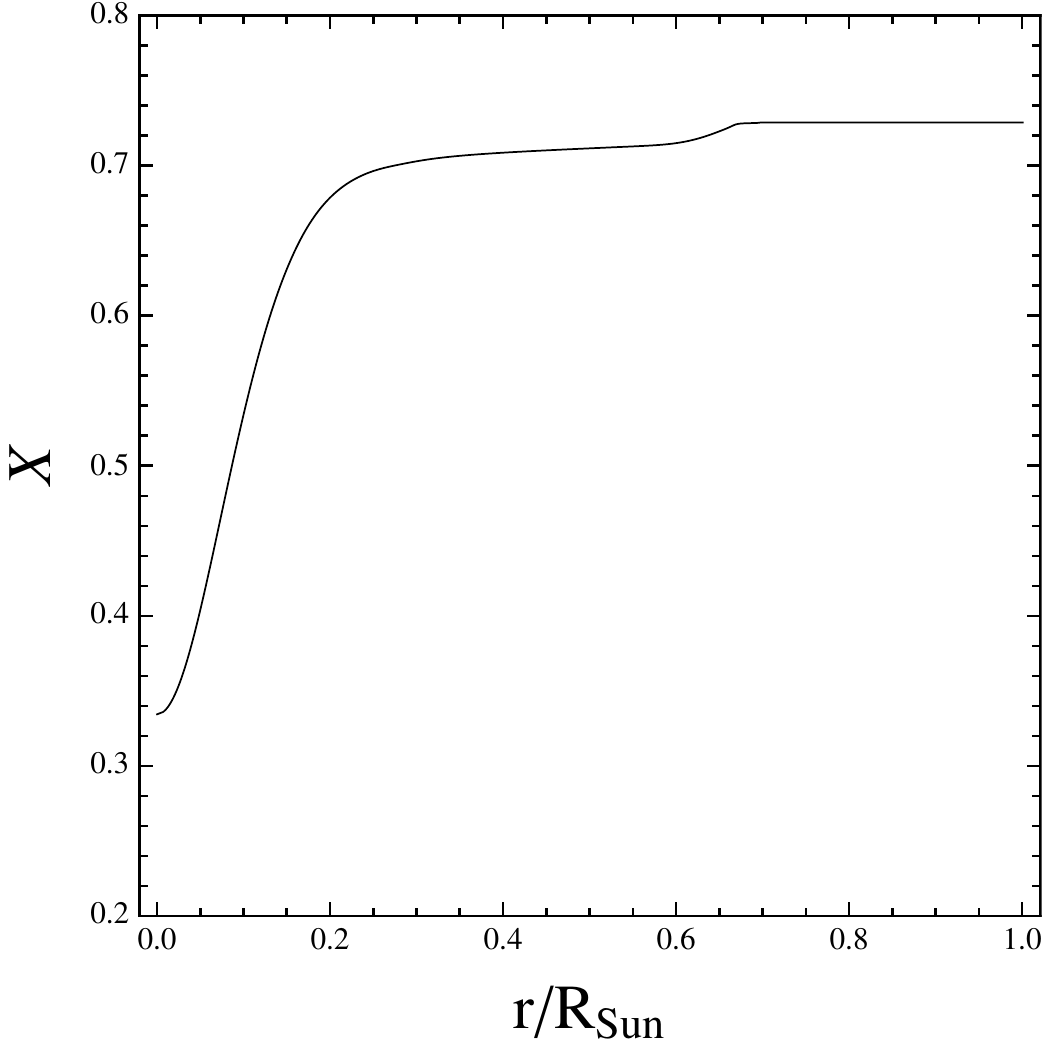}
\includegraphics[width=0.45\textwidth]{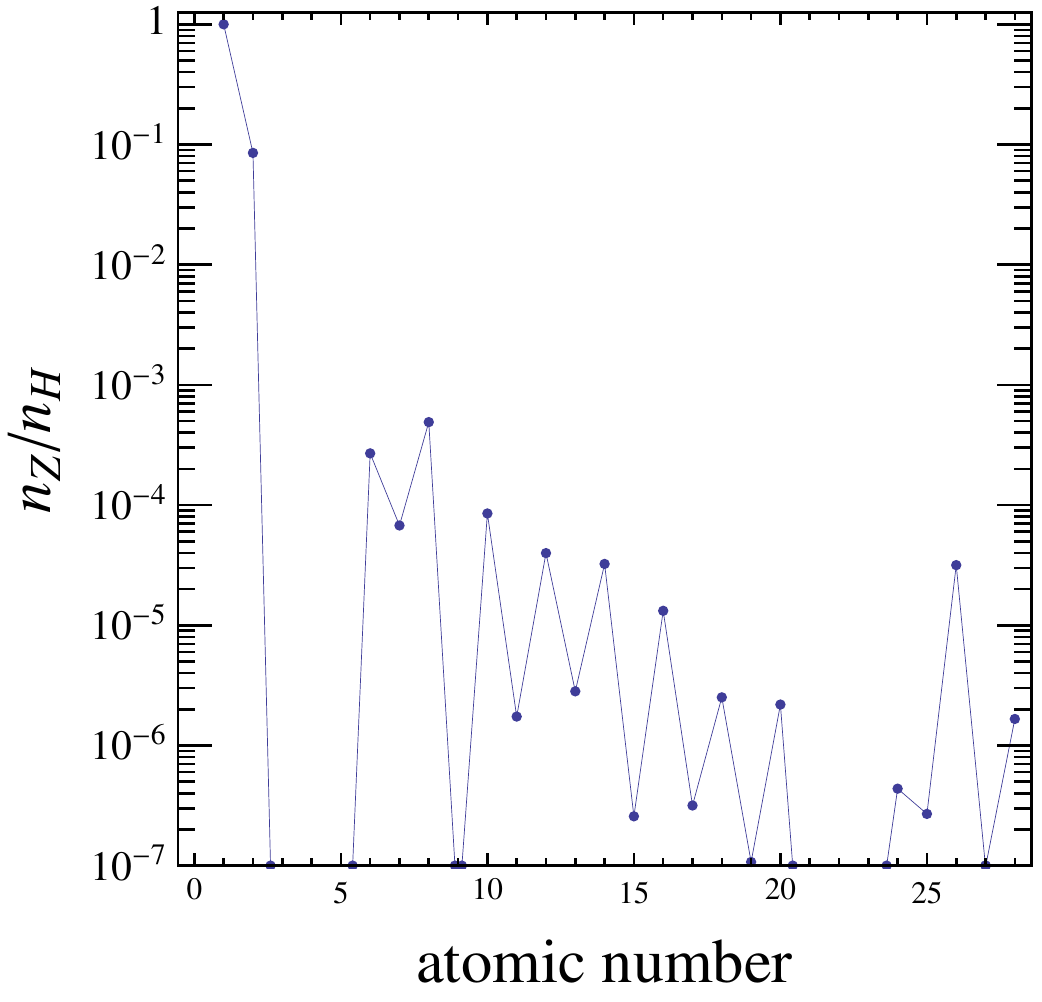}
\caption{LEFT: Hydrogen mass fraction of the solar model of ~\cite{saclay}. RIGHT: Chemical composition of the solar surface as function of the atomic number (number density of atoms of atomic number $Z$ normalised to that of hydrogen). }
\label{fig:solarcomposition}
\end{center}
\end{figure}

The surface chemical composition from~\cite{oai:arXiv.org:0909.0948} is shown in Fig.  \ref{fig:solarcomposition}. Hydrogen is $\sim 10$ times more abundant than helium, $\sim 1000$ times more abundant than CNO elements and more than $10^4$ than higher$-Z$ elements. 
The composition is relatively constant in the solar interior but we take into account diffusion from the solar model AGSS09~\cite{Serenelli:2009yc}, available at~\cite{AGSS09} (the solar model~\cite{saclay} does not provide the chemical abundance as a function of radius).  

The most relevant atomic species is of course, hydrogen. 
As we will see further on, the resonance region for low mass HPs happens close to the photosphere but still in the solar interior. 
In this region the temperature is so low that He atoms are not ionised (their first ionisation energy is $\sim 25$ eV) and their contribution to free-free transitions is negligible. 
Also, the main atomic transitions and ionisation thresholds lie in the far UV, and therefore do not affect visible light absorption. For the same reason, their contribution to refraction is also negligible. 
Note from \eqref{eq:Pibb} that the contribution to $\Pi_r$ of far UV resonant frequencies in the visible is $\propto -n_Z(\omega/\omega_r)^2$ and therefore those of helium are much less important than those of hydrogen. 
Higher $Z$ elements (often referred to as metals) are too scarce to compete with hydrogen in absorption in free-free collisions but contribute to bound-bound and bound-free absorption because they have resonant transitions in the visible and ionisation thresholds in the not-so-far-UV. 
The effects, if competitive with hydrogen have to be necessarily localised in frequency to very narrow intervals around the resonances and thresholds. 
Since we are interested in the continuum flux rather than a spectroscopically resolved spectrum we will neglect these contributions. 
The only effect in which metals have a leading role is to set the free electron density in the photospheric layers of the Sun where hydrogen is highly neutral. Only in this restricted sense we include metals in the solar refractive model. 

\subsubsection*{Partition functions and atomic ionisation}
The density of free electrons and of the 1s state of neutral hydrogen are the most influential factors that set the value of $m_\gamma^2$. Through this dependence, they determine the locus of the photon-HP resonant conversion region (and hence the temperature $T$ of the emission), and also the size of the conversion region, which sets the absolute value of the flux. 
Thus, for our purposes it is of capital importance to have a robust estimate of these quantities. 
The occupation probabilities, ${\cal Z}_n$, and the neutral H fraction cannot be accurately estimated with the partition function of an ideal gas and the Saha ionization equation. The first requires an {\rm ad hoc} cutt-off to the number of excited states, and we do not know a-priori if this is justified. Excited states could be very populated at large temperatures and contribute to $m_\gamma^2$ as much as the 1s. The second is well known to fail at large densities and not too large temperatures, predicting too little ionisation, even in the solar centre!. 
Fortunately, the calculation of the ionisation equilibrium is a fundamental ingredient of the equation of state of astrophysical plasmas and radiative opacity, so a large body of literature exists on the matter. 
The problems of the divergence of the partition function and the Saha equation are both cured by include interactions between the plasma species, which alter the bound states when the density gets closer to atomic density.  Here, we use the partition function of Hummer and Mihalas (HM)~\cite{HM88}, 
\be
{\cal Z}_n =2 n^2 w_n e^{\frac{E_n}{T}}/{\tilde {\cal Z}} \quad ;  \quad {\tilde {\cal Z}}=\sum_n 2 n^2 w_n e^{\frac{E_n}{T}}, 
\ee
where the factors $w_n$ are called ``occupation probabilities'' and encode the non-ideality of the gas. 
The most relevant effect to be included are the perturbations on the atomic states by the slow-varying electric fields of the plasma ions (protons in our case). Bound states suffer Stark shifts in the binding energies which drive ionisation very efficiently. In particular, there is a certain critical field above which a given bound state is has a lifetime smaller than its orbit period and it effectively destroyed. 
These effective occupation probabilities thus measure the probability of the atoms to be in a perturbing field such that a given bound state does still exist.  
Using a Holtsmark probability distribution for the magnitude of the fields, they find
\be
w_n = Q\(\frac{K_n}{4 Z}\frac{E_n^2}{\alpha^2}\(\frac{4 \pi n_p}{3}\)^{-2/3}\) 
\ee
where $Q(x)$ is the cumulative of the Holtsmark distribution, $n_p$ is the proton density and $K_n=16 n^2(n+7/6)/3(n+1)^2(n^2+n+1/2)$ for  $n\geq 3$ and $K_n=1$ for $n\leq 3$. 

The Saclay solar model~\cite{saclay} uses the OPAL equation of state (EOS) of Rogers and Iglesias~\cite{Iglesias1996}, which is not based on the HM chemical picture but on the physical picture (see~\cite{Rogers1986} for the derivation of occupation probabilities). 
The Opacity Project EOS is instead based on the much simpler HM picture and it has been shown to agree well with the OPAL~\cite{Trampedach2006}.   
Therefore, by using the HM formalism we simplify much our calculations without the danger of large inconsistencies on the chosen solar model. 
Note finally, that the chemical picture of HM was designed in principle to be valid at the small densities relevant  in stellar envelopes, $\rho\lesssim 10^{-2}$g/cm$^3$ but it was found to be valid up to much denser regions, and in particular, the whole Sun~\cite{christensen1988}. An update of the micro-field distribution including particle correlations showed an excellent agreement with OPAL up to the solar centre~\cite{Nayfonov:1999qd}, but for our purposes we do not need such levels of precision.     

\begin{figure}[h]
\begin{center}
\includegraphics[width=0.45\textwidth]{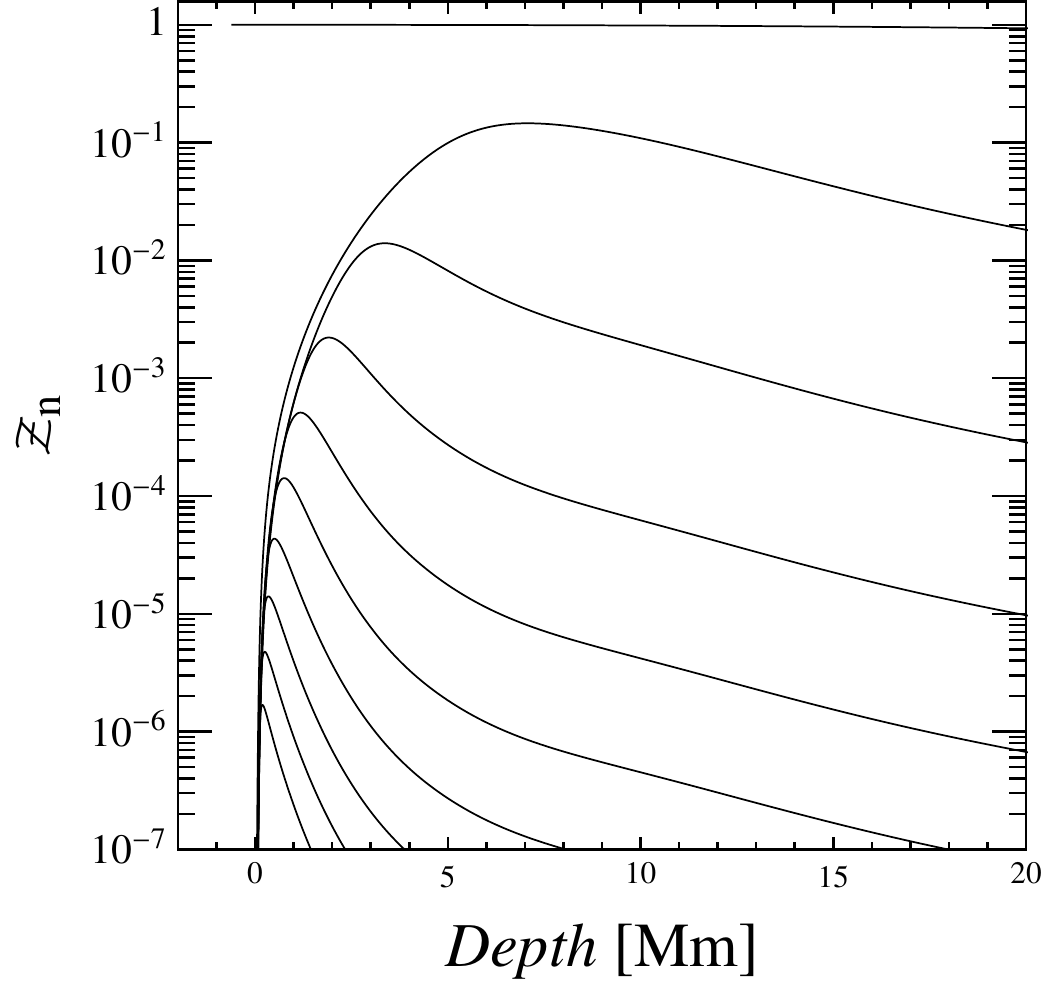}
\includegraphics[width=0.45\textwidth]{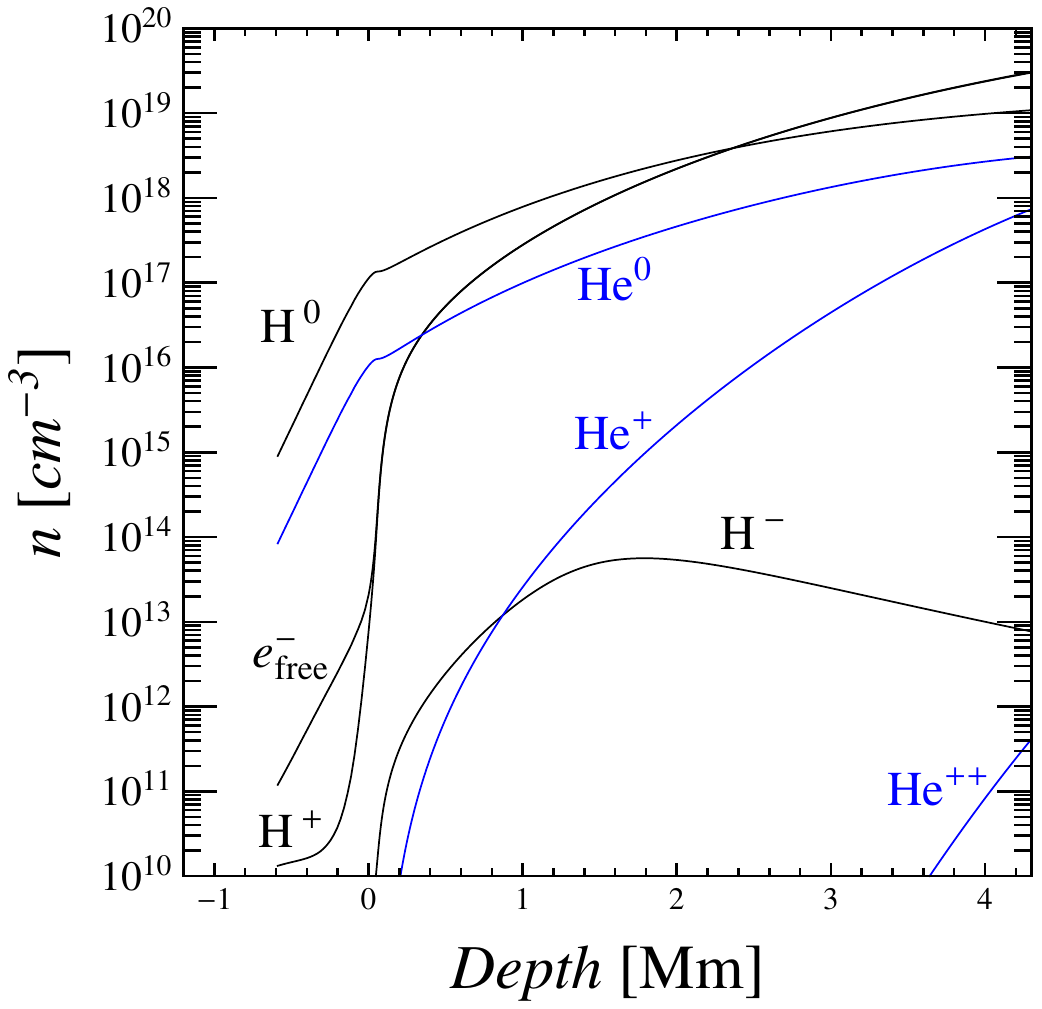}
\caption{LEFT: Probability of the bound (n=1,topmost) and excited states (n=2,3..,10, top to bottom) inside the solar plasma as function of depth inside the surface. 
RIGHT: Density of the different ionisation species relevant for this model of the solar refraction.  }
\label{fig:partfun}
\vspace{-.71cm}
\end{center}
\end{figure}

Following HM, the ionisation equilibrium is derived from the minimisation of the Helmholtz free-energy where every atomic state is treated as a different species. 
We consider the states of the hydrogen and helium and the first ionisation of metals. There are three states of H (H$^-$, H$_0$ and H$^+\equiv p$, i.e  protons), three of He (He$^0$, He$^+$ and He$^{++}$) and two for each metal. The effects of H bound states are only important in the surface, where excited states are scarce and therefore their correlations with the proton density irrelevant for the free energy. In this lucky situation, the minimisation of the free-energy, $F$, subject to the stoichiometric relations $\frac{\partial F}{\partial n_p}+\frac{\partial F}{\partial n_e}=\frac{\partial F}{\partial n_{\rm H}^0}$ from chemical equilibrium in the ionisation reaction $p^++e^-\leftrightarrow {\rm H}^0+\gamma$ leads to a Saha-like equation
\be
\frac{n_{{\rm H}^0}}{n_p n_e^{\rm free}}=\frac{{\bar {\cal Z}}}{2}\(\frac{m_e T}{2\pi}\)^{-3/2}e^{\frac{E_{\rm 1s-\alpha k_{\rm D}}}{T}}, 
\ee
where we have defined the usual partition function normalised to the ground state energy $\bar {\cal Z}=e^{E_{\rm 1s}/T}\tilde {\cal Z}$ (recall that $E_n>0$ in our convention) and the Debye correction involves the Debye screening scale defined before. The equations for H$^-$, He and metal ionisation are completely analogous. 
The 2 Saha equations of H, the 2 of He and the one for each metal are to be solved self-consistently with the constraints given by the solar abundances ($n_{\rm H, total}=n_{\rm H^+}+n_{\rm H^0}+n_{\rm H^-}$ for H and $n_{\rm He, total}=n_{\rm He^{++}}+n_{\rm He^+}+n_{\rm He^0}$ for He) and the expression of the free electron density
\be
n_e^{\rm free} = n_{\rm H^+}-n_{\rm H^-}+2 n_{\rm He^{++}}+n_{\rm He^{+}}+\sum_Z n_{\rm Z^{+}} . 
\ee
The resolution of this system outputs the required ion densities and bound-state probabilities ${\cal Z}_n$, which we show in Fig. \ref{fig:partfun}.  
  
Outside the photosphere, neutral H and He are the most abundant species. The free electron density has dropped much, although not as much as the proton density due to the electrons donated by metals (mostly by Mg, C, Si and Fe). 
Across the surface ($\rho\sim 3\times 10^{-7}$g/cm$^{-3}$) the ionised species have a sharp rise due to the sharp rise of the temperature. The ionisation of H is around 90\% around  $\rho\sim 10^{-3}$g/cm$^{-3}$ and that of helium around $\rho\sim 10^{-2}$g/cm$^{-3}$. 
Let us warn that the free electron density profile obtained by this whole procedure does not fit well with the one provided by the solar model~\cite{saclay} below $\rho \sim 10^{-6}$ g/cm$^3$ or so. 
Our calculations show a decrease of two orders of magnitude around $\rho = 3\times 10^{-7}$ g/cm$^3$, 
which is much milder in the solar model table provided in~\cite{saclayURL}. 
The reason is most likely that the available model used in~\cite{saclay} employs a basic Hopf atmosphere model, see~\cite{Couvidat:2003ba}. 
The results of our calculation agree perfectly with the solar atmosphere model of Kurucz~\cite{oai:arXiv.org:astro-ph/0405087}, which the authors of the Saclay solar models employ in the later publication~\cite{Couvidat:2003ba} to improve the agreement with Helioseismological data. This constitutes a reassuring test of our calculations.  

\subsubsection*{Results}

\begin{figure}[t]
\begin{center}
\includegraphics[width=0.45\textwidth]{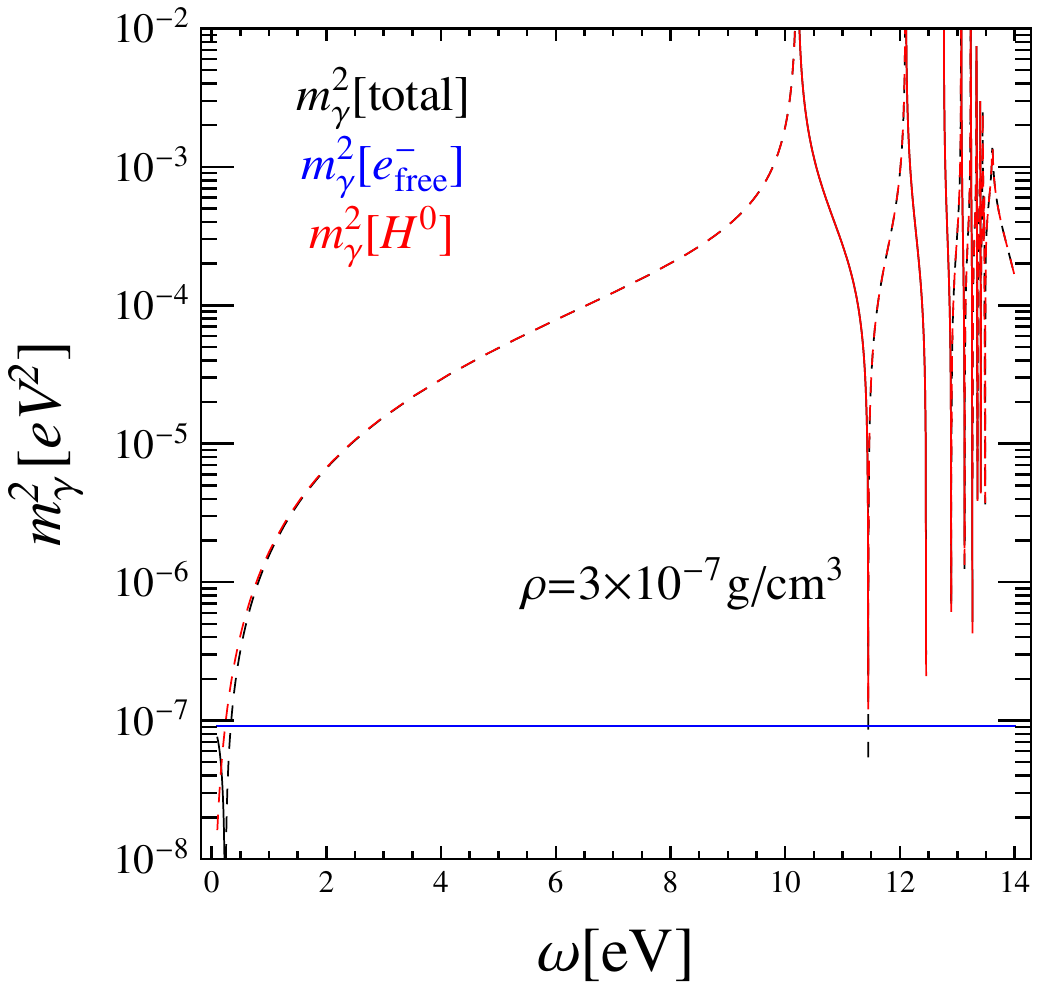}
\includegraphics[width=0.45\textwidth]{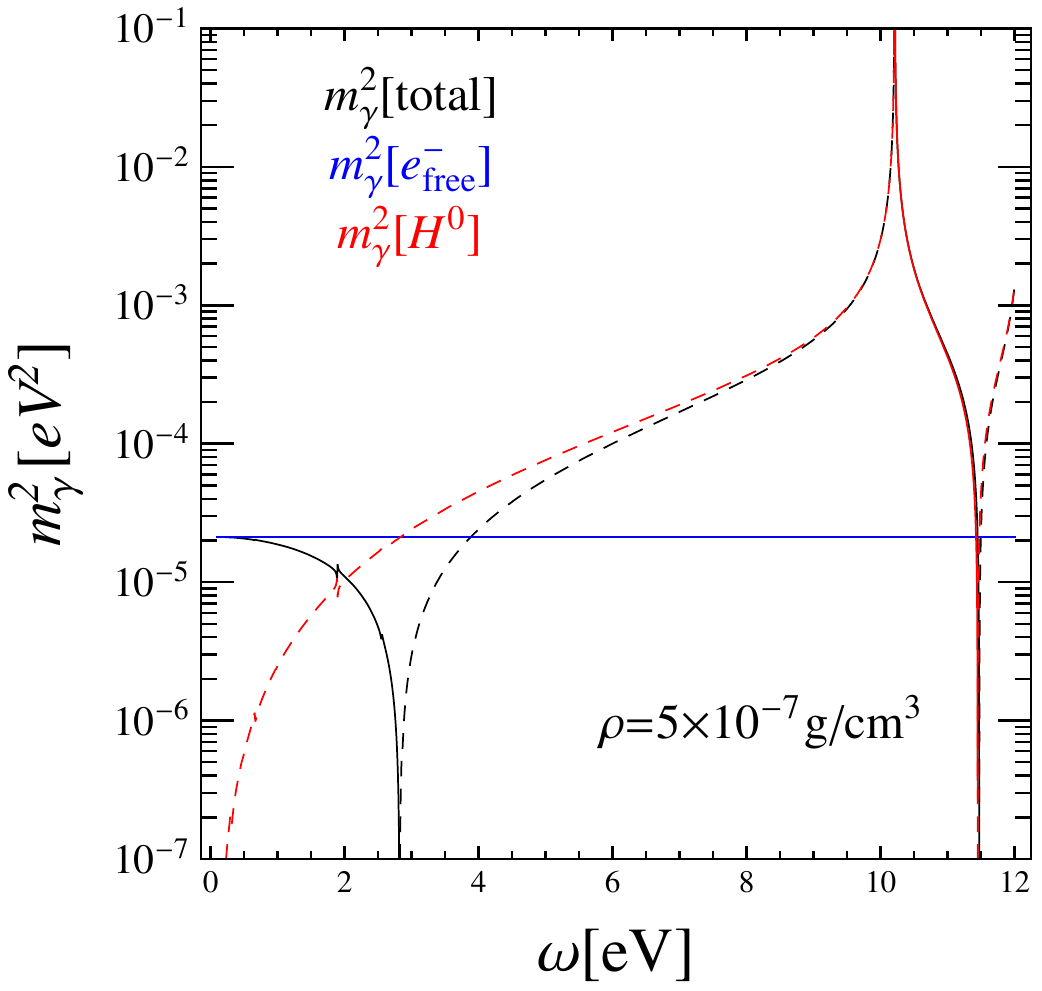}
\includegraphics[width=0.45\textwidth]{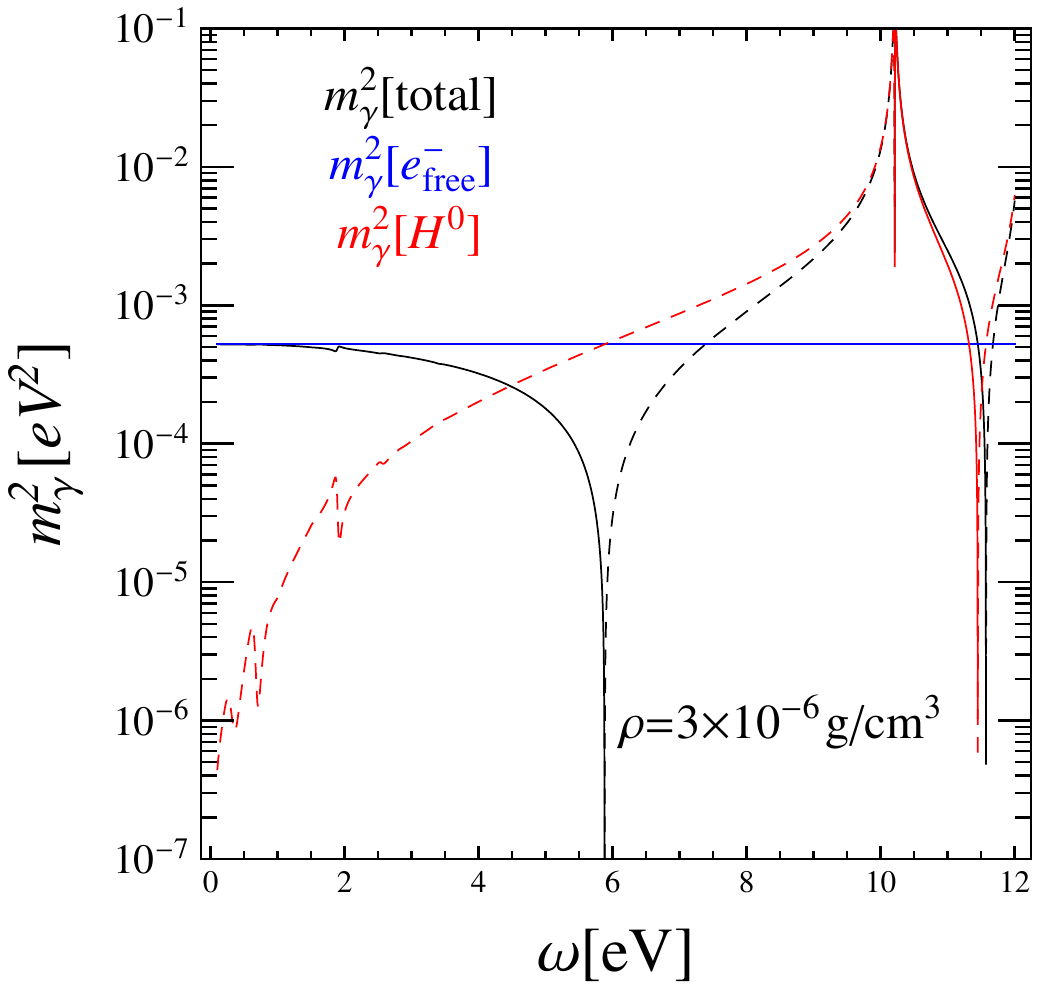}
\includegraphics[width=0.45\textwidth]{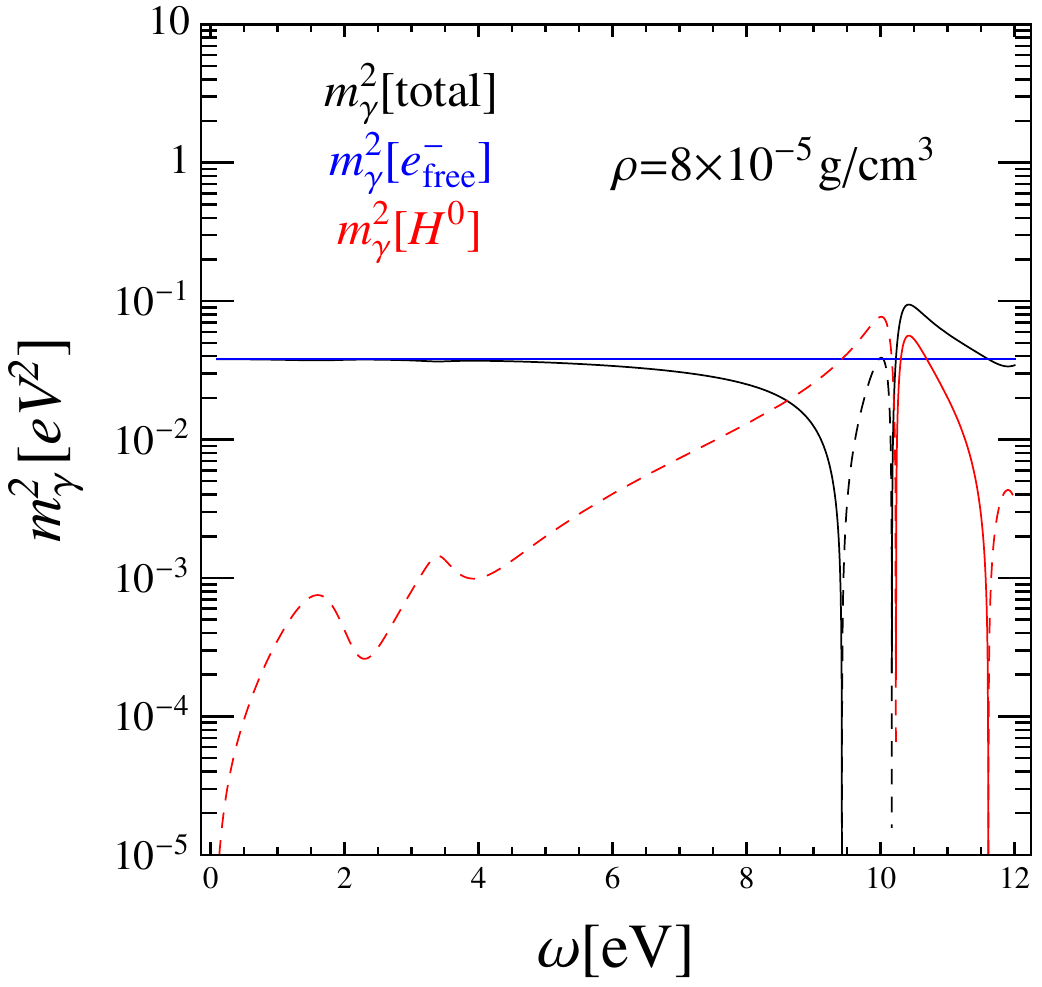}
\includegraphics[width=0.45\textwidth]{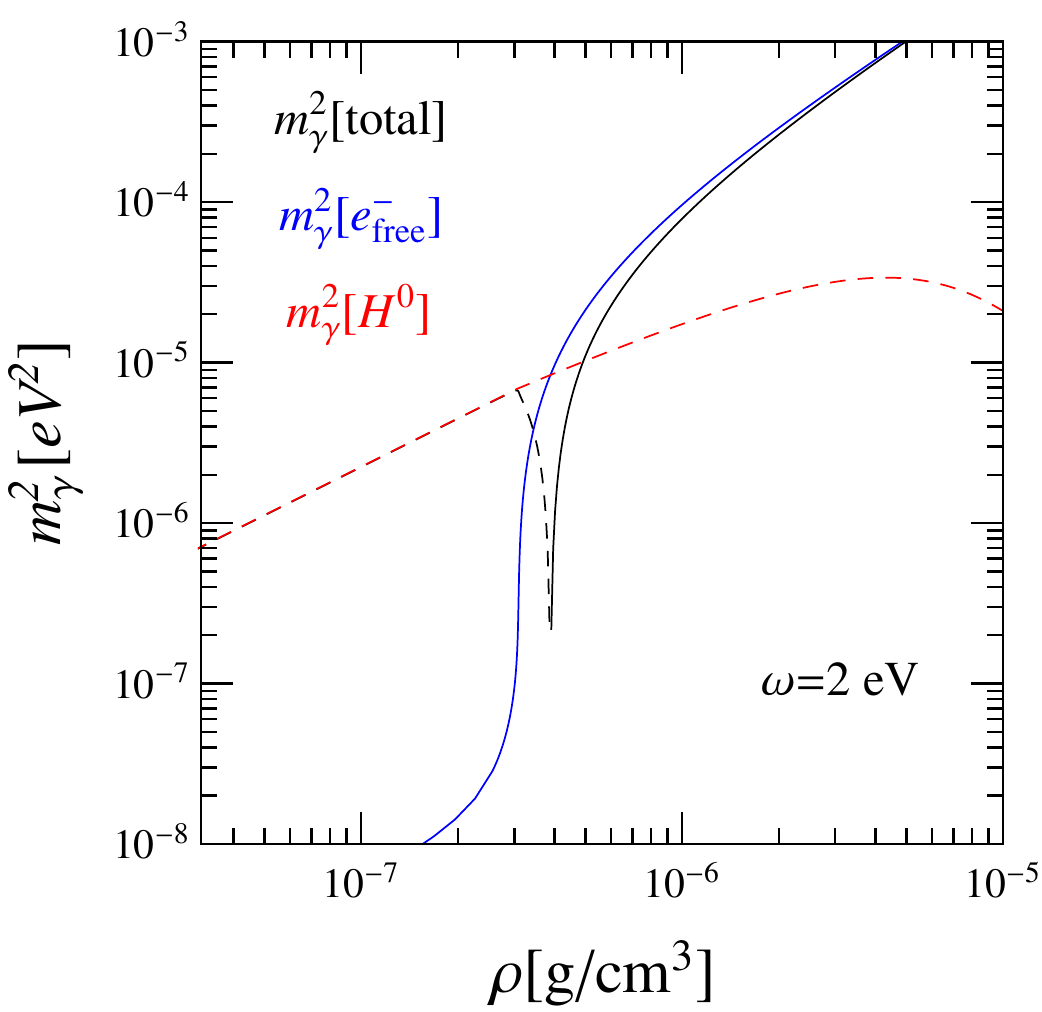}
\includegraphics[width=0.45\textwidth]{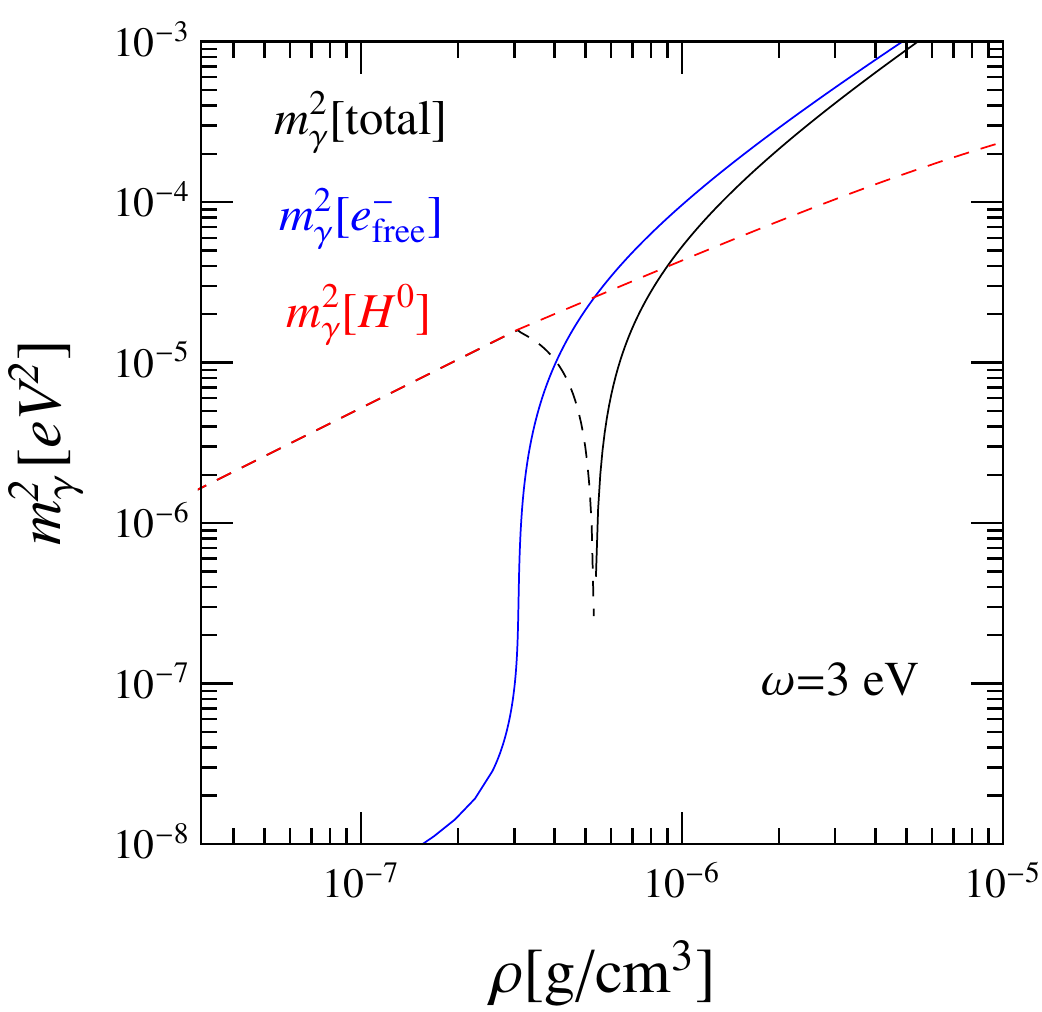}
\caption{The effective photon mass  $m_\gamma^2(\omega,t)$ in the solar model built for this paper (black lines). 
The contribution from free electrons is shown in blue and that of neutral H (through bound-bound and bound-free transitions) as red. 
When $m_\gamma^2(\omega,t)$ is negative we have plotted $-m_\gamma^2(\omega,t)$ as a dashed line.  
Upper and middle plots show the energy dependence for four positions inside the Sun. 
Lower plots show the dependence on the solar interior position, labeled by the mass density, for photon energies of $\omega=2,3$ eV (wavelength=$616,413$ nm, respectively). 
}
\label{fig:m2g}
\end{center}
\end{figure}

We can now construct the functions $m_\gamma^{2}(\omega,r)$ and $\Gamma(\omega,r)$ that we require for computing the solar HP flux. 
We show some relevant plots of the dependence of $m_\gamma^2(\omega,t)$ on $r$ and $\omega$ in Fig. \ref{fig:m2g}. 
In the upper plots we see the frequency dependence of the different contributions. Free electrons provide a frequency independent contribution (blue line) while atomic resonances at $\omega_r$ give positive contributions for $\omega>\omega_r$  and negative for $\omega<\omega_r$. 
Negative values are shown as dashed lines in the log-plot. 
The most notable line is the Ly-$\alpha$ at $\omega\sim 10.2$ eV although all the Lyman series has visible contributions. The value of $m^2_\gamma$ is determined thus not only by the density but very importantly by the ionisation fraction. 
In the surface, i.e. just above the fast drop of density visible in Fig. \ref{fig:partfun} ($\rho\sim 3\times 10^{-7}$g/cm$^3$) most of the H is neutral and $m_\gamma^2$ is negative in all the visible spectrum. 
Moving deeper in, at a density $\rho\sim 5\times 10^{-7}$g/cm$^3$ the ionisation fraction is already 10\%. 
Since the neutral H contribution is $\propto -\omega^2$ it becomes ineffective at low energies and such a small free electron density is able to make $m_\gamma^2$ positive in the red part of the spectrum.   
As we move inside the Sun, the ionisation fraction increases and neutral H becomes scarce. The region of negative $m_\gamma^2$ gets displaced deeper into the UV. At $\rho=8\times 10^{-5}$g/cm$^3$ it is only the $\sim 9-10.2$ eV range. 

\clearpage

These trends can be seen in the lower plots of  Fig. \ref{fig:m2g} where the dependence with the solar density is shown for a couple of frequencies. 
The positive free electron contribution drops suddenly at the photosphere with the ionisation fraction while the negative contribution from neutral H decreases softer and eventually dominates. Note that for these frequencies there is always a point in the Sun where $m_\gamma^2= 0$. Since the negative contribution is frequency dependent this point depends on frequency.  
This point will be very important for the discussion on photon-HP oscillation resonances. 
Note that as we consider smaller $\omega$ the neutral H contribution (red-dashed line) decreases as $\omega^2$ and  the point when it crosses the free electron (blue), i.e. the point where $m_\gamma^2=0$ moves out of the Sun. 
Higher energies have $m_\gamma^2= 0$ deeper in the solar interior because of their larger neutral H contribution.
Interestingly, due to the sharp drop of the free electron density at $\rho\sim 3\times 10^{-7}$g/cm$^3$ a sizeable range of frequencies will have $m_\gamma^2= 0$ around that region.

\begin{figure}[!h]
\begin{center}
\includegraphics[width=0.4\textwidth]{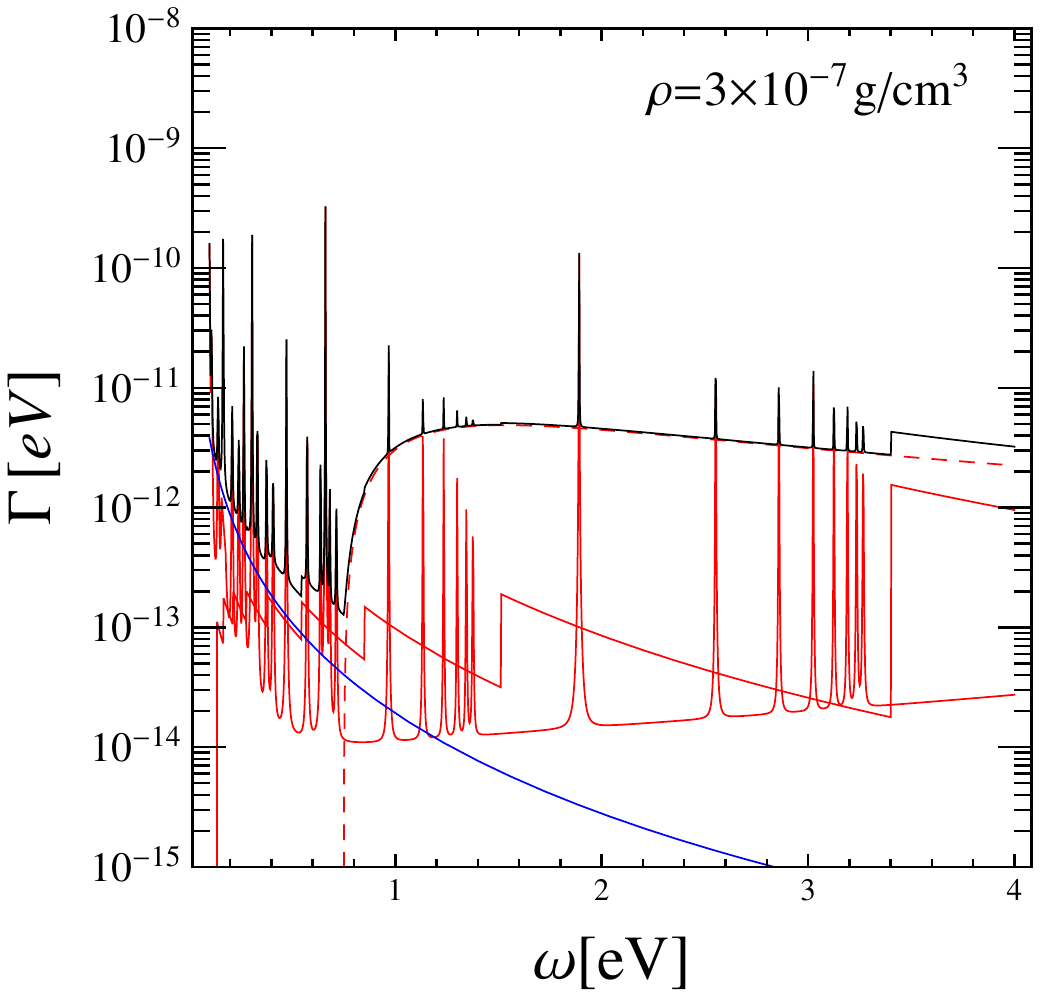}
\includegraphics[width=0.4\textwidth]{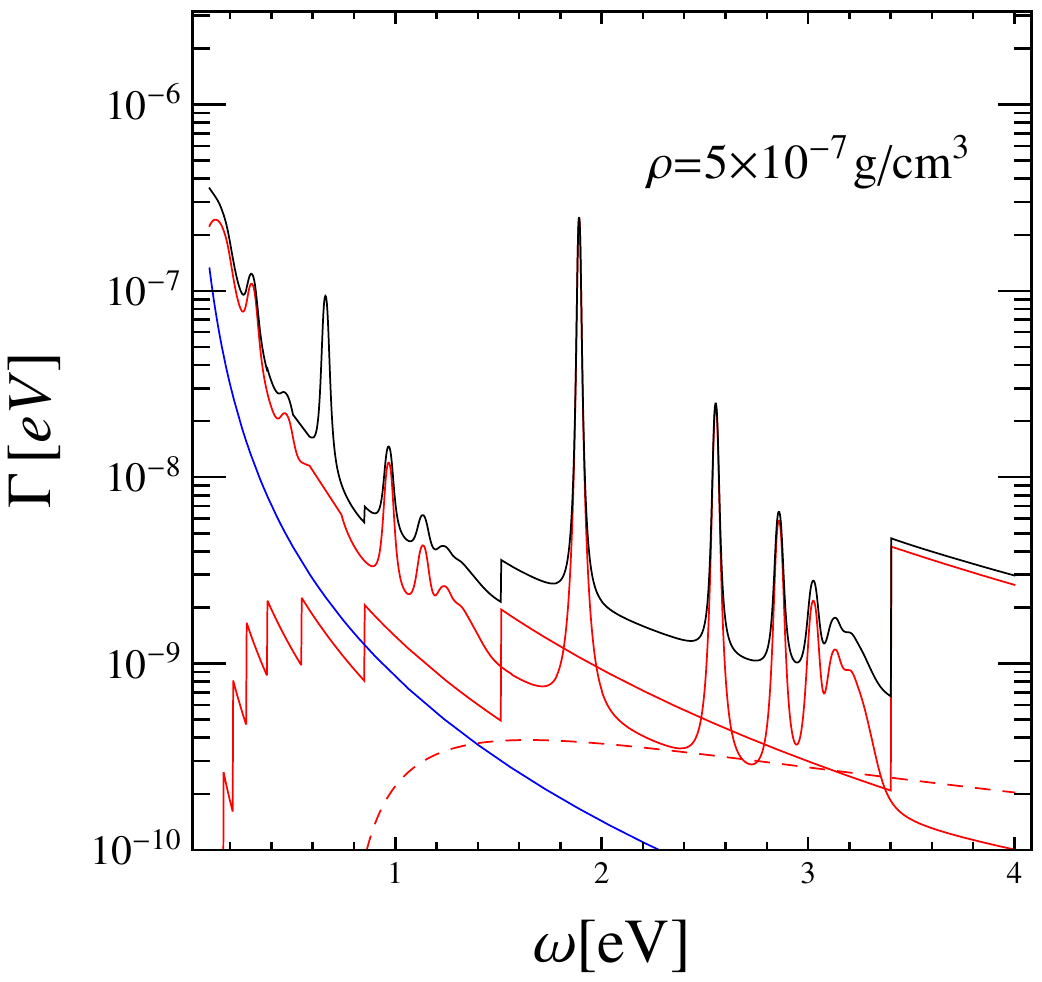}
\includegraphics[width=0.4\textwidth]{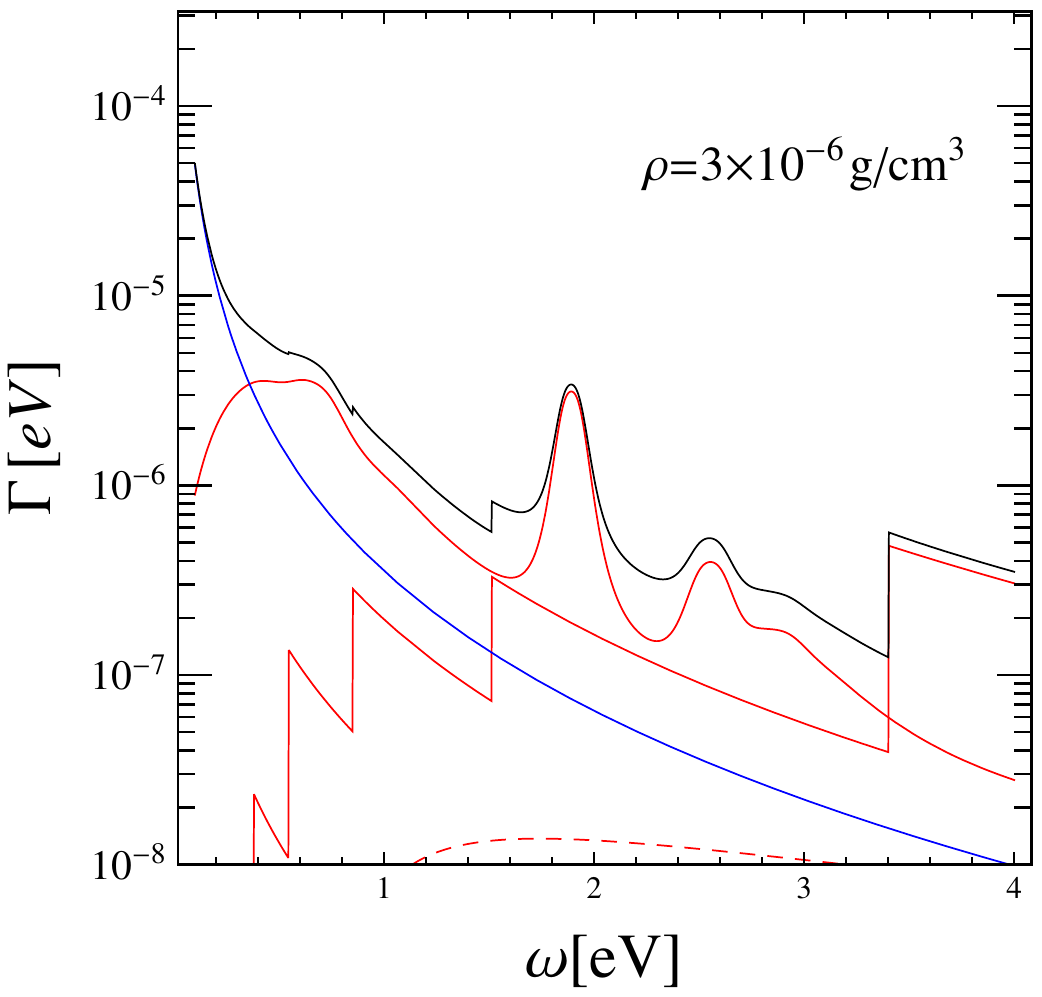}
\includegraphics[width=0.4\textwidth]{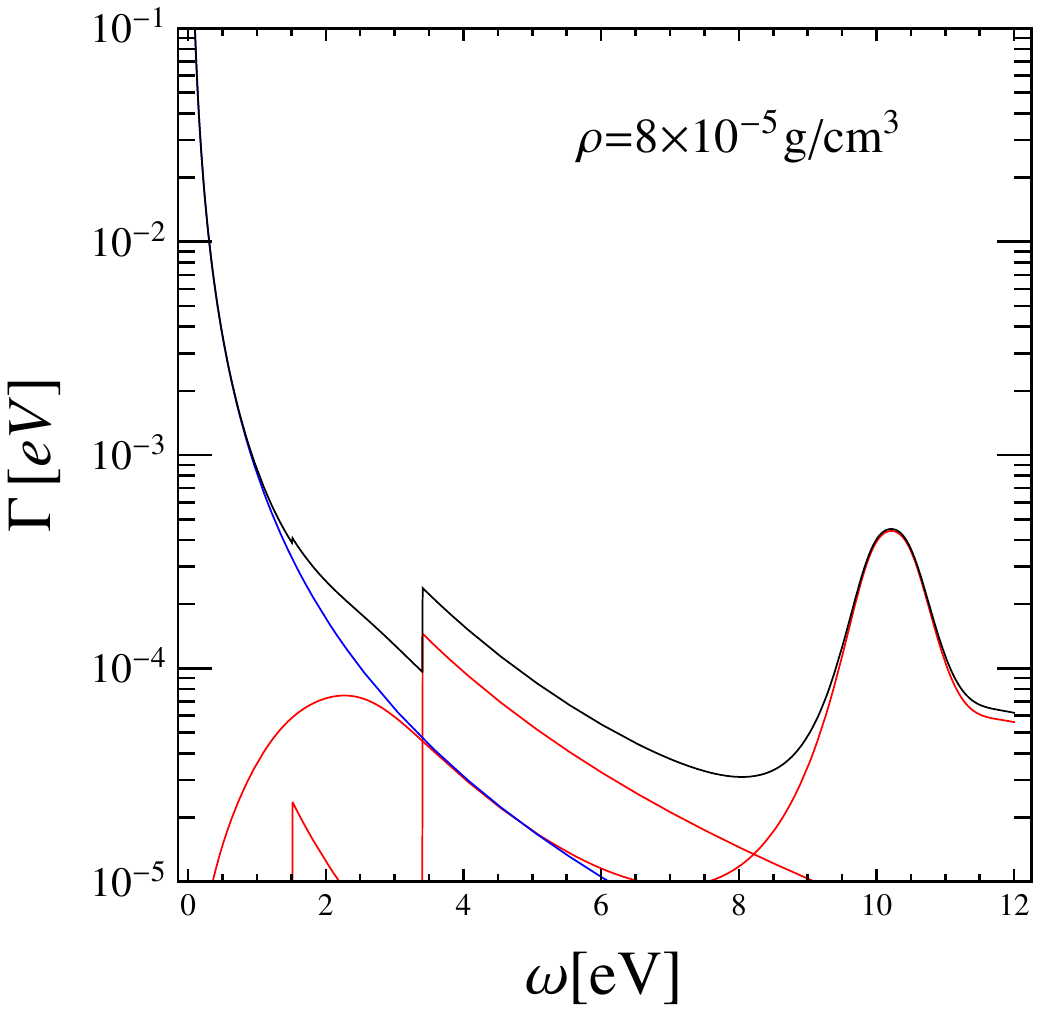}
\includegraphics[width=0.4\textwidth]{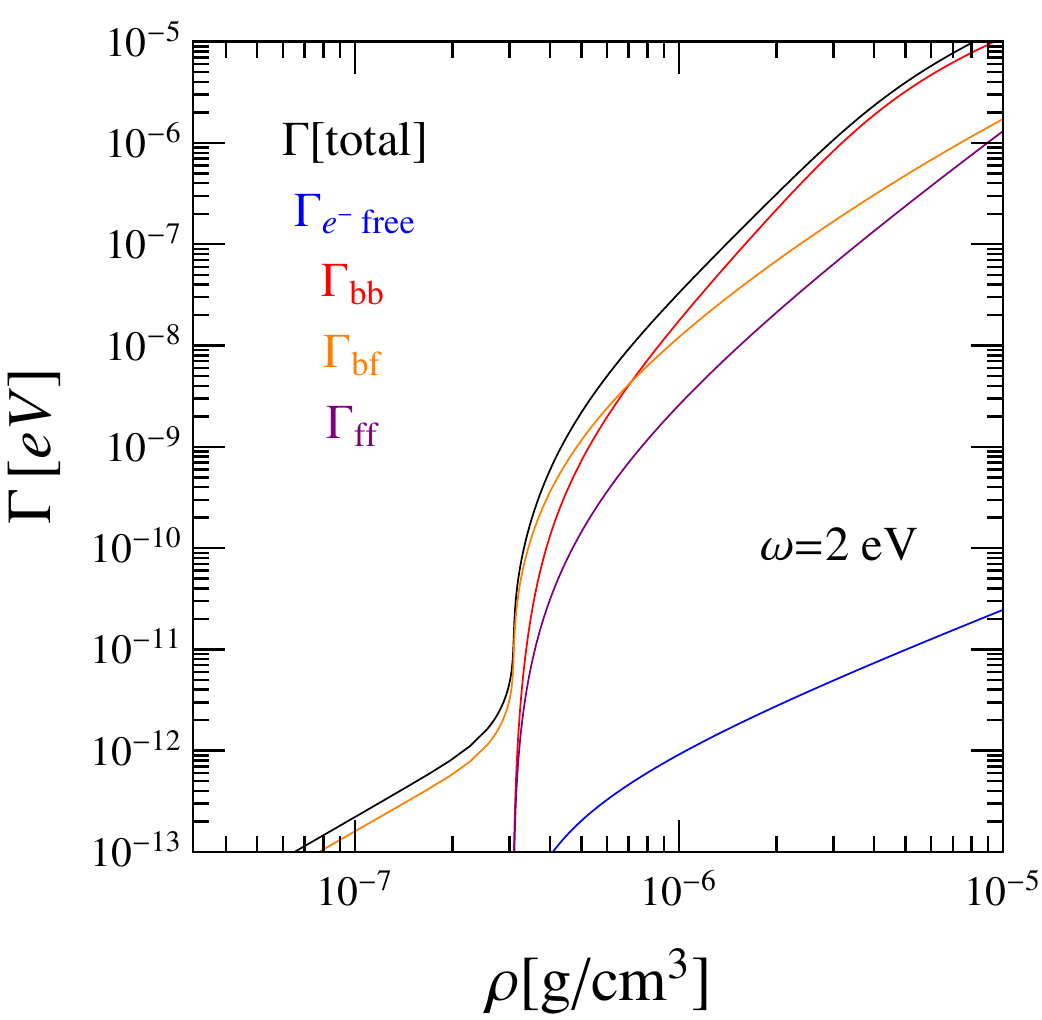}
\includegraphics[width=0.4\textwidth]{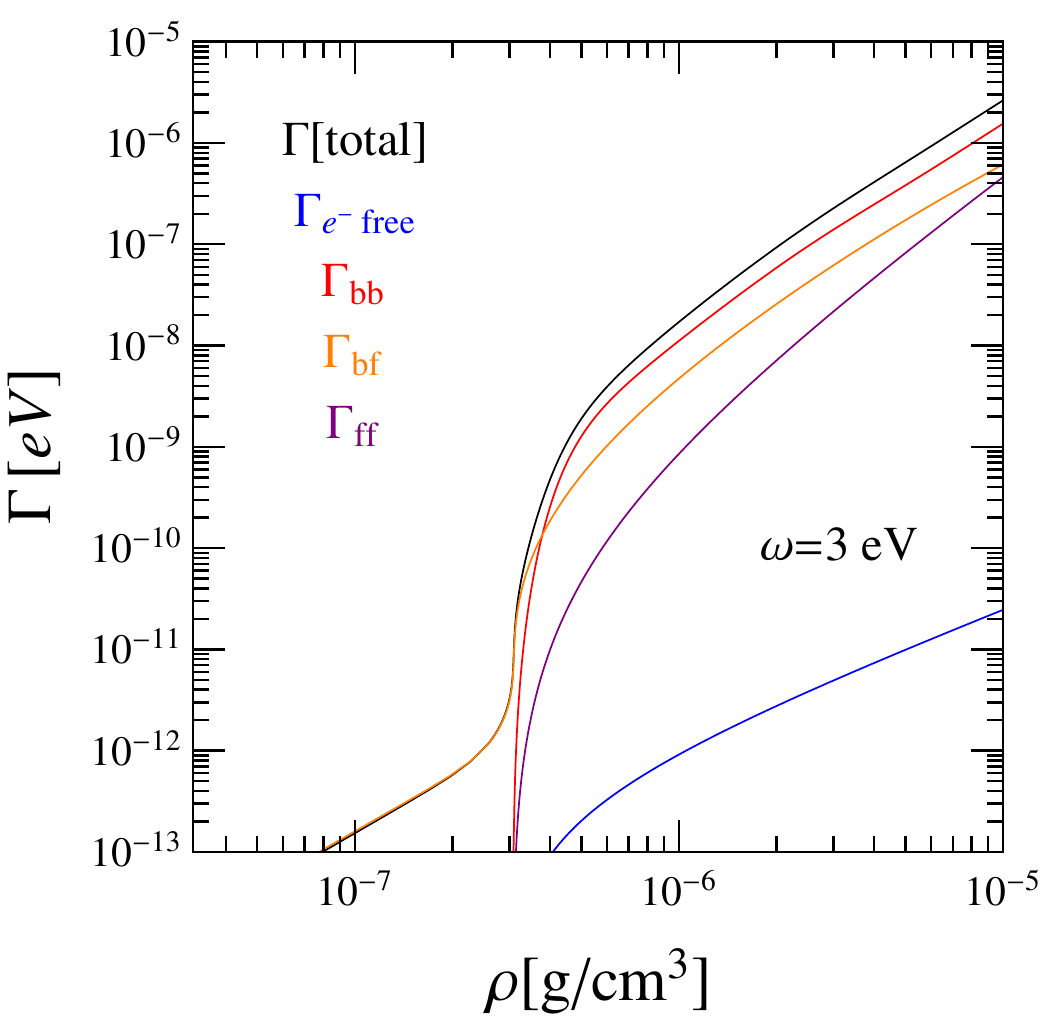}
\caption{Imaginary part of the self-energy divided by $\omega$, i.e. our parameter $\Gamma$ in the solar model built for this paper (black lines). 
The contribution from free-free absorption is shown in blue and bound-free and bound-bound absorption in red. 
Red dashed is the line for the H$^-$ bound-free contribution. 
Upper and middle plots show the energy dependence for four positions inside the Sun. 
Lower plots show the dependence on the solar interior position, labeled by the mass density, for photon energies of $\omega=2,3$ eV (wavelength=$616,413$ nm, respectively). 
}
\label{fig:gamma}
\end{center}
\end{figure}

The imaginary part of the self energy (the absorption coefficient up to a small correction) is shown in a few plots in Fig. \ref{fig:gamma}. In general it agrees well in the range of interest with the calculations of the OP interpolations and crosschecks with the LEDCOP database~\cite{LEDCOP,LEDCOPref,TOPS} at the low densities relevant for this work. 

\clearpage
\section{Solar hidden photon flux: Numerical results}
\label{sec:flux}

The model for refraction in the solar interior build in the last section allows us to compute the solar flux of HPs by direct integration of formula \eqref{eq:Ageneral}. This is a very time-consuming operation, which fortunately it is not necessary because the simple formula \eqref{eq:Prob1} is valid for most of the Sun, with the only exception of optically thin resonance regions where \eqref{eq:Prob2} can be used. 
The solar HP flux in the visible is dominated by resonant production, which we discuss in detail next.  
Later, we integrate \eqref{eq:Prob1} in the whole mass and energy range to compute the non-resonant contribution and present our atlas of solar HP emission. 
Finally, we estimate the corrections to the 1D solar models used by computing the  emission from a 3D time-dependent solar atmosphere model. 

\subsection{Resonance region contribution}

\begin{figure}[b]
\begin{center}
\includegraphics[width=0.47\textwidth]{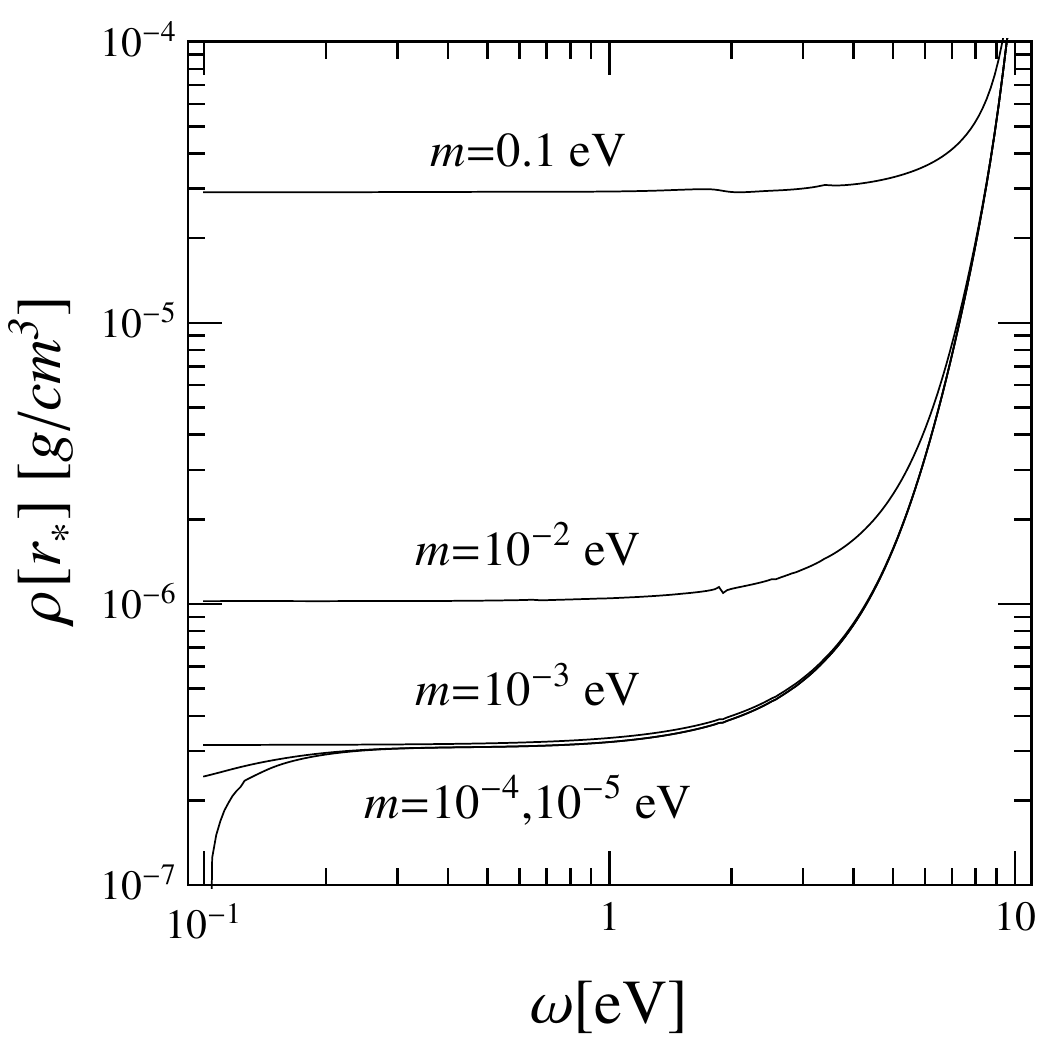}
\includegraphics[width=0.47\textwidth]{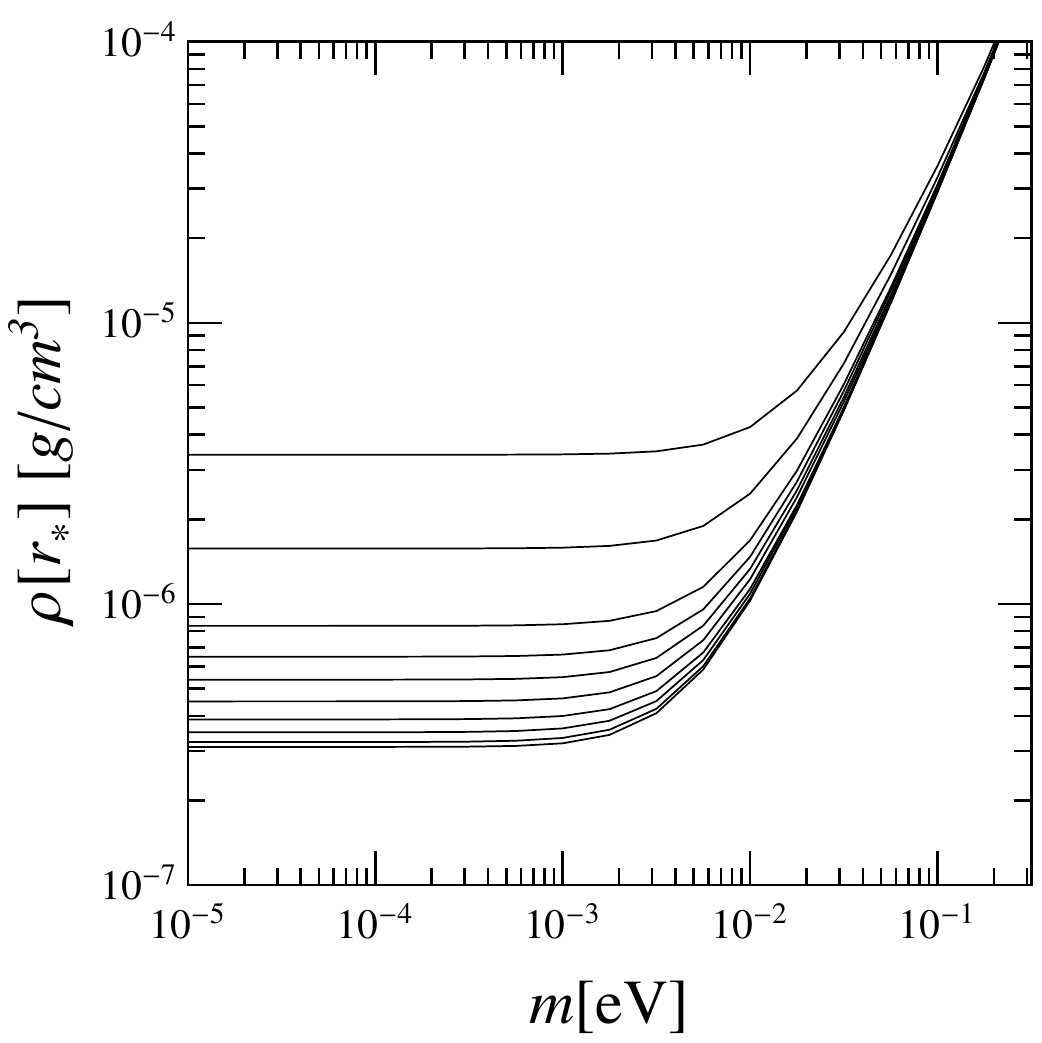}
\caption{Solar density at the radial distance from the solar centre, $r_*$, at which photon-HP conversions are resonant, $m^2=m_\gamma^2(\omega,r_*)$. 
LEFT: for different HP masses, $0.1,10^{-2},10^{-3},10^{-4},10^{-5}$ eV up to down, as a function of the HP energy.  
RIGHT: for different energies, $6,5,4,3.5,3,2.5,2,1.5,1,0.5$ eV up to down,  as a function of the HP mass. }
\label{fig:locus}
\end{center}
\end{figure}

In section \ref{sec:resonancedomination} we showed that the formula \eqref{eq:resonantflux} provides an accurate description of HP flux from the resonance region. 
In order to evaluate this expression we need the position of the resonance as a function of the frequency and the HP mass, 
$r_*=r_*(\omega,m)$. We have solved numerically the equation $m^2=m_\gamma^2(\omega,r_*)$ and present in Fig. \ref{fig:locus} our results. 
In the regions shown,  the resonance moves to low densities with decreasing HP mass and photon/HP energy. The first trend is obvious from the equation $m^2=m_\gamma^2(\omega,r_*)$ because $m_\gamma^2$ is proportional to the densities of charged particles. The second follows from the fact that, at low densities, $m_\gamma^2$ has a positive contribution from free electrons and a negative one from neutral H which decreases with $\omega^2$. A decrease in the negative contribution has to be balanced by a decrease in the positive one, i.e. displacement towards lower densities where the free electrons become much more scarce. 

\begin{figure}[t]
\begin{center}
\includegraphics[width=0.47\textwidth]{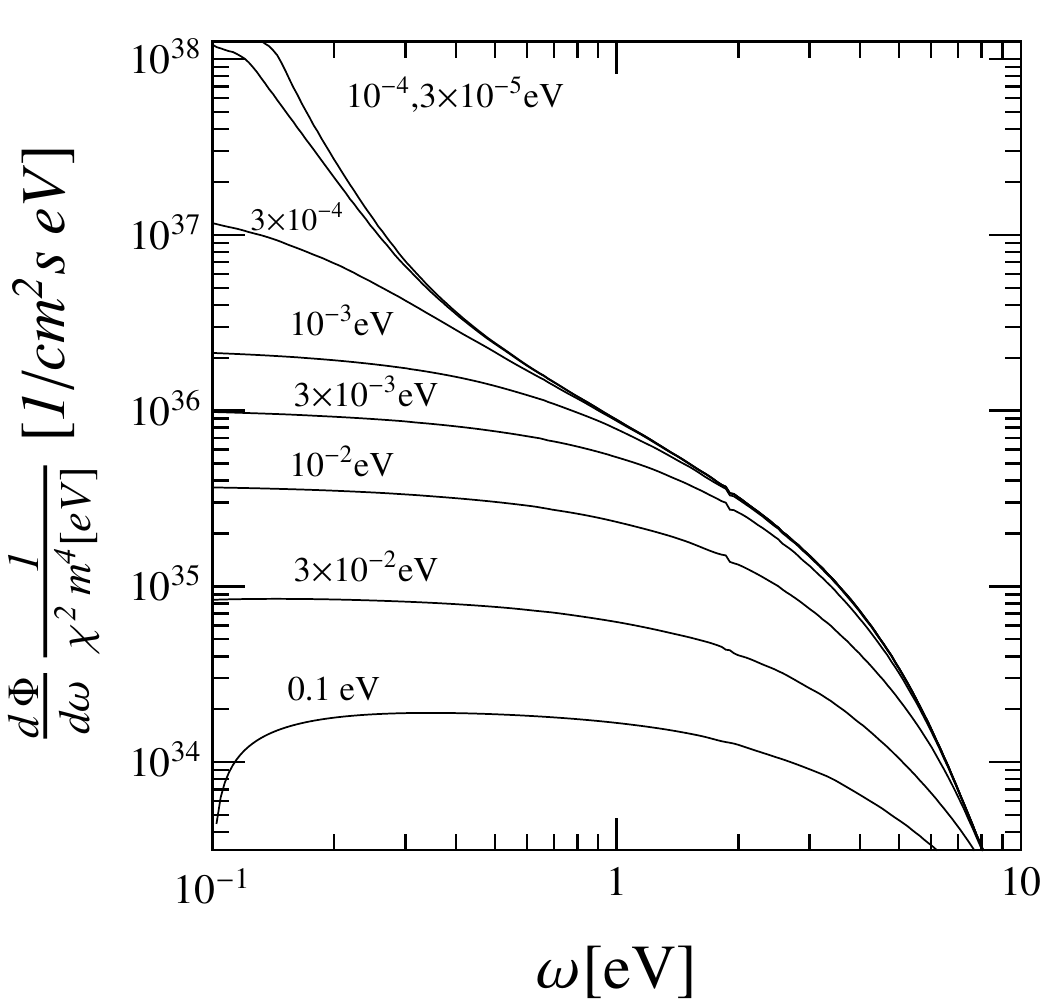}
\includegraphics[width=0.47\textwidth]{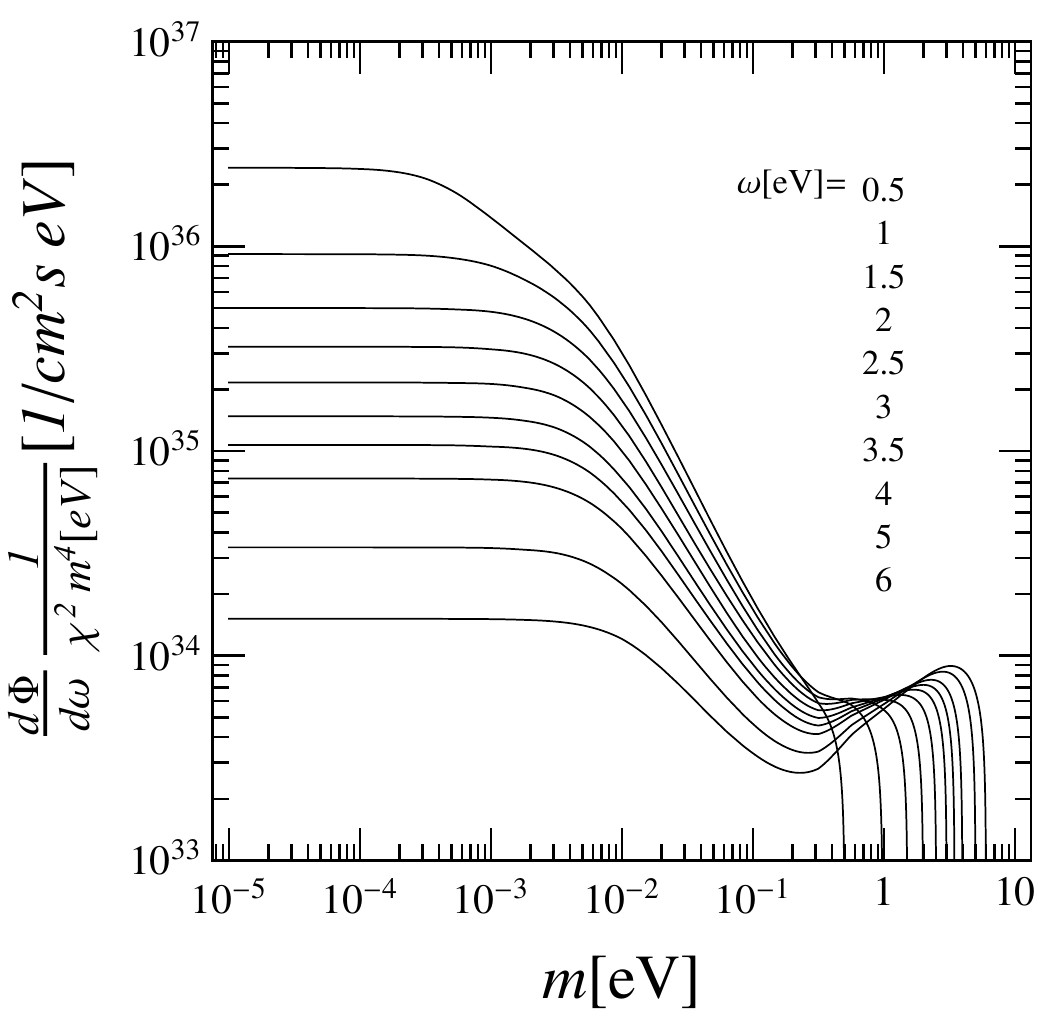}
\caption{Solar flux of hidden photons from the resonance region in HPs/cm$^2$s\,eV divided by the factor $\chi^2 (m/{\rm eV})^4$ for illustration purposes. 
LEFT: for different HP masses as a function of the HP energy.  
RIGHT: for different energies, $6,5,4,3.5,3,2.5,2,1.5,1,0.5$ eV up to down, as a function of the HP mass.}
\label{fig:fluxes}
\end{center}
\end{figure}

The flux of HPs can then be easily computed from \eqref{eq:resonantflux}, which we repeat here for convenience
\be
\frac{d\Phi}{d\omega}\approx \frac{r_*^2}{\pi R_{\rm Earth}^2}\frac{\chi^2 m^4\sqrt{\omega^2-m^2}}{e^{\omega/T(r_*)}-1} \left|\frac{d m_\gamma^{2}}{dr}\right|^{-1}_{r=r_*} \(1-\frac{e^{-\tau_*}}{2}\) ,  
\ee 
Our results are shown in Fig. \ref{fig:fluxes}. 
In the frequency range shown, the flux peaks at low frequencies and decreases strongly with the mass (note that the fluxes in Fig. \ref{fig:fluxes} are divided by $m^4$ for the sake of illustration). 
Let us first comment on the spectral shape. It is the result of the convolution of the factor $\sqrt{\omega^2-m^2}/(e^{\omega/T(r_*)}-1)$ and the derivative $|{d m_\gamma^{2}}/{dr}|^{-1}$ which gives the typical size of the resonant region. 
Both are larger at low energies. 
The first term is relatively flat because, at the masses and frequencies shown, $T(r_*)\sim O(1)$ eV. Taking $\omega\gg m$ we have $\sqrt{\omega^2-m^2}/(e^{\omega/T(r_*)}-1)\sim T(r_*)$. The exponentially suppressed regime only starts to be felt at the highest energies and the threshold at the largest masses. 
The second term can be understood if we write a very schematic model for $m_\gamma^2$ at low energies as  
\be
m_\gamma^2\sim\frac{4\pi \alpha}{m_e}n_t \(X_e-(1-X_e)\frac{\omega^2}{\omega_0^2}\)
\ee
where $n_t$ is the total number density of H atoms, $X_e=n_e^{\rm free}/n_t$ is the ionisation fraction and $\omega_0\sim 10.2$ eV is the resonant frequency of the Ly$-\alpha$ transition. 
The first term is due to free electrons and the second to neutral H. 
The derivative with respect to the radius can be written as 
\be
\label{eq:deriana}
\frac{dm_\gamma^2}{dr}\sim\frac{m_\gamma^2}{r}
\( \frac{d \log n_t}{d\log r} +\frac{d \log X_e}{d\log r}X_e\(1+ \frac{\omega^2}{\omega^2_0}\)\)
\ee
$n_t$ is a relatively smooth decreasing function of the solar radius and $X_e$ is essentially $=1$ flat in the interior and drops exponentially fast near the surface (see Fig. \ref{fig:partfun} right). 
Resonances happening in the deep Sun have  $d \log X_e/d\log r\sim 0$ and thus ${dm_\gamma^2}/{dr}$ independent of $\omega$. But for those happening near the surface, the ${d \log X_e}/{d\log r}$ term dominates. 
In that region,  the resonance condition $m_\gamma^2=m^2$ is realised despite a large value of $4\pi \alpha n_t/m_e>m^2$ by some degree of cancellation in the term $X_e-(1-X_e)\omega^2/\omega_0^2$ so that $X_e\sim \omega^2/\omega_0^2$ and the derivative becomes suppressed at low energies. The suppression is not proportional to $\omega^2$ because $d \log X_e/d\log r$ is very sensitive to $r_*$, which increases for lower $\omega$ but, nevertheless, the general trend of smaller $dm_\gamma^2/dr$ and thus bigger fluxes remains. 

Let us know shed some light on the dependence with the mass. At high masses, the resonances happen in the deep Sun and,  
according to \eqref{eq:deriana}, the derivative $d m_\gamma^2/dr$ is proportional to $m_\gamma^2\simeq m^2$ so the flux should go as $m^2|d \log n_t/d\log r|^{-1}$, which when properly evaluated at the resonance point gives something $\propto m^3$. This trend shows in the mass region $0.01-0.1$ eV in Fig. \ref{fig:fluxes} (right) and it is only stopped at higher masses because we have focused in the $\omega\sim O$(eV) region and the kinematic threshold cuts the curves. 
At masses below $0.01$ eV the resonances approach increasingly the photosphere and they cannot go further. 
This is because, for all the frequencies shown there is a point where $m_\gamma^2$ becomes 0. 
HP masses of arbitrarily small mass, will have their resonance arbitrarily close to that region, but not further. 
In that limit, $d m_\gamma^2/dr$ will be the same for all small masses (although it depends on the energy, as we have already explained). This explains the flattening of the low mass flux below $m\sim 0.01$ eV in Fig. \ref{fig:fluxes} (right).  

\subsubsection{The infrared rise}
If we consider sufficiently low energies and low masses the resonant region starts to move out of the photosphere out into the open atmosphere. 
In Fig. \ref{fig:locus} this is seen to happen in the region $m\lesssim10^{-4}$ eV and $\omega<0.4$ eV. 
In this region, the photon mass $m_\gamma^2$ decreases smoother than in the surface, (see the electron density at $\rho\sim 10^{-8}-10^{-7}$g/cm$^3$ displayed in Fig. \ref{fig:partfun} (right)) and consequently the resonance region is larger ($|d m_\gamma^2/dr|$ smaller) and the photon$\to$HP probability gets boosted. 
For $m=10^{-4}$ eV, the flux at $\omega\sim 0.2$ eV seems to be more than two orders of magnitude stronger than at $2$ eV. 
But it is not clear that we can claim these results to be true because our solar model for refraction is probably not very accurate here. There are a number of effects that can alter our calculation. Let us discuss some of them. 
First of all, where the rise starts depends very much on the precise determination of the profile of the free electron density. In the atmosphere, the contribution of metallic donors is crucial, and we have performed only a rough estimate. Second, even a small number of non-H atoms with resonances in the visible or IR could in principle contribute more than the UV resonances of H to the index of refraction here. These contributions are negative and push the resonances again deep into the solar interior where the gradient  $|d m_\gamma^2/dr|$ is larger and thus the flux smaller. 
A simple estimate tells us that this uncertainty grows very much at low energies. Imagine that every metal, but not He, contributes one electron with a resonant transition at frequency $\omega_m$, then in the far infrared 
\be
\left.m_{\gamma}^2\right|_Z \sim -\frac{4\pi \alpha}{m_e}\sum n_Z \frac{\omega^2}{\omega_{m}^2}
\sim \frac{\sum n_Z}{n_{\rm H}}\frac{\omega^2_{{\rm Ly}-\alpha}}{\omega_m^2}\left.m_{\gamma}^2\right|_{\rm H} 
\sim 10^{-3}\(\frac{10.2\rm \, eV}{\omega_m}\)^2\left.m_{\gamma}^2\right|_{\rm H},  
\ee
where $m_{\gamma,{\rm H}}^2$ is the estimation of the H contribution through the Lyman series. For $\omega_m$ in the visible, this effect is negligible but if there is considerable structure below $0.3$ eV the effect would be similar to the H. This rough estimate, shown with the aim of highlighting the typical orders of magnitude is enough to rise a voice of warning towards the refraction model validity in the IR. 
In the atmosphere, the density of H molecules like H$_2$, CH, OH, is extremely small but they have absorption lines  in the IR that one should in principle take into account. 
Another assumption which is prone to fail in the atmosphere is LTE, which we have used to compute all the abundancies and the radiation temperature. 
Finally, we will see that the spherically symmetric model fails to some extent to reproduce the true 3D behaviour of the solar surface. 
Due to these issues, in this paper we can make no claim of any rigour in the derivation of the IR flux produced in the solar atmosphere. Although it is clear that the typical flux grows if resonances happen in the shallower density profile of the atmosphere, the low energy limit of Fig. \ref{fig:fluxes} have to be understood as order of magnitude estimates.  

\subsubsection{Spectral lines and the UV region}
\label{sec:spectrallinesUV}
\begin{figure}[b]
\begin{center}
\includegraphics[width=0.46\textwidth]{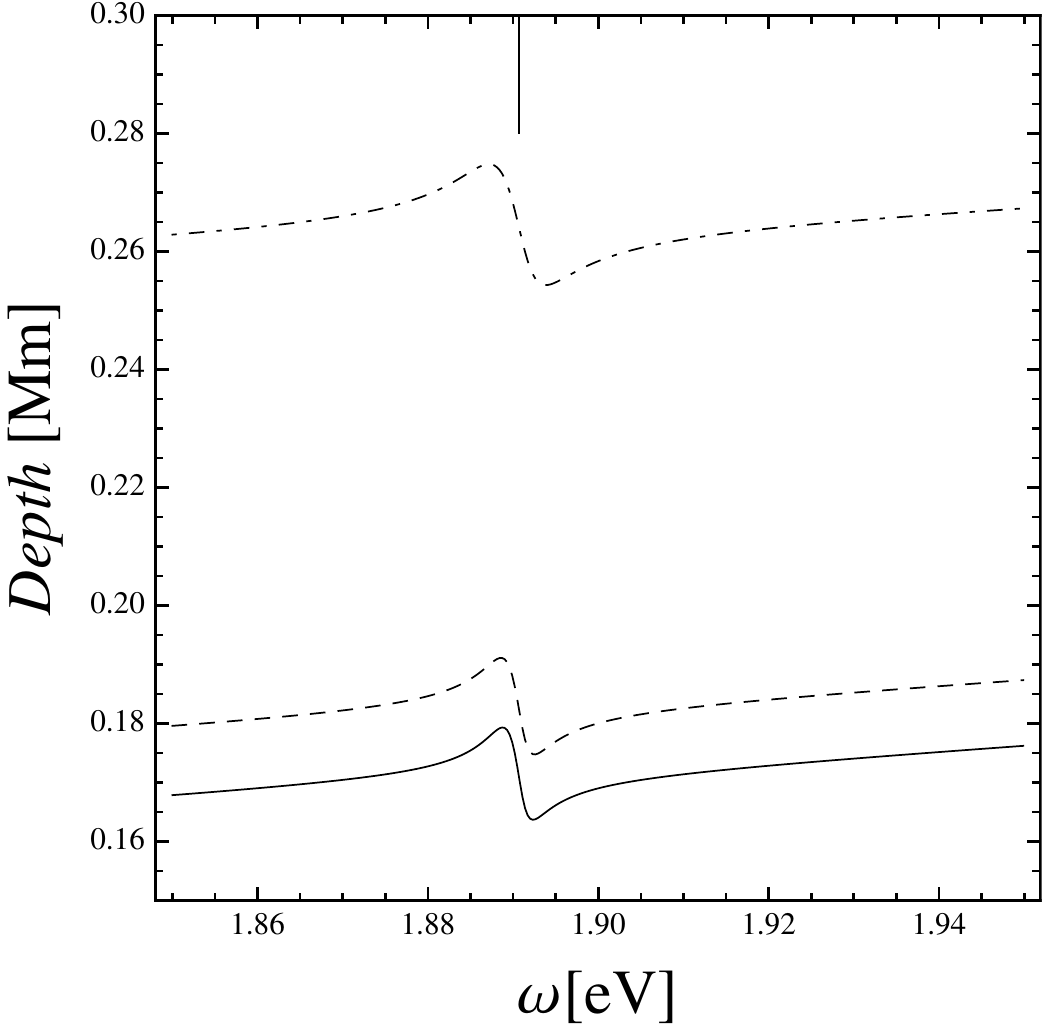}
\includegraphics[width=0.47\textwidth]{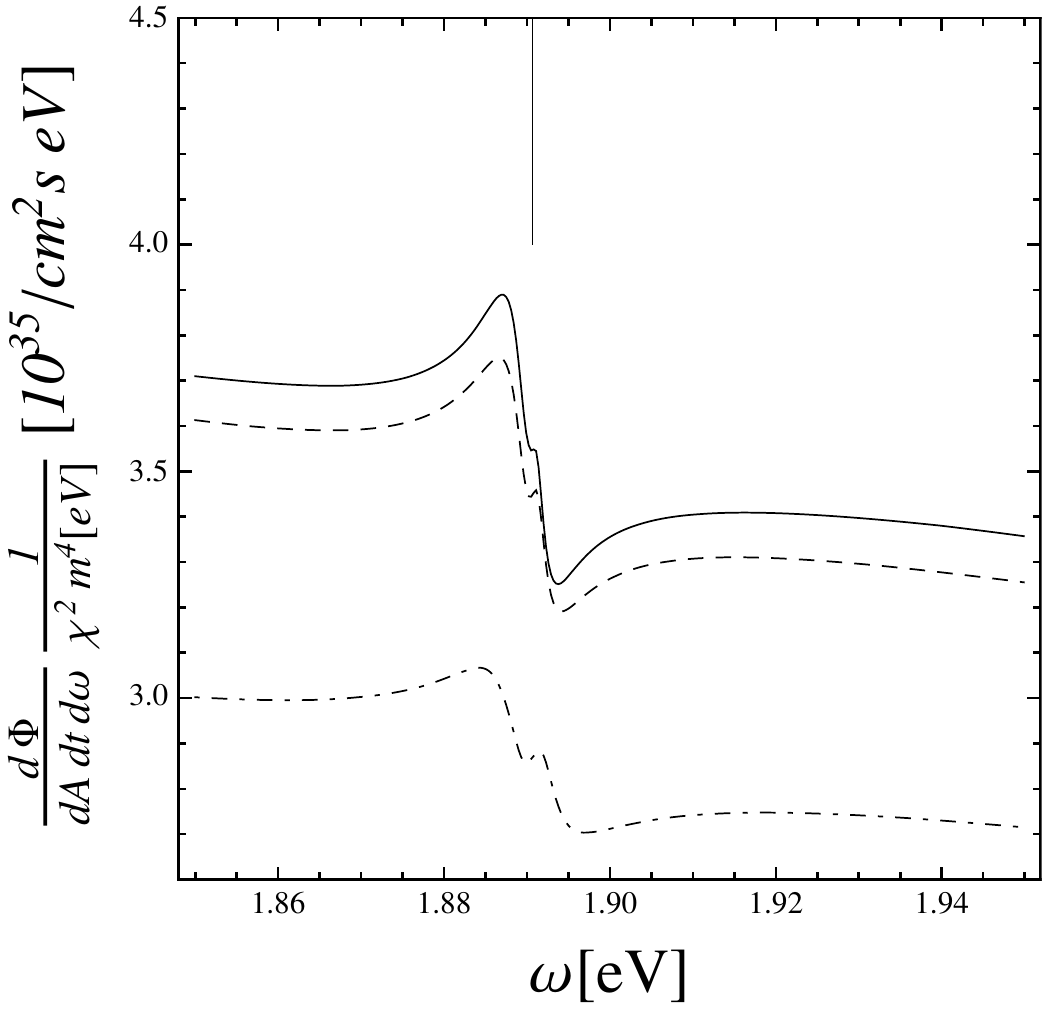}
\caption{Position of the resonance and resonance HP flux for $m=10^{-4},10^{-3},10^{-2}$ eV (solid, dashed,dot-dashed lines) near the H-$\alpha$ spectral line (shown as a vertical line).  }
\label{fig:Halpha}
\end{center}
\end{figure}

The results shown in Fig. \ref{fig:fluxes} show only tiny spectral features, which prove that the role of atomic resonances  in the visible is not crucial. 
Near a strong spectral line, the pattern of the flux is expected to change because, just below, the 
effective mass $m_\gamma^2$ is more negative and above, more positive. Resonances are displaced towards the solar interior in the red part and towards the surface in the blue and consequently, the red part is more luminous that the blue part. 
An example, corresponding to the H-$\alpha$ line is shown\footnote{The small feature inside in the flux probably corresponds to the broadening of line, which changes fast as a function of the rising proton density.} in Fig. \ref{fig:Halpha}. Even around this important line, the effects are at the 10 percent level. We will see later that O(1) fluctuations in the position of resonances  close to the photosphere and their associated fluxes arise due to convection so these effects are completely unobservable.  
One can also see that the spectral feature softens as the HP mass grows and the resonance region enters deeper into the Sun, where the line is broader. 

\begin{figure}[t]
\begin{center}
\includegraphics[width=0.95\textwidth]{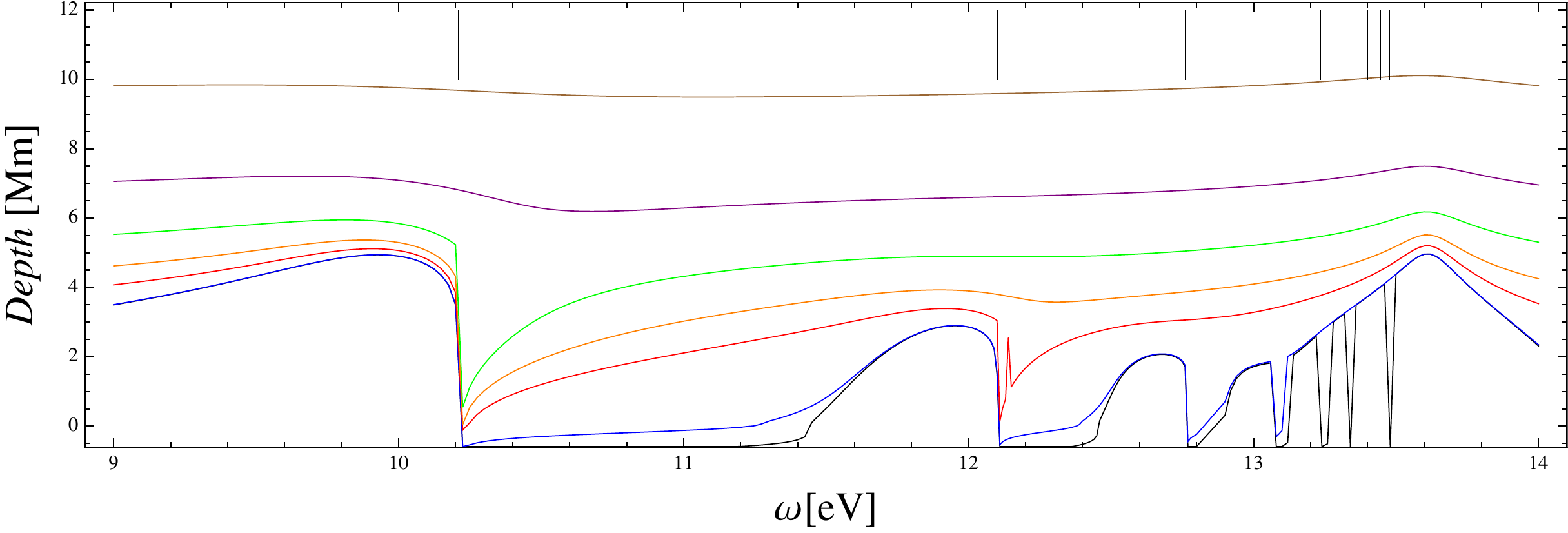}

\includegraphics[width=0.95\textwidth]{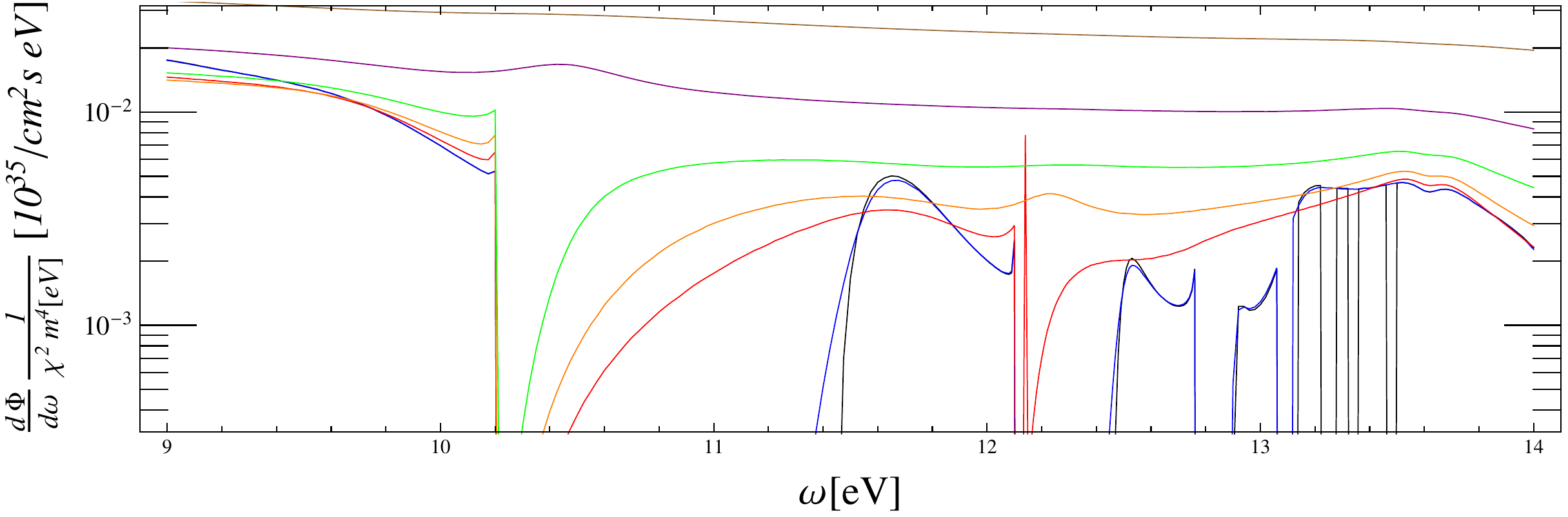}
\caption{Position of the resonance and resonance HP flux for $m=10^{-3},10^{-2},0.1,0.16,0.25,0.4,0.63$ eV (Black, blue, red, orange, green, purple, brown lines) in the spectral region of the Lyman series of absorption lines (shown as vertical lines from (1$\to$2) to (1$\to$10)).  }
\label{fig:Lyman}
\end{center}
\end{figure}

The behaviour around other lines is qualitatively similar to H-$\alpha$.  
Lines of the Lyman series give the most spectacular effects in the UV, shown in Fig. \ref{fig:Lyman}. 
The Ly-$\alpha$ line at $10.2$ eV is so strong that pushes the low mass HP resonances up to 5 Mm deep inside the Sun at $\omega\sim 10$ eV and expels them outside of the Sun in the $10.2-11$ eV range. The other Lyman lines have similar effects, bring the resonances some Mm inside of the Sun at their red sides and expel them to the solar surface in the blue sides. 
The resonant flux in the UV varies very violently with energy, dropping by many orders of magnitude in the blue sides of lines. Indeed, in these dips the resonance flux becomes subdominant with respect to the non-resonance flux as we shall see in section \ref{sec:nonresonant}. 
The situation softens progressively for increasing HP masses and it is very smooth above $m\sim 0.4$ eV.

\subsubsection{Width of the resonance region} 
\begin{figure}[h]
\begin{center}
\includegraphics[width=0.31\textwidth]{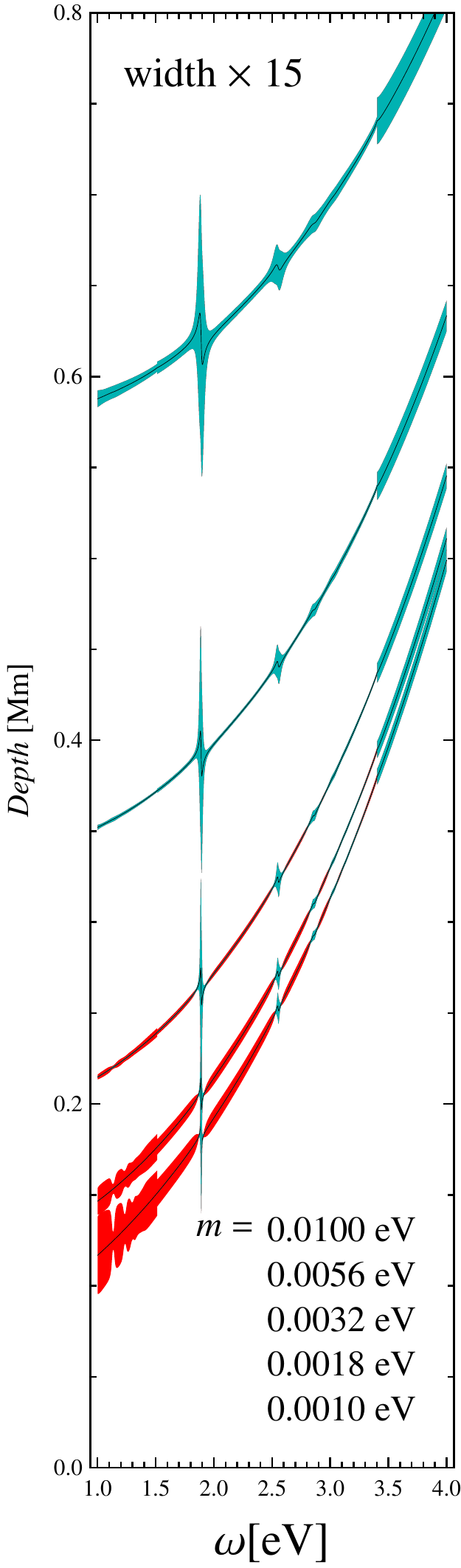}
\includegraphics[width=0.31\textwidth]{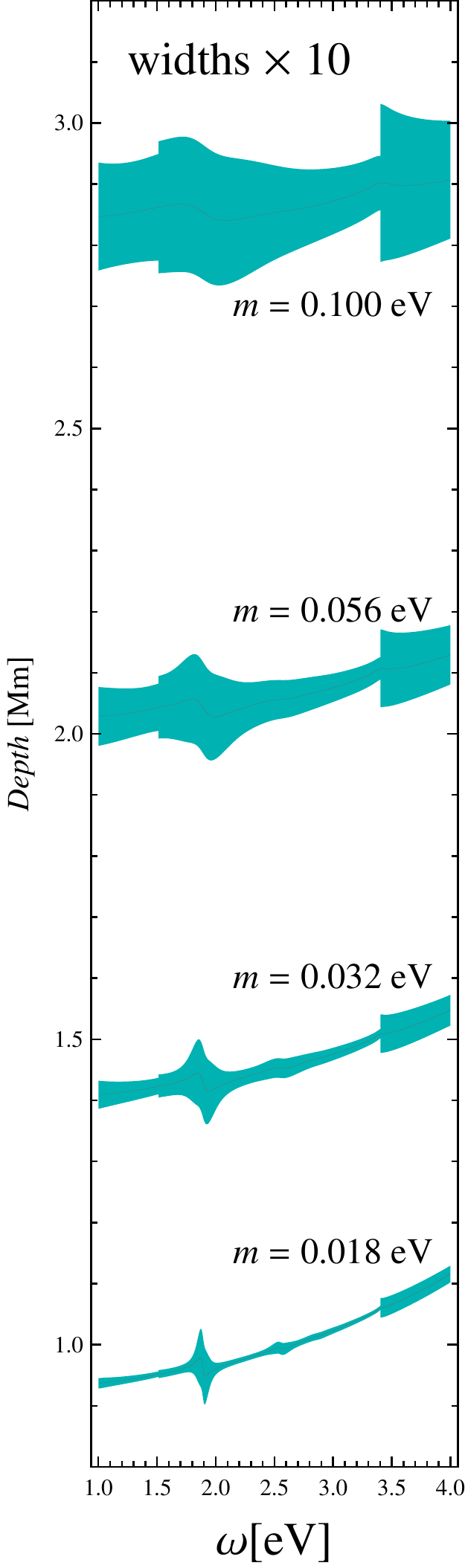}
\includegraphics[width=0.31\textwidth]{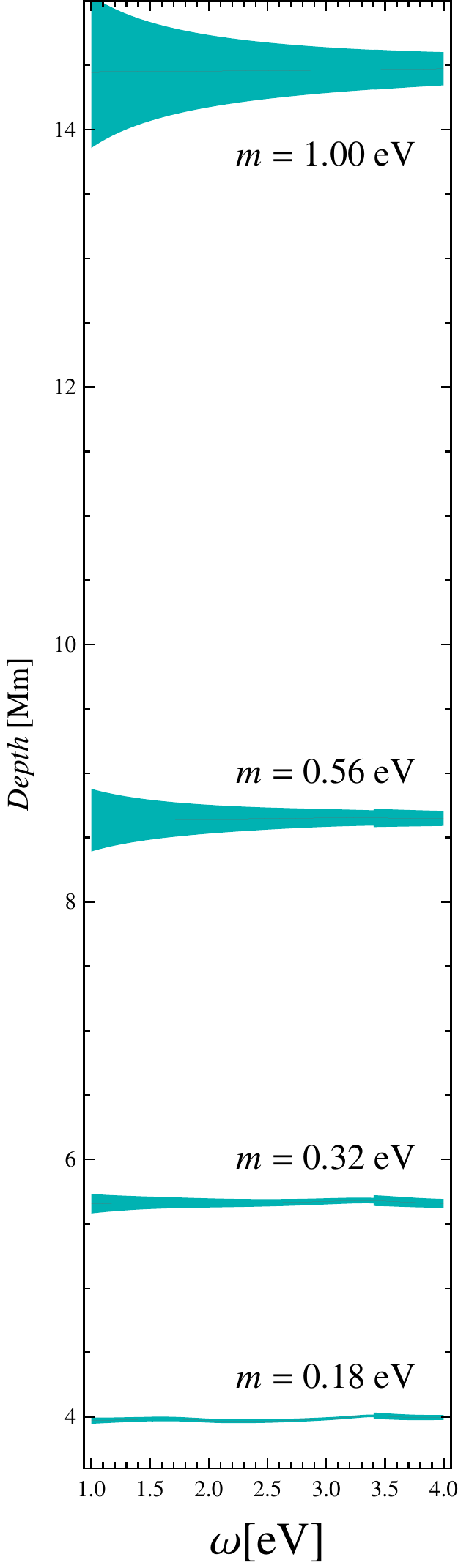}
\caption{Resonance region shown with the estimated width as a shadowed region around it in the visible energy range. The red and turquoise regions correspond to the optically thin and optically thick criteria. }
\label{fig:width}
\end{center}
\end{figure}

If the resonance is optically thick, it follows from \eqref{eq:Ithick} that the FWHM of the resonance is $\Delta r_{\rm thick}\sim \omega\Gamma(r_*)\left|\frac{d m_\gamma^2}{d r}\right|^{-1}_{r_*}$~\cite{Redondo:2008aa}. When it is optically thin, it is given by the mean-free-path around the saddle point, $\Delta l\sim 1/\Gamma(r_*)$ but since $l\sim \Delta r/\cos\theta$, $\Delta r_{\rm thin}\sim \cos\theta/\Gamma(r_*)$ it is azimuth dependent.  
We can then estimate the resonance width as 
\be
\Delta r \sim {\rm max}\left\{\frac{\cos\theta}{\Gamma(r_*)},\omega\Gamma(r_*)\left|\frac{d m_\gamma^2}{d r}\right|^{-1}_{r_*}\right\} . 
\ee
which is shown in Fig. \ref{fig:width} as a function of the HP energy. 
The red range around the resonance position represents $\Delta r_{\rm thin}$ for $\cos\theta=\pm1$ and the turquoise region $\Delta r_{\rm thick}$. 
In resonance regions close to the solar surface, scale heights are very short and thus $\Delta r_{\rm thick}$ decreases while mean-free-paths become longer. These resonance regions are optically thin, except near strong absorption lines. 
Note that in these resonances, trajectories with $\cos\theta\simeq0$ are thick because traveling in a direction perpendicular to the radial direction, the photon/HP sees the same properties of the Sun before being absorbed. So, in reality, the thickness of resonances in the red regions vary from the red to the turquoise range depending on $\cos\theta$. 
The size of the saddle point region, $\delta l\sim |\varphi''|^{-1/2}=\left|{d m_\gamma^2}/{d r 2\omega}\right|^{-1/2}$, is smaller than 100 meters in the region shown.  

Deeper than about 0.4 Mm, all the resonances are optically thick.  
Comparison of Fig. \ref{fig:width} with the temperature and density profile of Fig. \ref{fig:solarsurface} shows that  that in general, resonances are extremely thin with respect to the temperature scale height, even around absorption lines or phoionisation thresholds. 
Assuming constant temperature and density within the resonances seems a reasonable approximation in this energy range.

\subsection{Non-resonant contribution (bulk of the Sun)}
\label{sec:nonresonant}
So far we have focused on the resonance region and its contribution to the HP flux. 
We have advanced that it is the most relevant contribution in the visible range of energies. 
In this section we justify why this is so and then compute the non-resonant flux in the UV and above where it can dominate.  

Consider the integral over the Sun that gives the total HP flux on Earth for a given HP energy, \eqref{eq:generalflux}.  The locally homogeneous plasma approximation, \eqref{eq:Prob1}, is an extremely good approximation for the photon/HP conversion probability, except for optically thin regions, which however are so thin (like any optically thick) that do not make a difference in this discussion. Moreover, we know that once integrated over the Sun, the result is independent of the optical thickness of the resonance. 
After performing the now trivial $\cos\theta$ integral, the HP flux in this approximation is    
\be
\label{eq:fluxnonresonant}
\frac{d\Phi}{d\omega}= \frac{1}{R_{\rm Earth}^2}\frac{\omega\sqrt{\omega^2-m^2}}{\pi^2}  
\int_{\rm sun} r^2 dr  \frac{\Gamma(r)}{e^{\omega/T(r)}-1} \frac{\chi^2 m^4}{(m_\gamma^2(r)-m^2)^2+(\omega\Gamma(r))^2} . 
\ee
There are three qualitatively different regions in this integral. 
The most notable is the resonance region, of typical size $\Delta r \sim \omega\Gamma|d m_\gamma^2/dr|^{-1}\ll R_{\rm Sun}$ where the conversion probability is maximal $\sim \chi^2(m^2/\omega\Gamma)^2$ because typically $m_\gamma^2=m^2\gg \omega\Gamma$ (with exceptions being strong absorption lines, for which the resonance is proportionally wider). 

Outside the resonance region, $m_\gamma^2 \ll m^2$, the conversion probability is like in vacuum $\sim \chi^2$, the region is bigger than the resonance region but temperatures and production rates are very small so that we shall not expect competitive rates from here. 

Inside the resonance region $r\ll r_*$ we have $m_\gamma^2 \gg m^2$ and thus $(m_\gamma^2)^2+(\omega\Gamma)^2 \gg m^4$ and the conversion probability is suppressed with respect to the vacuum case $\sim O(\chi^2)$. 
However, this region has the typical size of the whole volume of the Sun, and the factor $\frac{\Gamma}{e^{\omega/T}-1}$ is larger because so are the temperature and the electron/proton densities that determine $\Gamma$. One could be tempted to think that, even if the probability is suppressed, the net flux is competitive with the resonance. What is stronger, the $\Gamma r^2/(e^{\omega/T}-1)$ enhancement or the $(m_\gamma^2)^2+(\omega\Gamma)^2$ suppression? 

\begin{figure}[h]
\begin{center}
\includegraphics[width=0.45\textwidth]{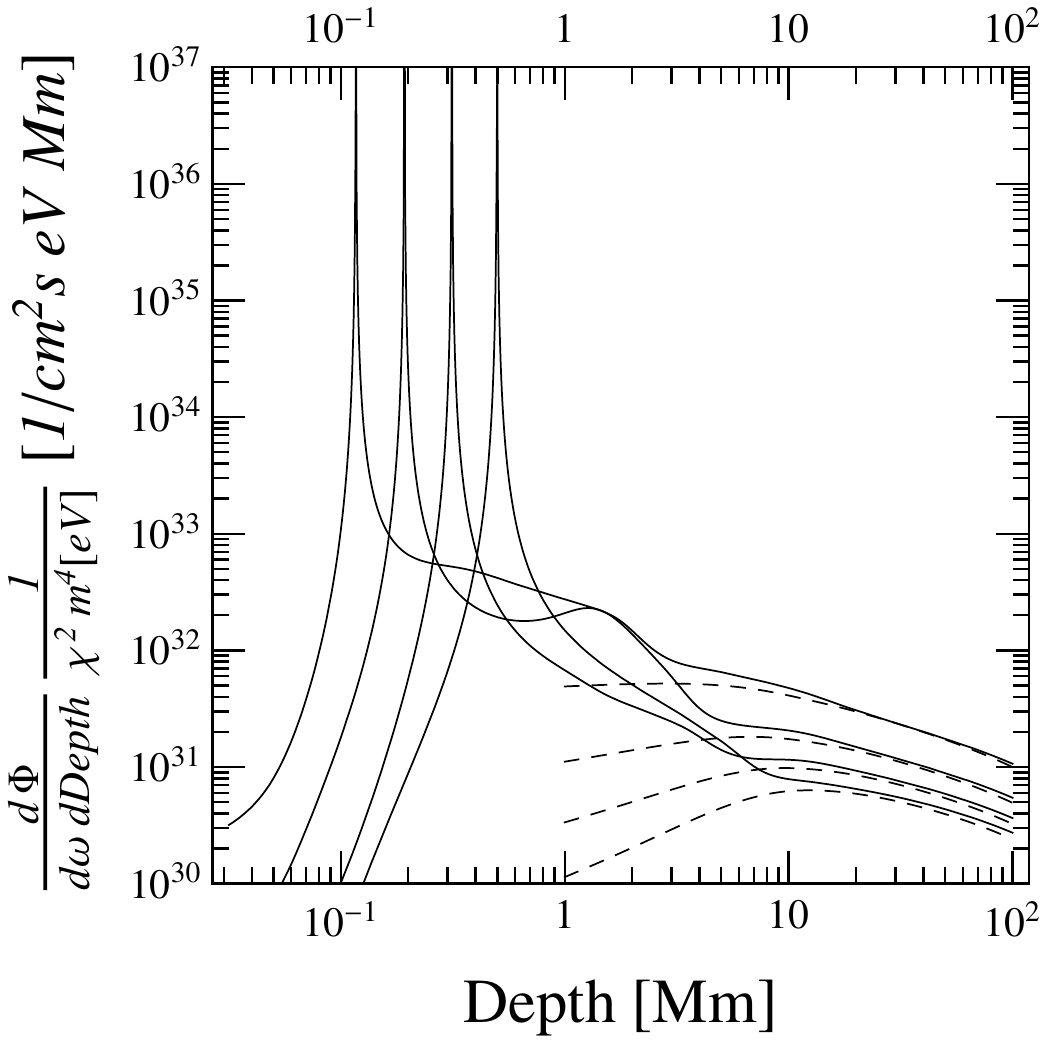}
\includegraphics[width=0.45\textwidth]{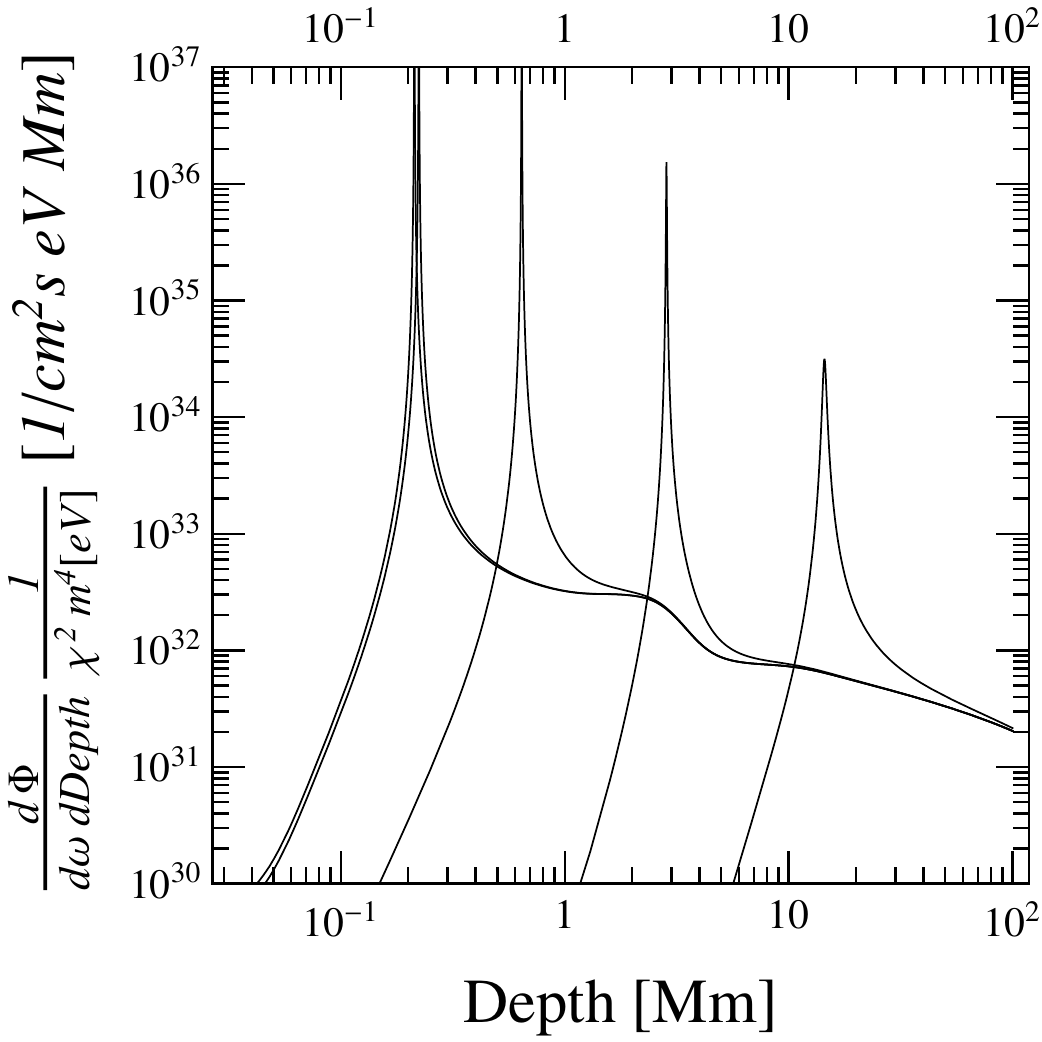}
\caption{ 
Solar HP production rate as a function of the depth inside the solar surface. 
LEFT: As a function of energy for $m=10^{-3}$ eV. Curves are for $\omega=1,2,3,4$ eV from left to right at the peak. The bump in the $\omega=2$ eV line corresponds to the H-$\alpha$ line, which is very close. RIGHT: As a function of HP mass for $\omega=2.3$ eV. Curves are for $m=10^{-4},10^{-3},10^{-2},0.1,1$ eV from left to right at the peak. 
 }
\label{fig:nonresonant}
\end{center}
\end{figure}

The situation at visible energies is depicted in Fig. \ref{fig:nonresonant}, where we show examples of the differential flux produced in the Sun as a function of the solar radius $\frac{d\Phi}{d\omega dr}$. Indeed, the flux outside of the resonance falls extraordinarily fast and can be neglected. 
The HP flux originating from the bulk Sun inside the resonance region drops dramatically below the resonance and then stabilises to a much smoother decreasing contribution. 
In the solar interior the refraction is dominated by free electrons ($m_\gamma^2=4\pi\alpha n_e^{\rm free}/m_e$) and the absorption by free-free transitions ($\Gamma\simeq 64\pi^2\alpha^3\sqrt{m_e}(1-e^{-\omega/T})n_e^{\rm free}n_p/(3 m_e^2\omega^3\sqrt{2\pi T})$ neglecting $F_{ff}$ and screening) from which we can already see that the density dependences cancel out in $\Gamma/m_\gamma^4$ ($n_e^{\rm free}\sim n_p$) and the resulting formula depends smoothly on the solar parameters. We find  
\be
\frac{d\Phi}{d\omega dr} \approx 1.2\times 10^{32} \chi^2\(\frac{m}{\rm eV}\)^4\(\frac{r}{R_{\rm Sun}}\)^2\frac{e^{-\omega/T}}{\sqrt{T}\omega} \frac{1}{{\rm cm}^2 {\rm s\, eV\, Mm} }
\ee
where $T,\omega$ in this formula are in eV units. These estimates are shown as dashed lines in Fig. \ref{fig:nonresonant} and fit very well the full formula at depths larger than 10 Mm.  
With $R_{\rm Sun}=695$ Mm, one could think that this contribution is comparable to the resonance contribution for the highest HP masses (O(eV) in this context of visible energies), however, the factor of $\sqrt{T}\omega$  in the denominator suppresses enough the contribution. 

\begin{figure}[h]
\begin{center}
\includegraphics[width=1\textwidth]{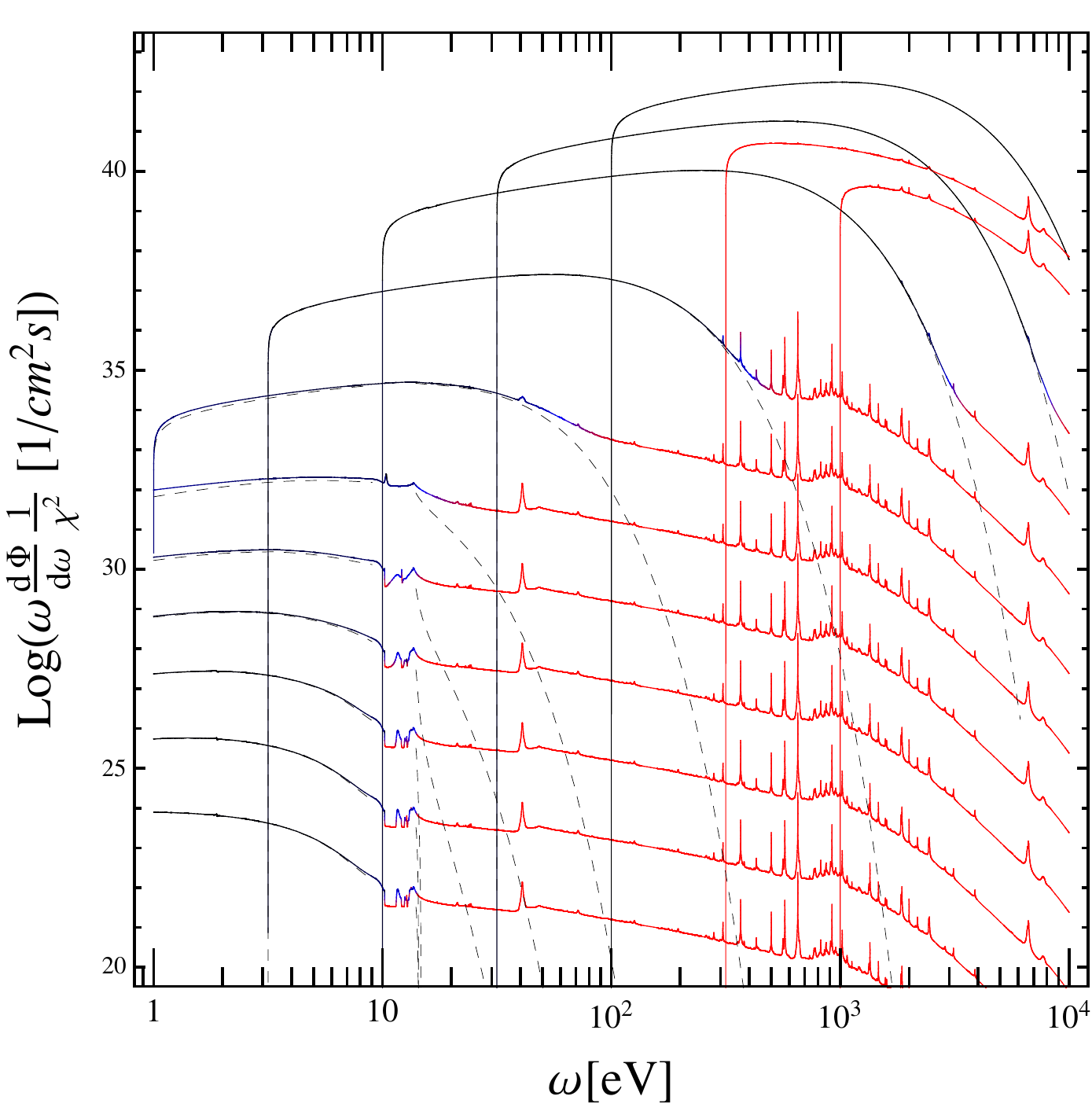}
\caption{ 
Solar flux of transversely polarised HPs computed in this paper as a function of HP energy for different HP masses. The lowest six lines correspond to $m=10^{-3},3.16\times 10^{-3},10^{-2},3.16\times 10^{-2},10^{-1},0.316$ eV from the bottom one up, and for the rest of the lines the mass is recognisable though the threshold, $m=1,3.16,10,31.6,100,316,10^3$ eV.  
Note that we show the energy flux, i.e $\omega (d\Phi/d\omega$).  
The curves are coloured according to the most relevant contribution: black for the resonance region, red for the bulk of the Sun and blue wherever the ratio of the two is in the 40-60\% range.   
 }
\label{fig:atlas}
\end{center}
\end{figure}

Finally, we have performed numerical integrations of \eqref{eq:fluxnonresonant} in the $1-10^4$ eV energy range in a generous range of masses to compare with the resonance estimate presented before. The atlas of HP solar emission is presented in Fig. \ref{fig:atlas}. 
For each HP energy and mass we show the full emission from the integral \eqref{eq:fluxnonresonant} (solid line) and the contribution from the resonance region estimated by \eqref{eq:resonantflux} (dashed line).  
The solid lines are coloured-coded according to the contribution that dominates the overall flux: black where the resonance accounts more than 60\%, red when it is less than 40\% and an interpolation in blue in between. The red colour reminds that these results depend on our model of $\Gamma$. 

In the visible, the results agree notably well with the resonance contribution. Only above $m>0.1$ eV the all Sun integration rises the prediction somehow, up to 40\% in some cases. 
However, most of this extra flux probably comes from the tails of the resonance region, particularly the inner one where the temperature $T$ and production rate $\propto\Gamma$ are larger and can rise somehow the flux. So this deviation has to be understood not as a contribution from the bulk of the Sun but as a correction to the thin-resonance condition that we have used. In conclusion, the resonance region dominates the production of solar HPs in the visible and provides a precise prediction independent on the uncertainties on $\Gamma$. 

In the UV, the situation changes very much. 
For low mass HPs (below $m\sim 10$ eV), if the resonance takes place, it does so in regions where the temperature is O(eV). The resonant production of this high energy HPs starts to be exponentially suppressed and thus the bulk contribution competes. 
In the near UV, below $\omega\sim $10 eV the overwhelming number of atomic transitions of metals rises the resonance prediction. In the Lyman region (10-14 eV) we see the structure advanced in section \ref{sec:spectrallinesUV}. In the blue part of the Lyman atomic transitions, the resonance is not possible and the flux is lower and dominated by the bulk (red dips), while in the red part there is still some resonance emission, which turns out to be of the same order than the bulk emission (probably also because the resonances are not thin). 
In the far UV the exponential suppression is huge and we only see the bulk emission. 
For instance, we can see the 1s-2p transition line of singly ionised helium at $\omega\sim 40$ eV and guess a multitude of smaller lines from metals. 
From $\omega\sim 300$ eV we see strong line emission in the soft X-ray region that decreases at 3 keV but shows up to 6-8 keV with the strongest Iron lines.

As we consider larger HP mass, the transition between resonance domination at low energies and bulk domination at the highest displaces towards higher energies because the resonance region is located deeper into the Sun and happens at higher temperatures. 
For instance, HPs with masses around and above 10 eV have resonance regions where the temperature is high enough to dominate the full emission up to $\omega\sim 3$ keV.    
Finally, above $m\sim 300$ eV, there is no resonance region in the Sun (298 eV is the highest photon effective mass in our solar model) and the HPs are always produced non-resonantly. 
These high mass HPs are produced mostly in the core, where most of the elements are fully ionised and therefore one does not see many atomic lines, with Iron being the obvious exception.  

\subsection{3D dynamic solar surface emission}

So far we have assumed that the properties of the solar plasma, temperature, density and composition, are spherically symmetric. 
The solar interior is radiative and spherically symmetric to a good enough precision but, from $r\sim 0.7R_{\rm Sun}$ the outgoing energy transport becomes convective and inhomogeneities start to grow towards the solar surface. Energy is transported by hot upflows that expand progressively as they ascend, cool radiatively in the photosphere, and descend in a turbulent dense downflow. Hot diverging upflows bounded by cool lanes of downflows form the characteristic pattern of solar granules observed in the solar photosphere, see~\cite{nordlund2009} for a review. 
The density and temperature profiles do not only depend on the solar radial coordinate. We have a full time-dependent 3D-problem. 

Notwithstanding the complexity, the situation for the production of HPs has not changed much. 
HPs can be produced resonantly around a region where $m_\gamma^2(\omega,{\bf x})=m^2$ and for frequencies in the visible and low mass HPs, this region is in very close to the solar surface. When we considered the Sun as spherically symmetric, the resonance region was a perfect spherical shell. Now we consider the inhomogeities due to surface granulation and we expect to have a textured surface which displaces slightly out of a perfect sphere with the hot upflows (because ionisation is larger) and in with the downflows. 
Solar granules have a characteristic extent of the order of 1 Mm, and evolve in times of the order of 10 minutes. These magnitudes are very large compared with the width of the resonance region and therefore it is reasonable to assume that our resonant flux equation \eqref{eq:resonantflux} still holds locally. 
Since now the resonance region is not perpendicular to the radial vector we have to substitute the derivative $|d m_\gamma^2/dr| $ by the gradient $|\nabla m_\gamma^2|$ and average over the surface. We thus propose  
\bea
\label{eq:resonantflux3D}
\frac{d\Phi}{d\omega}&\approx& \chi^2 m^4\sqrt{\omega^2-m^2}\frac{R^2_{\rm Sun}}{\pi R_{\rm Earth}^2}\int \frac{dS}{4\pi R_{\rm Sun}^2} \frac{1}{{e^{\omega/T}-1}}\frac{1}{|\nabla m_\gamma^2|}
\eea
where the integral is over the surface (or surfaces) where $m_\gamma^2({\bf x})=m^2$. We have dropped the term involving $e^{-\tau_*}$ because resonances in the visible lie typically in the optically thick Sun. 

We have made use of the 3D solar atmosphere model of~\cite{oai:arXiv.org:0909.0948} to estimate this 3D flux and compare it with the 1D idealisation shown before. 
The model covers a small 6Mm$\times$6Mm patch of the surface, extends from -0.8 Mm to 0.5 Mm in depth and a period of 45 solar min, and thus gives a sensible description of the surface in length and time scales where granulation evolves. It was used to interpret the solar  spectrum and determine atomic abundances from the shape and intensity of absorption lines. 
In general, the agreement between the fits and observations is excellent, which builds strong confidence in this model as a trustable description of the solar surface, at least in the line forming region. The model was used in ~\cite{Vincent:2012bw} to assess the effect of HPs in the determination of solar abundances, which was found to be negligible. 
The model does not include magnetic fields and thus includes no sunspots or faculae. Since they are relatively scarce, we can neglect them. 
  
\begin{figure}[htbp]
\begin{center}
\includegraphics[width=\textwidth]{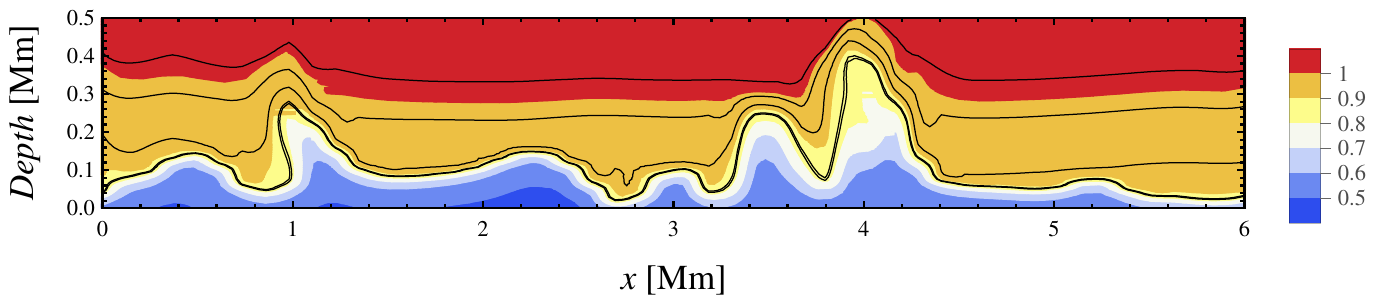}
\includegraphics[width=\textwidth]{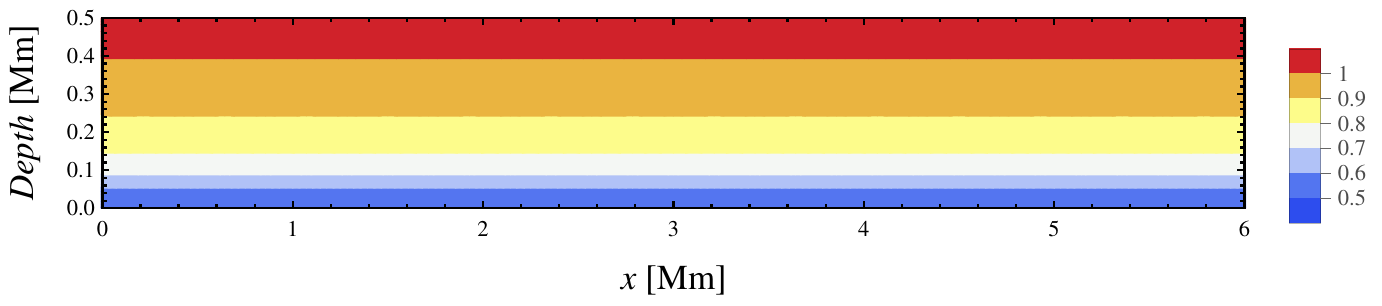}
\caption{UP: 2D slice of the solar atmosphere model of~\cite{oai:arXiv.org:0909.0948} showing the temperature profile in eV as a function of depth and transverse extent along the solar surface ($x$). Superimposed are the resonance loci, $m_\gamma^2(\omega,{\bf x})=m^2$, where photon$\to$HP conversions are resonant for $m=10^{-4},10^{-3},3.2\times 10^{-3},4.5\times 10^{-3},5.5\times 10^{-3}$ eV from bottom to top. 
DOWN: For comparison we show the temperature profile of the spherically symmetric solar model of~\cite{saclay}. Note that the reference level to measure the depth is different in both cases. }
\label{fig:surface}
\end{center}
\end{figure}

A 2D time-slice of the situation is depicted in Fig. \ref{fig:surface}. We see the corrugated shape of the resonance locus for several HP masses (black lines) with a temperature background in colours. The function $m_\gamma^2(\omega,{\bf x})$ follows very closely the temperature because the ionisation fraction is most sensitive to it. 
HP masses below $10^{-3}$ eV have resonance regions which are indistinguishably close in the picture to the region where $m_\gamma^2(\omega,{\bf x})=0$.  

\begin{figure}[htbp]
\begin{center}
\includegraphics[width=0.45\textwidth]{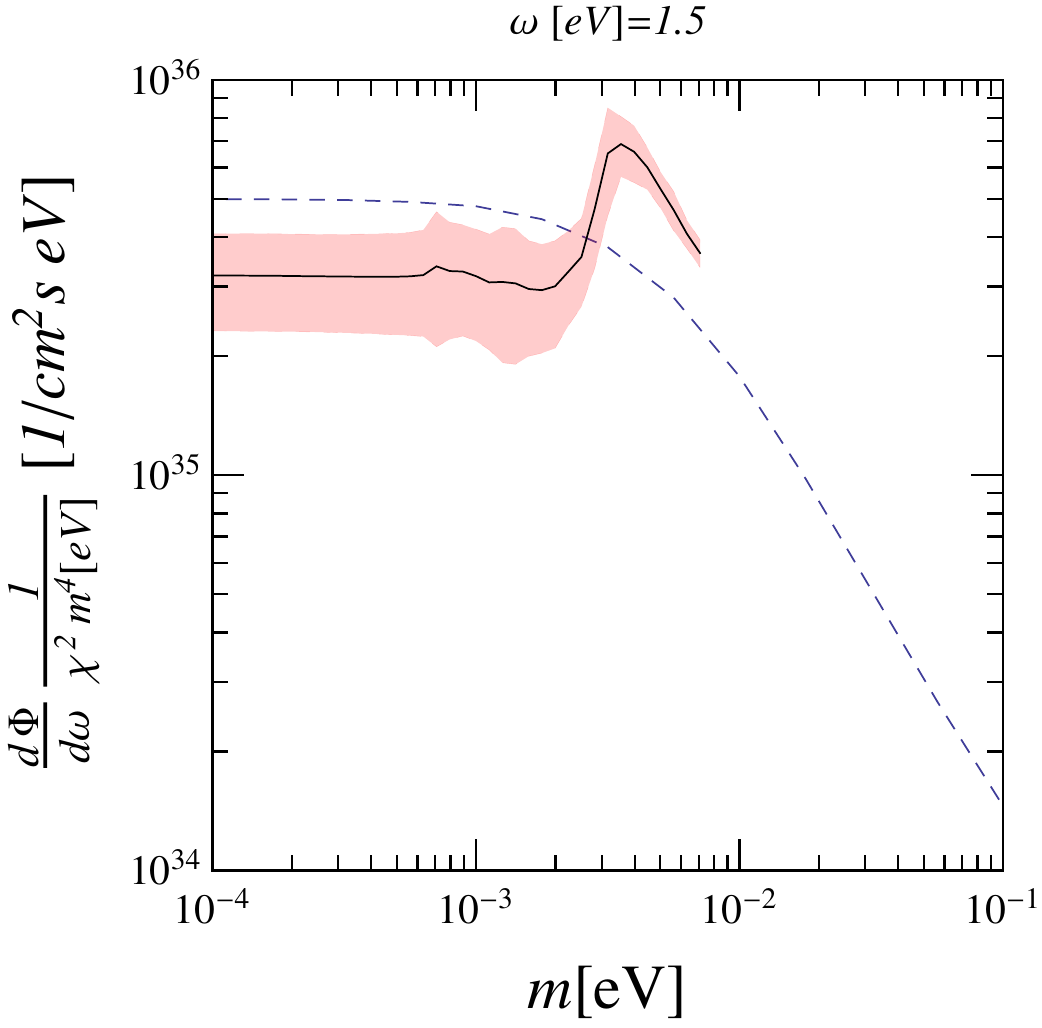}
\includegraphics[width=0.45\textwidth]{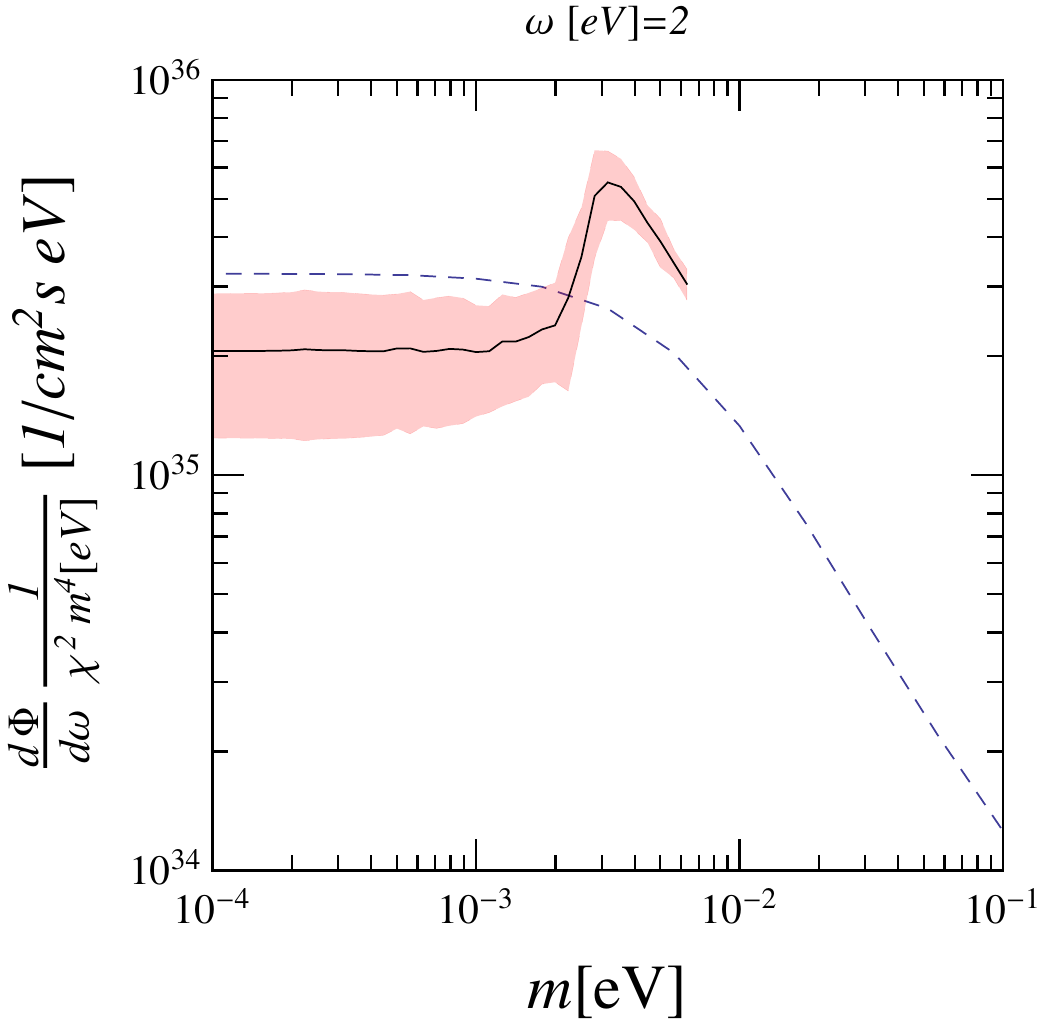}

\includegraphics[width=0.45\textwidth]{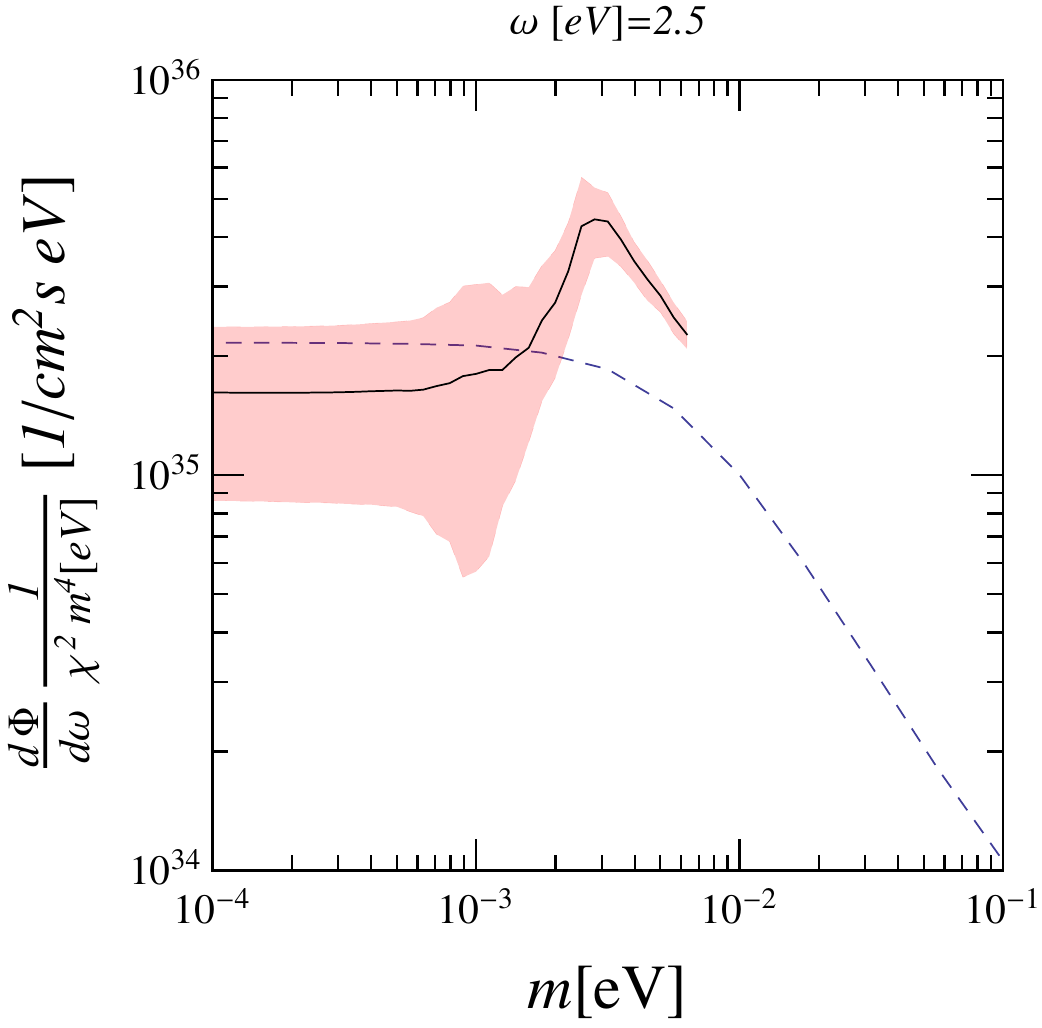}
\includegraphics[width=0.45\textwidth]{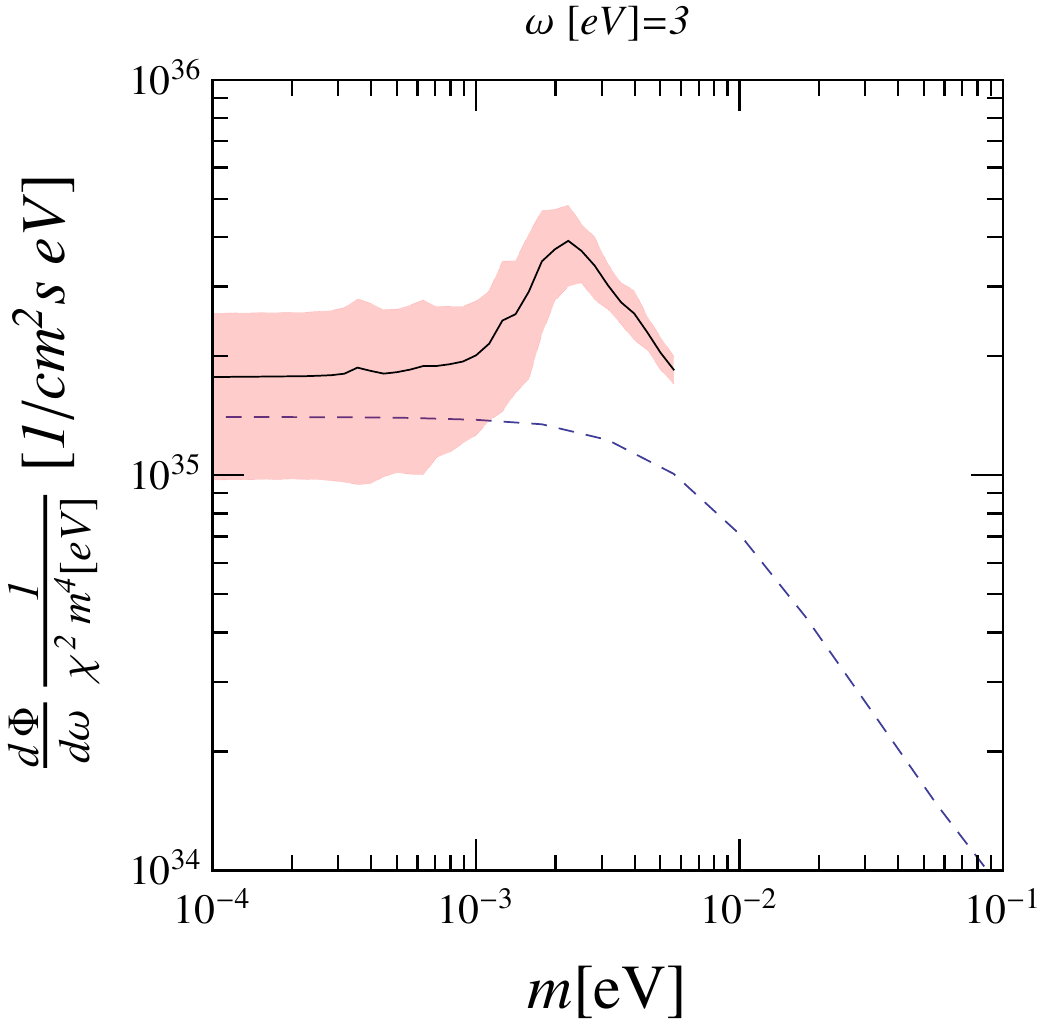}

\includegraphics[width=0.45\textwidth]{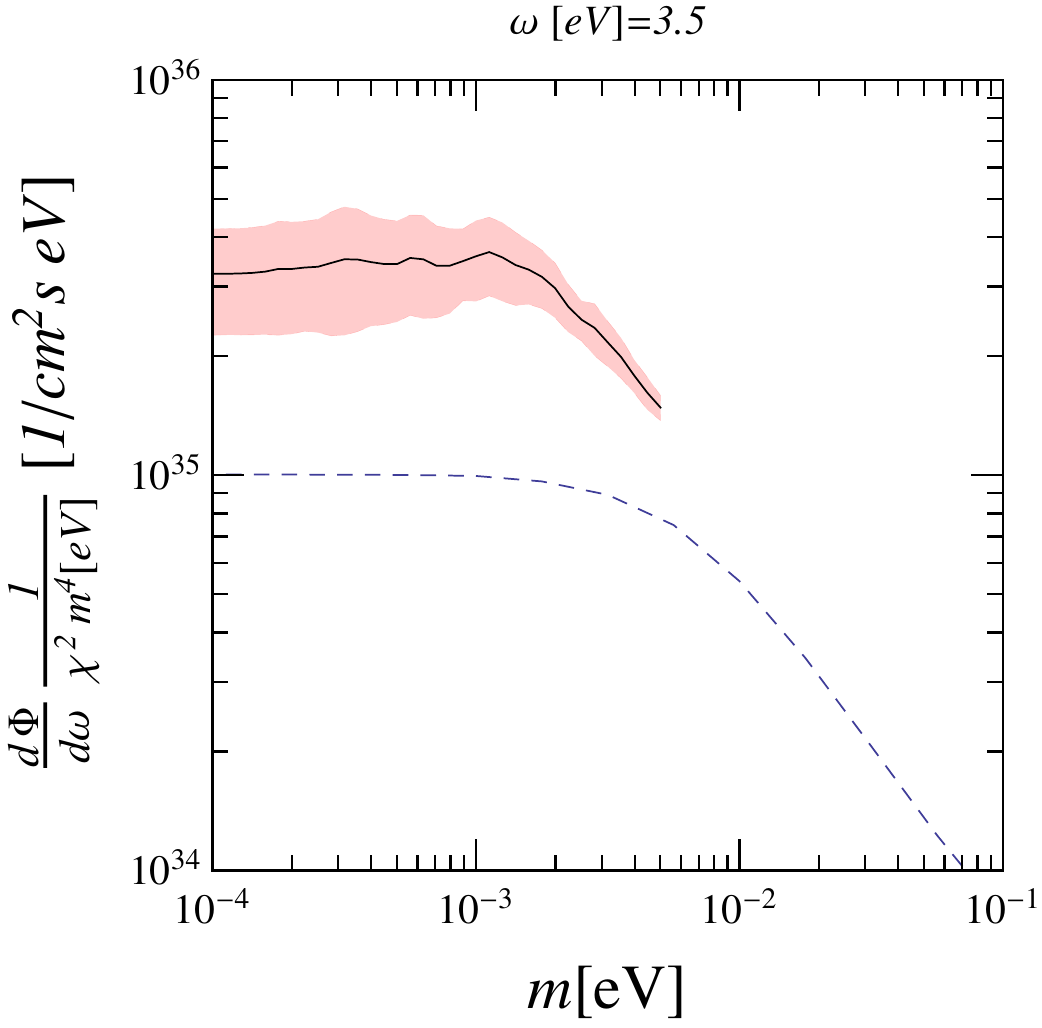}
\caption{Flux of HPs in the 3D surface model of \eqref{eq:resonantflux3D} computed from \eqref{eq:resonantflux3D} (solid lines) compared with the result of the 1D model (dashed)}
\label{fig:3Dflux}
\end{center}
\end{figure}

We have estimated the integral \eqref{eq:resonantflux3D} over a range of HP masses and energies and show our results in Fig. \ref{fig:3Dflux}. 
Due to computational limitations, we have not integrated over the whole 3D model but just over one central 2D slice and then extrapolated the results to 3D. 
The procedure is as follows. For each of the 89 time-steps of the simulation we have a 2D (depth vs surface coordinate $x$) temperature and density map from which we can compute $m^2_\gamma(d,x)$. 
For each HP mass and energy we find the line (or lines) where $m^2_\gamma(d,x)=m^2$. 
If we assume that the properties do not depend on the orthogonal surface coordinate $y$ we would compute the flux simply as 
\be
\frac{d\Phi}{d\omega}\approx\chi^2 m^4\sqrt{\omega^2-m^2}\frac{R^2_{\rm Sun}}{\pi R^2_{\rm Earth}}
\left\langle\sum_{\rm lines}\int_{\rm res}\frac{d l}{X}
\frac{1}{{e^{\omega/T}-1}}\frac{1}{|\nabla m_\gamma^2(d,x)|} 
\right\rangle , 
\ee 
where $X=6$ Mm and the brackets denote a time average 
However, in the real 3D calculation we have to correct for the corrugation in the coordinate $y$ as well as the fact that the gradient will in general be larger than the 2D version. 
In these integrals, the number of points where the 2D-gradient is really low and thus the contribution to the integral is large is relatively scarce. These contributions are mostly suppressed by the third component of the gradient in the 3D case, but since they don't contribute much already we simply neglect them and approximate (on average) the inverse 3D-gradient with the inverse 2D gradient. All in all, we use 
\be
\frac{d\Phi}{d\omega}\approx \chi^2 m^4\sqrt{\omega^2-m^2}\frac{R^2_{\rm Sun}}{\pi R_{\rm Earth}}
\left\langle
\frac{L_t}{X}
\sum_{\rm lines}\int_{\rm res}\frac{d l}{X}
\frac{1}{{e^{\omega/T}-1}}\frac{1}{|\nabla m_\gamma^2(d,x)|}
\right\rangle. 
\ee 
We remark that estimating the 3D flux by a $x-$average over radial profiles (i.e. 1D gradients) leads to very erroneous conclusions. The reason can be understood looking at Fig. \ref{fig:surface}. Around $x\sim 1$ Mm there are regions where $m_\gamma^2$ depends mostly on the depth $x$ but not on $d$, the 1D gradient goes to zero in some cases and the flux would blow up. In the 2D calculation, the resonance width is much more controlled because there comparatively much less points where the 2D gradient goes to zero.

We show the time-averaged flux as a solid line and the $1-\sigma$ contours due to time variability, which shall go to zero when a larger surface is integrated. 
The results are compared with the spherically symmetric estimate, shown as a dashed black line. 
Since the 3D simulation does not extend very deep in the Sun,  we cannot cover the resonance region for the highest masses $m>5\times 10^{-3}$ eV in this study and our results stop around that mass. 
At large depths $d\gtrsim 0.5$ Mm, the inhomogeneities are relatively small, variations shrink and the calculations seem to converge to the $\propto m^3$ trend observed in the $10^{-2}-0.1$ mass range derived before with the 1D averaged atmosphere.  The last points of the 3D data lie typically a factor of $\sim$1.5 above the 1D calculations and we expect that the 3D and 1D lines converge before $m\sim 0.2$ eV because resonances would happen deeper than 1 Mm where temperature and density fluctuations are small and 3D and 1D profiles should agree very well. 

In general, the results agree well with our previous calculation in order of magnitude and trend but there are some O(1) differences. For HP masses in the meV, the flux is larger by a factor of a few and for smaller masses the flux is lower by a similar factor, although with some frequency dependence.  
Actually, there is a very simple interpretation of the differences. 
In general, 3D models of the solar surface have temperature profiles with are shallower in the solar interior and steeper in the outside when compared with the averaged 1D versions, see for instance Fig. 13 of~\cite{nordlund2009} and compare our Fig. \ref{fig:surface} up and down. 
This translates into the average profile for $m_\gamma^2$ near the solar surface. 
Therefore, resonance regions which lie deeper in the Sun are larger and those lying outside are smaller than in the averaged 1D model used and the same translates into the HP fluxes. 
Since decreasing the mass and the frequency, the resonance region gets displaced towards the solar exterior, the HP fluxes are expected to decrease for low masses and low energies and increase for relatively high masses and high energies. This is precisely what we observe in Fig. \ref{fig:3Dflux}.    
The 1D averaged profiles might be very useful for solar modelling but the formula for resonant HP emission \eqref{eq:resonantflux3D} is highly non-linear in the temperature and density and thus it is clear that using them introduces new uncertainties. A precise determination of the solar HP flux below $m\sim 0.2$ eV thus requires a full 3D model of the solar atmosphere. 

Finally, note that the fluxes increase slightly also because, due to corrugation, the resonance region is larger. We estimate this effect to be at most 40\%. 
The corrugation also influences the angular distribution of the flux, discussed in section \ref{sec:angular}.  For HP masses below 0.02 eV, the resonance width $\Delta\psi_*$ gets broadened on average to a value 
$\Delta\psi_*\sim 0.1$Mm/$R_{\rm Earth}\equiv 0.14$ arcseconds. 
This is of the same order than the widening of  $\Delta\psi_*$ that an experiment would see if measuring in the whole visible range at the same time because of the energy dependence of $r_*$, see Fig. \ref{fig:width}. 
 
\clearpage

\section{Summary and conclusions}
\label{sec:conclusions}

In this paper we have reviewed the phenomenon of photon$\leftrightarrow$HP oscillations in a plasma, focusing on a stellar hydrogen-dominated medium. 
We have shown how to build the polarisation tensor (index of refraction) of such a plasma 
and we have performed calculations for a spherically symmetric standard solar model~\cite{saclay} and 
a 3D model of the solar atmosphere~\cite{oai:arXiv.org:0909.0948}. 
The most complete atlas of transverse HP emission to date is presented in Fig.~\ref{fig:atlas}, 
and it is corrected at low energies and HP masses by the 3D calculations shown in Fig.~\ref{fig:3Dflux}. 
Tables with all the results are available and can be obtained by contacting with the author.  

We have studied in depth resonant photon$\leftrightarrow$HP oscillations in the conditions relevant for the solar plasma, including the effects of inhomogeneities. We identified the regions of the Sun where resonant conversions take place integrated the HP production from these regions. 
At low energies and HP masses, this contribution dominates the flux. 
For this task we were helped by two facts that we have also discussed: 1) that resonance regions are very thin with respect to temperature and density scale heights inside of the Sun and 2) the integral of the HP production across a resonance region is independent of the value of the mean-free-path of photons around it, i.e. it is the same for optically thick and thin resonances (as long as the mean free path is non-zero and the temperature is reasonably constant across its width).   
The HP flux from resonance regions only, in the 1D and 3D models is presented in Fig. \ref{fig:fluxes} and Fir. \ref{fig:3Dflux}. 
In general the flux grows towards lower energies, until overcome by the production threshold. 
For very low masses, where the threshold is not of concern, our calculations are limited by ou solar refraction model. Here, resonances happen in the solar atmosphere, where atomic transitions of metals and molecules might be relevant but were not included. 
In the ultraviolet range, the flux is smaller and has a non-trivial energy dependence, cf. Fig. \ref{fig:Lyman}, which we have discussed at length. 
In the visible, the energy dependence is very smooth and the effects of hydrogen lines are small, cf. Fig.~\ref{fig:Halpha}.  
Differences between the 1-D and 3-D models have been found to be substantial, cf.~Fig.\ref{fig:3Dflux}.   Further refined work is necessary to improve the precision here, because our 3-D models did not extend much inside the Sun.  Our calculations are uncertain by a factor of $\sim 2$ below $m\sim 0.01$ eV which decreases up to $20\%$ around $m\sim$ eV. 

We have found the visible flux very similar to previous estimates~\cite{oai:arXiv.org:1202.4932,oai:arXiv.org:1010.4689}, which were however based on shaky grounds. 
The current calculation settles, to the authors knowledge, all the issues left previously undiscussed.   

The computed HP flux at visible energies can be used to search for these particles with helioscopes in regions of parameter space where they might arise in string theories~\cite{Goodsell:2009xc,Cicoli:2011yh} and even constitute the dark matter of the universe~\cite{Arias:2012az}. 
Such a future helioscope should be much more powerful than present-day CAST, SUMICO or SHIPS to be more sensitive than the solar precision constraint~\cite{Vinyoles:2015aba} or the absence of ionisation events in XENON10~\cite{An:2013yua}, both based on the much larger flux of longitudinally polarised HPs from the Sun.  
Due to the relatively modest flux of transversely polarised HPs, the magnitude of such an experiment is considerably big, although perhaps not unrealistic. The International Axion Observatory (IAXO)~\cite{Armengaud:2014gea} is a proposed axion helioscope that seems to fall in the required category, large aperture and low background~\cite{Irastorza:2011gs}, although it mainly aims at the X-ray energies relevant for axions and deploys a huge superconducting toroid, which would not be necessary to search for solar hidden photons.

\section*{Acknowledgements}
The author is gratefully indebted to Pat Scott and Martin Asplund for making available their 3D atmosphere model for the results of this paper. The input and inspiration from illuminating conversations with S. Troiksky, J. Jaeckel and G. Raffelt cannot be underestimated. Finally, it is a real pleasure to acknowledge the help of Davide Cadamuro during the early developments of this project. 
This work was supported by the the Alexander von Humboldt Foundation, Deutsche Forschungsgemeinschaft through grant No. EXC 153, the European Union through the Initial Training Network Invisibles, grant No. PITN-GA-2011-28944 and Spanish ministry of Economy and Competitively (where is the science?) through the Ramon y Cajal fellowship 2012-10597. 

\appendix

\section{Between thin and thick}\label{sec:prueba}

We start with \eqref{eq:Ageneral} and expand $m_\gamma^2$ around the resonance point $m_\gamma^2=m^2+m_\gamma^{2'}(r-r_*)$. Now consider the integral 
\bea
\nonumber
I(\theta) &=& \frac{r_*^2\Gamma_*}{e^{\omega/T_*}-1}\int_{-\infty}^\infty dr P(r,\cos\theta)  \\
&=& \frac{r_*^2\Gamma_*}{e^{\omega/T_*}-1}\int_{-\infty}^\infty dr \left|\int^\infty_r \frac{dr_1}{\cos\theta}\frac{m^2}{2\omega}
e^{i \frac{m_\gamma^{2'}(r_1^2-r^2-2r_*(r_1-r))}{4\omega\cos\theta}-\frac{\Gamma_*(r_1-r)}{2|\cos\theta|}}\right|^2  \\
\nonumber
&=& \frac{r_*^2\Gamma_*}{e^{\omega/T_*}-1}\(\frac{m^2}{2\omega}\)^2
\int_{-\infty}^\infty dr 
\int^\infty_r\int^\infty_r
 \frac{dr_1dr_2}{\cos^2\theta}e^{i \frac{m_\gamma^{2'}(r_1^{2}-r_2^2-2r_*(r_1-r_2))}{4\omega\cos\theta}-\frac{\Gamma_*(r_1+r_2-2r)}{2|\cos\theta|}} \\
\nonumber
&=& \frac{r_*^2\Gamma_*}{e^{\omega/T_*}-1}\(\frac{m^2}{2\omega}\)^2
\int_{-\infty}^\infty dr 
\int^\infty_0\int^\infty_0
 \frac{dx_1dx_2}{\cos^2\theta}e^{i \frac{m_\gamma^{2'}(x_1^{2}-x_2^2-2(r_*-r)(x_1-x_2))}{4\omega\cos\theta}-\frac{\Gamma_*(x_1+x_2)}{2|\cos\theta|}} \\
\nonumber
&=& \frac{r_*^2\Gamma_*}{e^{\omega/T_*}-1}\(\frac{m^2}{2\omega}\)^2
\int^\infty_0\int^\infty_0
 \frac{dx_1dx_2}{\cos^2\theta}e^{i \frac{m_\gamma^{2'}(x_1^{2}-x_2^2)}{4\omega\cos\theta}-\frac{\Gamma_*(x_1+x_2)}{2|\cos\theta|}}2\pi \delta\(\frac{2 m_\gamma^{2'}}{4\omega\cos\theta}(x_1-x_2)\) \\
\nonumber
&=& \frac{2\pi r_*^2\Gamma_*}{e^{\omega/T_*}-1}\(\frac{m^2}{2\omega}\)^2\left|\frac{m_\gamma^{2'}}{2\omega\cos\theta}\right|^{-1}
\int^\infty_0
 \frac{dx_1}{\cos^2\theta}-e^{-\frac{\Gamma_*x_1}{\cos\theta}} \\
\nonumber
&=& \frac{\pi r_*^2}{e^{\omega/T_*}-1}\frac{m^4}{\omega}\left|m_\gamma^{2'}\right|^{-1}. 
\eea

\end{document}